\documentclass[a4paper,11pt]{article}

\usepackage{geometry}
\usepackage[titletoc,page]{appendix}            
\usepackage{afterpage}
\usepackage[nottoc,notlot,notlof]{tocbibind}
\usepackage{pdflscape}                          
\usepackage{dsfont} 
\usepackage{eurosym}                            
\usepackage{xcolor}                             
\usepackage{url}                       			
\usepackage{hyperref}                       	
\usepackage{authblk}                            
\usepackage[english]{babel}
\usepackage{csquotes}                           
\usepackage{multicol}
\usepackage{textcomp}
\usepackage{comment}

\usepackage{amsmath,amssymb,amsthm,amsfonts}    
\usepackage{mathrsfs,bm}                        
\usepackage{mathtools}

\graphicspath{{./Figures/}}                     
\usepackage{float}                              
\usepackage{caption}                  			
\usepackage{subcaption}   
\usepackage{booktabs,multirow,longtable}        
\usepackage[dvipdfmx]{graphicx}
\usepackage{psfrag}                             
\usepackage{epstopdf}                           
\usepackage{epsfig}                             
\usepackage{tikz}                               
\usetikzlibrary{shapes,arrows}
\usetikzlibrary{arrows.meta}
\usepackage{algorithm,algorithmic}              

\begin{document}
%

\newcommand{\ie}{i.e.~} 	                    
\newcommand{\eg}{e.g.~} 	                    
\newcommand{\rhs}{r.h.s.~} 	                    
\newcommand{\wrt}{w.r.t.~} 	                    
\newcommand{\old}[1]{\textcolor{blue}{#1}}      
\newcommand{\new}[1]{\textcolor{red}{#1}}       
\newcommand{\textbox}[1]{\mbox{\textit{#1}}} 

\newtheorem{theorem}{Theorem}[section]
\newtheorem{definition}[theorem]{Definition}
\newtheorem{definitions}[theorem]{Definition}
\newtheorem{proposition}[theorem]{Proposition}
\newtheorem{remark}[theorem]{Remark}
\newtheorem{corollary}[theorem]{Corollary}
\newtheorem{example}[theorem]{Example}
\newtheorem{lemma}[theorem]{Lemma}

\newenvironment{system}{\left\lbrace\begin{array}{@{}l@{}}}{\end{array}\right.}

\newcommand{\Parenthesis}[1]{\left( #1 \right)}         
\newcommand{\Brack}[1]{\left\lbrack #1 \right\rbrack}   
\newcommand{\Brace}[1]{\left\lbrace #1 \right\rbrace}   
\newcommand{\Abs}[1]{\left\lvert #1 \right\rvert}       
\newcommand{\Norm}[1]{\left\lVert #1 \right\rVert}      
\newcommand{\Mean}[1]{\left\langle #1 \right\rangle}    
\newcommand{\QuoteDouble}[1]{``#1''}                    
\newcommand{\QuoteSingle}[1]{`#1'}                      

\newcommand{\beq}{\begin{equation}}         
\newcommand{\eeq}{\end{equation}}           
\newcommand{\Max}{\mbox{Max}}                                   
\newcommand{\Min}{\mbox{Min}}                                   
\newcommand{\Derp}[2]{\frac{\partial #1}{\partial #2}}          
\newcommand{\DerpXX}[2]{\frac{\partial^2 #1}{\partial #2 ^2}}   
\newcommand{\DerpXY}[3]{\frac{\partial^2 #1}{\partial #2 \partial #3}}    
\newcommand{\half}[0]{\frac{1}{2}}                                                  
\newcommand{\ind}[1]{\mathbf{1}_{#1}} 	                                            
\newcommand{\E}[1]{\mathbb{E}\left[#1\right]}                                       
\newcommand{\Econd}[2]{\mathbb{E}\left[#1\;\middle \vert\;\mathcal{F}_{#2}\right]}  
\newcommand{\argminA}{arg\,min}       		

\newcommand{\Nset}{\mathbb{N}}      
\newcommand{\Zset}{\mathbb{Z}}      
\newcommand{\Qset}{\mathbb{Q}}      
\newcommand{\Rset}{\mathbb{R}}      

\newcommand{\Black}{\mbox{Black}}				    

\newcommand{\Depo}{\mbox{\textbf{Depo}}}			
\newcommand{\FRA}{\mbox{\textbf{FRA}}}			    
\newcommand{\Futures}{\mbox{\textbf{Futures}}}      
\newcommand{\ZCB}{\mbox{ZCB}}					    
\newcommand{\Swap}{\mbox{\textbf{Swap}}}			
\newcommand{\Swaplet}{\mbox{\textbf{Swaplet}}}		
\newcommand{\IRS}{\mbox{\textbf{IRS}}}			    
\newcommand{\IRSlet}{\mbox{\textbf{IRSlet}}}        
\newcommand{\OIS}{\mbox{\textbf{OIS}}}              
\newcommand{\OISlet}{\mbox{\textbf{OISlet}}}        
\newcommand{\BSwap}{\mbox{\textbf{BSwap}}}          
\newcommand{\BSwaplet}{\mbox{\textbf{BSwaplet}}}    
\newcommand{\IRBS}{\mbox{\textbf{IRBS}}}            
\newcommand{\IRBSlet}{\mbox{\textbf{IRBSlet}}}      
\newcommand{\CCS}{\mbox{\textbf{CCS}}}              
\newcommand{\CCSwap}{\mbox{\textbf{CCSwap}}}        
\newcommand{\CCSwaplet}{\mbox{\textbf{CCSwaplet}}}  
\newcommand{\CCSlet}{\mbox{\textbf{CCSlet}}}        
\newcommand{\Caplet}{\mbox{\textbf{Caplet}}}        
\newcommand{\Floorlet}{\mbox{\textbf{Floorlet}}}    
\newcommand{\cf}{\mbox{\textbf{cf}}}                
\newcommand{\CAP}{\mbox{\textbf{Cap}}}              
\newcommand{\Floor}{\mbox{\textbf{Floor}}}          
\newcommand{\CF}{\mbox{\textbf{CF}}}                
\newcommand{\Swaption}{\mbox{\textbf{Swaption}}}    
\newcommand{\CMSlet}{\mbox{\textbf{CMSlet}}}        
\newcommand{\CMS}{\mbox{\textbf{CMS}}}              
\newcommand{\CMScf}{\mbox{\textbf{CMScf}}}          
\newcommand{\FXFwd}{\mbox{\textbf{FXFwd}}}          

\title{Everything You Always Wanted to Know About XVA Model Risk but Were Afraid to Ask}

\author[1]{Lorenzo Silotto}
\author[1]{Marco Scaringi}
\author[1,2,*]{Marco Bianchetti}

\affil[1]{Financial \& Market Risk Management, Intesa Sanpaolo, Milan, Italy}
\affil[2]{Department of Statistical Sciences \enquote{Paolo Fortunati}, University of Bologna, Italy}
\affil[*]{Corresponding author, \texttt{marco.bianchetti@unibo.it}}

\date{14 June 2023}

\maketitle

\begin{abstract}
	Valuation adjustments, collectively named XVA, play an important role in modern derivatives pricing to take into account additional price components such as counterparty and funding risk premia. They are an exotic price component carrying a significant model risk and computational effort even for vanilla trades. 
	\par
	We adopt an industry-standard realistic and complete XVA modelling framework, typically used by XVA trading desks, based on multi-curve time-dependent volatility G2++ stochastic dynamics calibrated on real market data, and a multi-step Monte Carlo simulation including both variation and initial margins. We apply this framework to the most common linear and non-linear interest rates derivatives, also comparing the MC results with XVA analytical formulas.
	\par
	Within this framework, we identify the most relevant model risk sources affecting the precision of XVA figures and we measure the corresponding computational effort. In particular, we show how to build a parsimonious and efficient MC time simulation grid able to capture the spikes arising in collateralized exposure during the margin period of risk. 
	As a consequence, we also show how to tune accuracy vs performance, leading to sufficiently robust XVA figures in a reasonable time, a very important feature for practical applications. 
	Furthermore, we provide a quantification of the XVA model risk stemming from the existence of a range of different parameterizations according to the EU prudent valuation regulation.
	\par
	Finally, this work also serves as an handbook containing step-by-step instructions for the implementation of a complete, realistic and robust modelling framework of collateralized exposure and XVA.
\end{abstract}

\newpage

\tableofcontents

\vspace{2cm} 

\vspace{0.5cm}

\vspace{0.5cm}
\noindent \textbf{Acknowledgements}: the authors contributed equally to this work. The authors acknowledge fruitful discussions with many colleagues in Intesa Sanpaolo Risk Management and Front Office Departments. A. Principe and M. Terraneo collaborated to the early stage of this work. 

\vspace{0.5cm}
\noindent \textbf{Disclaimer}: the views expressed here are those of the authors and do not represent the opinions of their employers. They are not responsible for any use that may be made of these contents.

\vspace{0.5cm}
\noindent \textbf{Availability}: this paper was published as Silotto, L., Scaringi, M. Bianchetti, M. XVA modelling: validation, performance and model risk management. Annals of Operations Research 336, 183–274 (2024). \url{https://doi.org/10.1007/s10479-023-05323-4}. The Matlab code is available at \url{https://github.com/Lsilotto}.

\newpage

\section{Introduction} 
\label{sec:introduction}  

The credit crunch crisis started in August 2007 forced market practitioners and academics to review the methodologies used to price over-the-counter (OTC) derivatives consistently with the available market quotations. In particular, basis spreads between interest rate instruments characterised by different underlying rate tenors (e.g.~IBOR 3M, IBOR 6M, overnight\footnote{IBOR denotes a generic Interbank Offered Rate, such as EURIBOR. 3M, 6M, etc. denote the rate tenor, i.e. the time period used to compute the interest amount. Overnight rates have a one-day tenor.}, etc.) exploded from few to hundreds of basis points. 
This change of regime led to the adoption of a ``multi-curve'' valuation  framework, based on distinct yield curves to compute forward rates with different tenors\footnote{i.e. yield curves built from market quotations of homogeneous interest rate instruments with the same underlying rate tenor.}, and discounting curves consistent with the collateral remuneration rate\footnote{Collateral agreements are used to mitigate the counterparty default risk of derivatives transactions. They are typically based on the Credit Support Annex (CSA), a section of the International Swaps and Derivatives Association (ISDA) master agreement used to contractualise OTC derivatives. Collateral remuneration rates are typically overnight rates. Discounting curves are typically built from Overnight Indexed Swaps, based on (compounded) overnight rates. Hence the names ``CSA discounting'' or ``OIS discounting''. This post-crisis valuation framework is different from the previous ``single-curve framework'', characterized by a single yield curve, used to compute both forward and discount rates, built from inhomogeneous market instruments with mixed tenors, e.g. Deposits, Futures, Forward Rate Agreements, Swaps, etc.}.
The multi-curve framework is extensively discussed in the literature, a non-exhaustive list of references includes \cite{Hen07,AmeBia09,Mer09,Hen09b,Bia10,Ken10a,Pit12}. 
Such framework was recently simplified by the interest benchmark reform and the progressive replacement of many IBOR rates with the corresponding overnight rates. We refer to e.g. \cite{ScaBia20} and references therein for a discussion.  
\par 
The credit crunch crisis also forced market participants to extend the multi-curve valuation framework to include additional risk factors, such as counterparty and funding risk, leading to a set of valuation adjustments collectively named XVA, for which we refer to the wide existing literature (see e.g.\cite{BriMor13,Kja18,Gre20}). In particular, Credit and Debt Valuation Adjustments (CVA and DVA, respectively) take into account the bilateral counterparty default risk premium affecting derivative transactions, and are also required by international accounting standards (in EU since the introduction of IFRS13 in 2013, see \cite{IFRS11}). 
\par 
After the crisis, regulators pushed to mitigate counterparty default risk for OTC derivatives. With regard to non-cleared OTC derivatives, in 2015 the Basel Committee on Banking Supervision (BCBS) and the International Organization of Securities Commissions (IOSCO) finalized a framework (\cite{BCBSIOSCO15}), introduced progressively from 2016, which requires derivatives counterparties to bilaterally post Variation Margin (VM) and Initial Margin (IM) on a daily basis at netting set level. 
VM aims at covering the current exposure stemming from changes in the value of the portfolio by reflecting its current size, while IM aims at covering the potential future exposure that could arise, in the event of default of the counterparty, from changes in the value of the portfolio in the period between last VM exchange and the close-out of the position. 
In particular, in 2016 ISDA published the Standard Initial Margin Model (SIMM) (see \cite{ISDA13,ISDA18}), with the aim to provide market participant with a uniform risk-sensitive IM model, and to prevent both potential disputes and the overestimation of IM requirements due to the use of the BCBS-IOSCO non-risk-sensitive standardized model. The ISDA-SIMM is a parametric VaR model based on Delta, Vega and Curvature (i.e.~``pseudo'' Gamma) sensitivities, defined across risk factors by asset class, tenor and expiry, and computed according to specific definitions. 
\par 
In general, XVA pricing is subject to a significant \textbf{model risk}, since it depends on the many assumptions made for modelling and calculating the relevant quantities. Since computational constraints impose to reduce the number of floating-point calculations, model risk arises principally from the need to find an acceptable compromise between accuracy of the XVA figures and computational performance. 
In the EU pricing model risk is envisaged in \cite{EUParlPrd13} (art. 105.10), and \cite{EUComPrd16} (art. 11), and refers precisely to the valuation uncertainty of fair-valued positions linked to the ``\emph{potential existence of a range of different models or model calibrations used by market participants}''. 
This may occur when a unique model recognized as a clear market standard for computing the price of a certain financial instrument does not exist, or when a model allows for different parameterizations or numerical solution algorithms leading to different model prices. We stress that this measure of model risk does not refer to the universe of possible pricing models and model parameterizations, which is virtually illimited, but, on the contrary, it does intentionally focus the range to those alternatives effectively used by market participants. 
While the pricing models and their parameterizations used by market participants are not, in general, easily observable, the situation for XVA pricing is slightly different, since market participants may occasionally infer some information from the XVA prices observed in the case of competitive corporate auctions, novations\footnote{A novation occurs when a bank A is called to step in an existing trade between another bank B and a client C. Since typically the two banks A and B have a collateral agreement, while the client trade is not collateralized, the XVA exit price is observed in the transaction.}, negotations of collateral agreements, and also systematically from consensus pricing services (e.g. Totem).
\par
In light of the considerations above, our paper is intended to answer to the following three interconnected \textbf{Research Questions}:  
\begin{itemize}  
\item[Q1] which are the most critical model risk factors, to which exposure modelling and thus XVA are most sensitive?   
\item[Q2] How to set the XVA calculation parameters in order to achieve an acceptable compromise between accuracy and performance?  
\item[Q3] How to quantify the model risk affecting XVA figures and estimate the AVA Model Risk required by the regulatio?  
\end{itemize}   
\par 
We address these questions as follows. We identify all the relevant calculation parameters involved in the Monte Carlo simulation used to compute the exposure and the XVA figures under different collateralization schemes. For each parameter we quantify its relevance in terms of impacts on XVA figures and computational effort required. 
Putting all these results together, we may identify the parametrization which allows a compromise between accuracy and performance, i.e. leading to sufficiently robust XVA figures in a reasonable time, a very important feature for practical applications. 
As a consequence, we are also able to provide a quantification of the XVA model risk stemming from the existence of a range of different pricing model calibrations, numerical methods and their related parameterizations according to the EU provisions.  
\par 
To these purposes we adopt an industry-standard realistic and complete modelling framework, typically adopted by XVA trading desks, including both VM and ISDA-SIMM dynamic IM, based on real market data, i.e.~distinct discounting and forwarding yield curves, CDS spread curves, and swaption volatility cube. 
We apply this framework to the most diffused derivative financial instruments, i.e. interest rate Swaps and physically settled European Swaptions, both with different maturities and moneyness. 
The stochastic dynamics of the underlying risk factors is modelled with a multi-curve two-factors G2++ short rate model with time-dependent volatility, which, including 19 parameters (see tab. \ref{tab:calibrations_parameters}), allows a richer yield curve dynamics and a better calibration of the market swaption cube. 
Our XVA numerical implementation is based on a multi-step Monte Carlo simulation with nested exposures calculated by means of analytical formulae for Swaps and semi-analytical formulas for Swaptions. Since the G2++ model allows for an analytical expression for the transition probability under the $T$-forward measure, we may use a parsimonious time simulation grid\footnote{Our Monte Carlo simulation framework does not depend neither on the specific stochastic dynamics of the risk factors nor on the length of the time simulation steps, and could be used with more complex stochastic dynamics requiring short time simulation steps. Obviously, the corresponding pricing formulas should be plugged in the framework.} able to capture the spikes arising in collateralized exposure during the margin period of risk.  
The Monte Carlo XVA figures are also compared against the results obtained through analytical XVA formulas available for Swaps. The latter require the valuation of a strip of co-terminal European Swaptions for which we used both the G2++ and the SABR models, calibrated to the same market swaption cube.
\par 
Regarding collateral modelling, since VM and IM determine important mitigations of XVA figures, it is crucial to correctly model their dynamics taking into account the most important collateral parameters, i.e. the margin threshold, the minimum transfer amount, and the margin period of risk. 
On the one hand, extensive literature exists regarding dynamic VM modelling (see e.g.~\cite{BriMor13,Bri18,BriFraPal19}). 
On the other hand, dynamic IM modelling of the ISDA-SIMM involves the simulation of several forward sensitivities and their aggregation according to a set of predefined rules, imposing difficult implementation and computational challenges. 
Different methods have been proposed to overcome such challenges: approximations based on normal distribution assumptions (see \cite{Gre16,AndPyk17}), approximated pricing formulas to speed up the calculation (see e.g.~\cite{ZenRui18,MarPal21}), adjoint algorithmic differentiation (AAD) for fast sensitivities calculation (see e.g.~\cite{CapGil12,HugSav20}) and regression techniques (see e.g. \cite{AnfAzi17,CasGli17,CreDix20}. We focus on the implementation of ISDA-SIMM avoiding as much as possible any approximation, computing forward sensitivities by a classic finite-difference (``bump-and-run'') approach which, although computationally intensive, may be used when performances are not critical.
\par 
The choice of the risk factors dynamics is a crucial aspect and should be based on a careful balance between the model sophistication and the corresponding unavoidable calibration and computational constraints. 
Even though our G2++ model model does not embed advanced features like stochastic volatility (see \cite{BorBri18}), stochastic basis (see \cite{KonMcc19}) or stochastic credit process (see \cite{GlaXu14}), it is commonly preferred by financial institutions because of several reasons, as extensively argued in \cite{Gre15} (secs. 16.1.3, 16.3, 19.1.2) and \cite{Gre20} (sec. 15.4.2), which we summarize here: 
i) simpler models, like the G2++ adopted in this work, allow (semi-)analytical pricing formulas for the most diffused plain vanilla instruments, like Swaps and Swaptions; 
ii) more sophisticated models typically do not allow for (semi)-analytical formulas for transition probabilities and/or for the price of plain vanilla instruments, results much more computationally demanding and easily become unsustainable;
iii) more sophisticated models introduce additional model parameters which are typically more difficult to calibrate to market data, particularly if one takes into account the complex covariance structure associated to multiple stochastic risk factors; 
iv) since we deal with trades under collateral, which reduces and possibly neutralize the corresponding exposures, the sophistication of the stochastic dynamics chosen for risk factors simulation is dominated in importance by the modelling choices adopted for the collateral dynamics, which we extensively discuss in this work; 
v) in particular, for interest rates derivatives, the adoption of a stochastic basis between discounting and forward rates is not crucial since the sensitivity w.r.t. discounting rates is much smaller, and, historically, the volatility of the basis is typically much smaller that the volatility of the corresponding rates. A stochastic basis would play a role only in the case of basis swaps, as discussed in \cite{KonMcc19}, which are typically traded on the OTC interbank market for hedging purposes. Moreover, the financial benchmark reform and the IBOR cessation reduced the importance of such kind of instruments (only EURIBOR single-currency basis swaps survived). 
\par 
Finally, it’s worth to notice that, while the multi-curve single-factor G1++ model is commonly used, and may also be found in commercial software packages, the multi-curve two-factors G2++ model with time-dependent volatility parameters used in this work is less straightforward and, to the best of our knowledge, less diffused. Because of this reason, we report all the relevant G2++ equations in app. \ref{app:Multi-curve G2++ model}. As a consequence, this work also serves as a handbook containing step-by step instructions for the implementation of a complete, realistic and robust modelling framework of collateralized exposure and XVA.  
\par 
The aforementioned considerations led us to the choice of our G2++ framework, in order to provide an XVA model risk investigation based on a realistic XVA pricing architecture, typically adopted by XVA trading desks, and consistent with the prescriptions of the EU prudent valuation framework. 
\par 
This paper is organized as follows. In sec.~\ref{sec:XVA Pricing Framework} we briefly remind the XVA framework and the numerical steps involved in the calculation. In sec.~\ref{sec:XVA numerical calculations} we show the results both in terms of counterparty exposure and XVA figures for the selected financial instruments and collateralization schemes using the target parameterization of the framework, allowing an acceptable compromise between accuracy and performance. In sec.~\ref{sec:XVA model validation} we report the analyses conducted on model parameters in order to answer the research questions n. 1 and n. 2 above. In sec.~\ref{sec:Model Risk} we describe the calculation of the AVA Model Risk (MoRi), answering to research question n. 3 above. In sec.~\ref{sec:conclusions} we draw the conclusions.  Finally, the four apps. \ref{app:theoretical framework} to \ref{app:Market Data} reports many details related to the corresponding main sections.

\section{XVA Pricing Framework}
\label{sec:XVA Pricing Framework}
The framework required for XVA calculation including Variation and Initial Margins is a complex combination of many theoretical and numerical approaches that we summarize in the following list.
\begin{enumerate}
\item The general no-arbitrage pricing formulas for financial instruments subject to XVA, discussed in app. \ref{app:pricing approach}.

\item The description of the financial instruments that we wish to test in our XVA calculations, i.e. interest rate Swaps and European Swaptions, discussed in app. \ref{app:financial instruments and pricing formulas}.

\item The G2++ model adopted to describe the evolution of forward and discount curves, discussed in app. \ref{app:Multi-curve G2++ model}, including: i) the multi-curve, time-dependent volatility G2++ stochastic dynamics (app. \ref{app:short rate dynamics}), ii) the corresponding G2++ pricing formulas for Swaps and European Swaptions (app. \ref{app:G2++pricing}), iii) the calibration procedure of G2++ model parameters to the available market data (app. \ref{app:G2++ calibration procedure}), and iv) the G2++ dynamics under the forward measure, suitable for efficient Monte Carlo simulation (app. \ref{app:G2++ MC simulation}). 

\item The model adopted to describe the XVA, discussed in app. \ref{app:XVA pricing}, including: i) the XVA definition and pricing formulas (app. \ref{app:XVA formulas}), ii) the discretized XVA formulas suitable for Monte Carlo simulation (app. \ref{app:XVA numerical formulas}), and iii) the analytical XVA formulas applicable to single, uncollateralized linear derivatives (app. \ref{app:XVA analytical formulas}).

\item The model adopted to describe the collateral dynamical evolution, discussed in app. \ref{app:collateral modelling}, including: i) the formulas for the collateralised exposure with both VM and IM (app. \ref{app:collateral management}), ii) the formulas to dynamically compute VM (app. \ref{app:collateral VM}), and iii) the formulas to dynamically compute IM (app. \ref{app:collateral IM}) according to the ISDA Standard Initial Margin Model formulas (app. \ref{app:SIMM Formulas}).

\item The market data set used to calibrate the G2++ model parameters and to compute the XVA, discussed in app. \ref{app:Market Data}.
\end{enumerate}

The framework described above requires a precise sequence of calculation steps to compute XVA for the selected instruments, that can be summarized as follows.

\begin{enumerate}
\item Calibration of the G2++ model parameters to market data (see sec.~\ref{app:G2++ calibration procedure}).

\item Construction of the parsimonious time grid $\left\{t_i\right\}_{i=0}^N$ for MC simulation, which includes both the \emph{primary time grid} $\left\{\bar{t}_i\right\}$ and the \emph{collateral time grid} $\left\{\hat{t}_i\right\}$ (see sec. \ref{sec:Parsimonious Time Grid}).

\item Simulation of the processes $[x_m(t_i), y_m(t_i)]$ for each time step $t_i, i=1,\cdots,N$ and Monte Carlo path $m = 1,\dots,N_{MC}$ according to the G2++ dynamics under the forward measure (see app.~\ref{app:G2++ MC simulation}). 

\item Calculation of the collateralized exposure $H_m(t_i)$ for each time step $t_i$ and simulated path $m$ (see eq.~\ref{eq:exposure vm_im}). This requires to:
	\begin{itemize}
		
	\item[a.] compute the instrument's future mark-to-market values $V_{0,m}(\bar{t}_i)$ at time $\bar{t}_i$ on the primary time grid using the G2++ pricing formulas (see app.~\ref{app:G2++pricing});
	
	\item[b.] compute the Variation Margin $\text{VM}_m(\bar{t}_i)$ available at time $\bar{t}_i$ on the primary time grid, which is a function of $V_{0,m}(\hat{t}_i)$ at the previous time $\hat{t}_i = \bar{t}_i-l$ on the collateral time grid (see eq.~\ref{eq:vm});
	
	\item[c.] compute the ISDA-SIMM dynamic Initial Margin $\text{IM}_m(\bar{t}_i)$ available at time $\bar{t}_i$ on the primary time grid, which is function of instrument Delta $\Delta_m^c(\hat{t}_i)$ and Vega $\nu_m(\hat{t}_i)$ sensitivities at the previous time $\hat{t}_i = \bar{t}_i-l$ on the collateral time grid (see eq.~\ref{eq:im}).
	
	\end{itemize}

\item Calculation of EPE $\mathcal{H}^{+} (t;t_i)$ and ENE $\mathcal{H}^{-} (t;t_i)$ for each time step $t_i$ (see eqs.~\ref{eq:epe_ene} and \ref{eq:epe_ene_MC}).

\item Calculation of survival probabilities for each time step $t_i$ from default curves built from market CDS quotes.

\item Calculation of CVA and DVA (see eqs.~\ref{eq:CVA_disc} and \ref{eq:DVA_disc}). 
\end{enumerate}

As discussed in the introduction, our multi-curve, time-dependent volatility G2++ model has a number of important characteristics for XVA calculation: i) it allows perfect calibration of the market term structures of both discount and forward curves, ii) it allows a better calibration of the market term structure of volatility, iii) it allows a volatility skew that can be fitted, at least partially, to the market volatility skews, iv) it allows, under the forward probability measure, an efficient Monte Carlo simulation, iv) it allows analytical pricing formulas for European Options (Caps/Floors/Swaptions).

\section{XVA Numerical Calculations}
\label{sec:XVA numerical calculations}

In this section we report the results for exposure profiles and XVA figures for the financial instruments and the collateralization schemes considered in this work. In order to do so, we use the XVA pricing framework discussed in the previous sec. \ref{sec:XVA Pricing Framework}, and the set of parameters meeting our acceptable compromise between accuracy and performance, discussed in the next sec. \ref{sec:XVA model validation} and summarized in sec.~\ref{sec:accuracy vs performance}.
\par 
We consider the most diffused derivative instrument, i.e. interest rate Swaps, which are typically traded in at least two very common situations: i) between banks and their corporate clients, frequently without collateralization, and ii) between banks for hedging purposes, collateralized with variation margin and (frequently) with initial margin. We include both spot and forward starting Swaps with different maturities and moneyness. 
Besides linear derivatives, we also consider another common interest rate option, i.e. the physically settled European Swaption, with different moneyness. 
All the instruments considered are listed in app. \ref{app:financial instruments} (tab.~\ref{tab:instruments}).
Accordingly, we consider three different collateralization schemes: without collateral, with Variation Margin (VM) only, and with both VM and Initial Margin (IM). The case with IM only is not analyzed since IM is typically associated to VM.
\par 
Overall, we consider 36 different cases, as shown in the following section reporting the XVA figures (tab. \ref{tab:XVA_results}).

\subsection{Exposure Results}
\label{sec:exposure results}

\begin{figure}[H]

	\begin{subfigure}[t]{0.3\textwidth}
		\caption{15Y Swap OTM.}
    		\includegraphics[width=\linewidth]{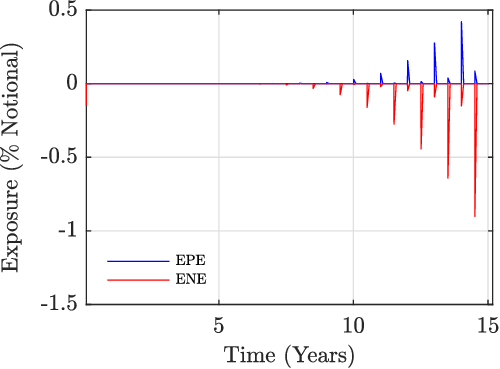}
		\label{fig:irs_15Y_OTM_VM_IM}
	\end{subfigure}
	\hfill
	\begin{subfigure}[t]{0.3\textwidth}
		\caption{15Y Swap ATM.}
  		\includegraphics[width=\linewidth]{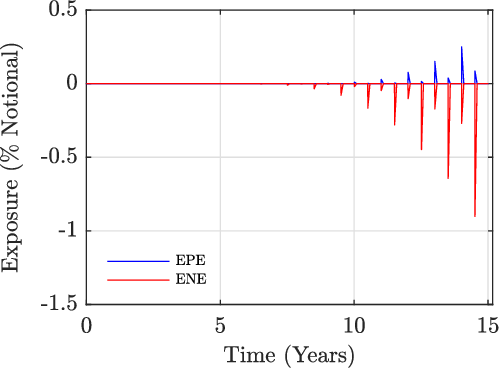}
		\label{fig:irs_15Y_ATM_VM_IM}
	\end{subfigure}
	\hfill
	\begin{subfigure}[t]{0.3\textwidth}
		\caption{15Y Swap ITM.}
    		\includegraphics[width=\linewidth]{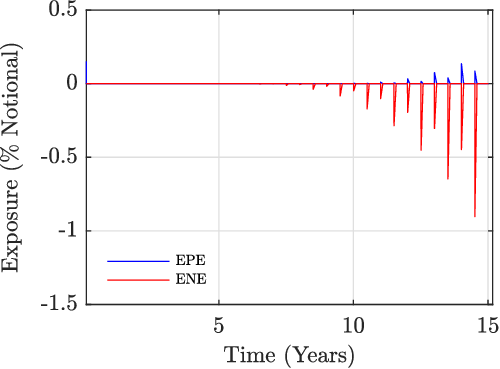}
		\label{fig:irs_15Y_ITM_VM_IM}
	\end{subfigure}
\\
	\begin{subfigure}[t]{0.3\textwidth}
		\caption{30Y Swap OTM.}
    		\includegraphics[width=\linewidth]{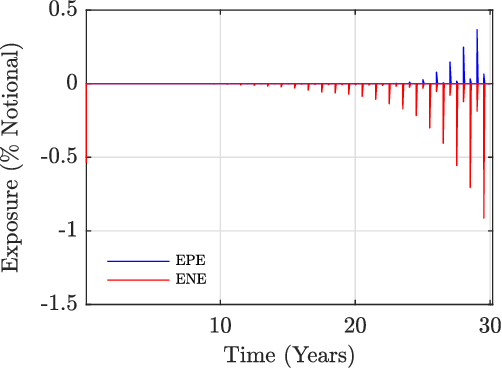}
		\label{fig:irs_30Y_OTM_VM_IM}
	\end{subfigure}
	\hfill
	\begin{subfigure}[t]{0.3\textwidth}
		\caption{30Y Swap ATM.}
  		\includegraphics[width=\linewidth]{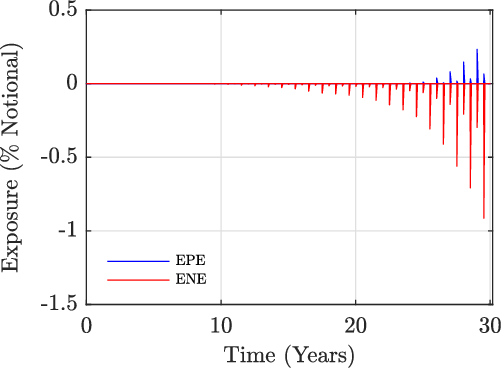}
		\label{fig:irs_30Y_ATM_VM_IM}
	\end{subfigure}
	\hfill
	\begin{subfigure}[t]{0.3\textwidth}
		\caption{30Y Swap ITM.}
    		\includegraphics[width=\linewidth]{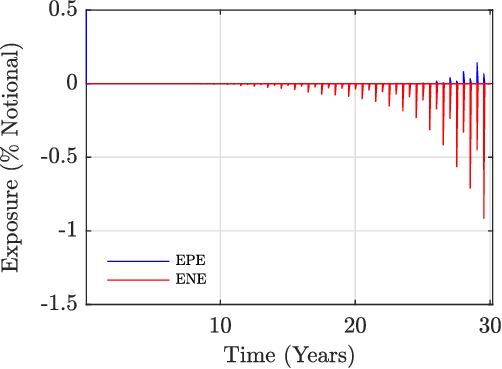}
		\label{fig:irs_30Y_ITM_VM_IM}
	\end{subfigure}
\\
	\begin{subfigure}[t]{0.3\textwidth}
		\caption{5x10Y Fwd Swap (R) OTM.}
    		\includegraphics[width=\linewidth]{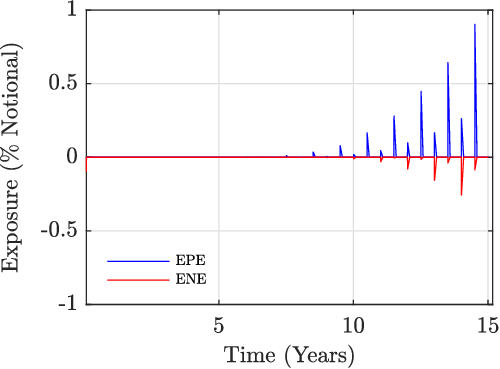}
		\label{fig:irs_fwd_receiver_OTM_VM_IM}
	\end{subfigure}
	\hfill
	\begin{subfigure}[t]{0.3\textwidth}
		\caption{5x10Y Fwd Swap (P) ATM.}
  		\includegraphics[width=\linewidth]{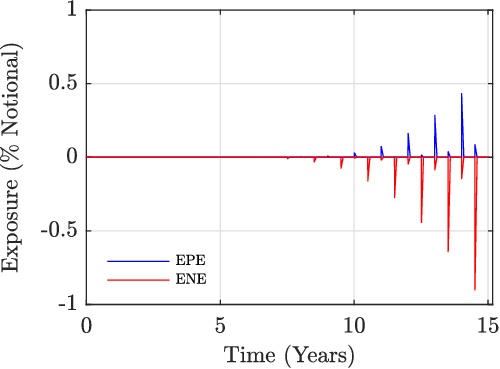}
		\label{fig:irs_fwd_ATM_VM_IM}
	\end{subfigure}
	\hfill
	\begin{subfigure}[t]{0.3\textwidth}
		\caption{5x10Y Fwd Swap (P) OTM.}
   		 \includegraphics[width=\linewidth]{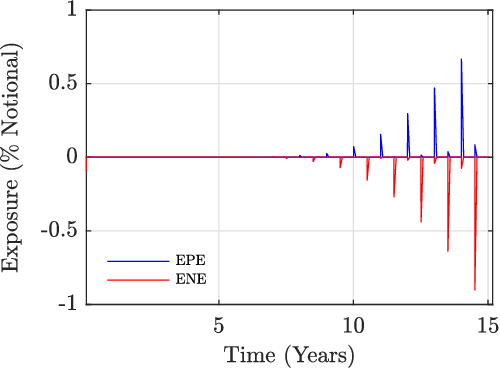}
		\label{fig:irs_fwd_payer_OTM_VM_IM}
	\end{subfigure}
\\
	\begin{subfigure}[t]{0.3\textwidth}
		\caption{5x10Y Swaption (R) OTM.}
    		\includegraphics[width=\linewidth]{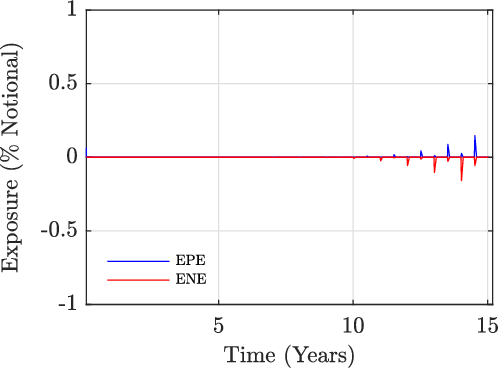}
		\label{fig:swpt_receiver_OTM_VM_IM}
	\end{subfigure}
	\hfill
	\begin{subfigure}[t]{0.3\textwidth}
		\caption{5x10Y Swaption (P) ATM.}
  		\includegraphics[width=\linewidth]{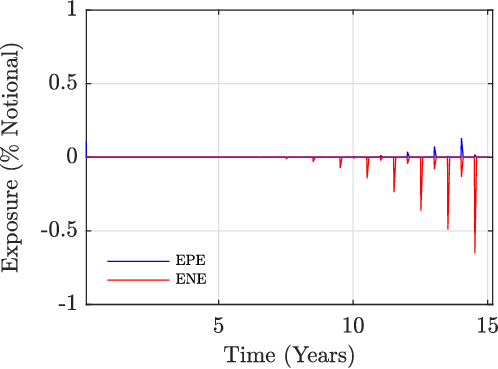}
		\label{fig:swpt_ATM_VM_IM}
	\end{subfigure}
	\hfill
	\begin{subfigure}[t]{0.3\textwidth}
		\caption{5x10Y Swaption (P) OTM.}
    		\includegraphics[width=\linewidth]{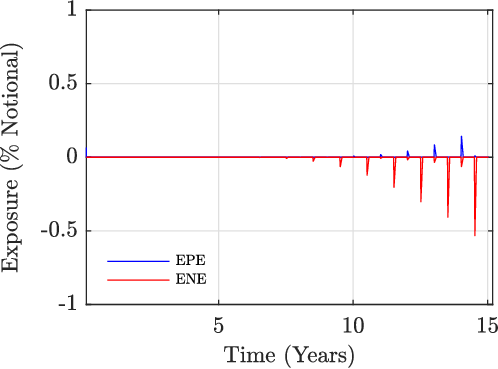}
		\label{fig:swpt_payer_OTM_VM_IM}
	\end{subfigure}
\caption{EPE and ENE profiles (blue and red solid lines respectively) for the 12 instruments in tab.~\ref{tab:instruments}, full collateralization scheme with both VM and IM. To enhance plots readability, we excluded the initial time step $t_0$ showing a very high exposure not yet mitigated by the collateral exchanged MPoR days later.  All the quantities expressed as a percentage of the nominal amount. Letters P and R distinguish between payer and receiver instruments, respectively. Model setup as in sec.~\ref{sec:accuracy vs performance}.}
\label{fig:VM_IM_results}
\end{figure}

We show in fig. \ref{fig:VM_IM_results} the most general and complex case of exposures with VM and IM, and we report in app.~\ref{app:exposure} the complete set of results for the instruments listed in tab.~\ref{tab:instruments} and the three collateralization schemes mentioned above, along with the corresponding detailed comments. 
\par 
Overall, we observe that the expected exposure profiles are broadly consistent with those found in the existing literature (see e.g.~\cite{BriMor13,Gre20}). 
A closer inspection reveals that our approach is able to capture the detailed and complex shape of the exposure mitigated by VM and IM, an important feature for XVA calculation. 
In particular we notice in fig. \ref{fig:VM_IM_results}, for all the instruments and moneynesses considered, that EPE/ENE profiles display short and medium term flat shapes with spikes appearing with increasing magnitude close to maturity, suggesting that IM turns out to be inadequate to fully suppress the exposure because of its decreasing profile.
These spikes originate from the semi-annual jumps in Swaps' future simulated mark-to-market values at cash flow dates, due to the different frequency of the two legs, captured by VM with a delay equal to the length of the margin period of risk (MPoR, see sec.~\ref{sec:Spikes analysis} for details).

\subsection{XVA Results} 
\label{sec:XVA results}

We report in tab.~\ref{tab:XVA_results} the XVA figures for the full set of 36 cases considered in this work. Notice that we compute XVA from the point of view of the instruments' holder (i.e. with positive nominal amount $N$). Accordingly, uncollateralized physically settled Swaptions show non-zero DVA figures\footnote{From the point of view of the holder, uncollateralised cash-settled Swaptions have zero ENE and DVA. In the presence of collateral, small ENE and DVA figures may appear because of MC scenarios where the received collateral exceeds the Swaption's price.}.
\par 
Regarding uncollateralized XVA, we observe that Swaps display larger CVA (DVA) figures for ITM (OTM) instruments due to the greater probability to observe positive (negative) future simulated mark-to-market values, also reflected in lower Monte Carlo errors. 
Moreover, CVA is larger than DVA except for the OTM 15Y Swap due to the simulated forward rates structure which causes expected floating leg values greater than those of the fixed leg. Finally, analysing the results for different maturities, the higher risk of 30 years Swaps leads to larger adjustments compared to those maturing in 15 years. 
Analogous results are obtained for forward Swaps. In this case, the asymmetric effect of simulated forward rates on opposite transactions causes larger CVA and smaller DVA for the OTM payer forward Swap with respect to the OTM receiver one. 
Slightly lower (absolute) CVA values are observed for the corresponding physically settled European Swaptions, since OTM paths are excluded after the exercise, while DVA values are considerably lower as negative exposure exists only after the expiry.

Regarding XVA with VM only, we observe that the adjustments are reduced on average by approx.~two orders of magnitude with respect to the uncollateralized case, and are widely driven by spikes in exposure profiles. In general, DVA figures are greater compared to CVA ones, since for payer instruments the magnitude of the spikes in ENE becomes larger by approaching the maturity, where the default probability increases.

Finally, regarding XVA with VM and IM, we observe that the adjustments are reduced on average by approx.~four orders of magnitude with respect to the uncollateralized case. Here, XVA figues are entirely driven by the spikes closest to maturity which are not fully suppressed by IM, thus confirming the importance of a detailed simulation of the collateralized exposure.

\begin{table}[!htbp]
  \centering  
    \begin{tabular}{l l l r r r}
    \toprule
    \multicolumn{1}{c}{Collateral} & \multicolumn{2}{c}{Instrument} &  \multicolumn{1}{r}{$\omega$} &\multicolumn{1}{c}{CVA (\euro)}&  \multicolumn{1}{c}{DVA (\euro)} \\
    \midrule
	\multirow{12}[2]{*}{None} & 15Y Swap  & OTM   & 1     & -779085  ($\pm6$\%) &  884382 ($\pm 6$\%) \\
	& 15Y Swap  & ATM   & 1     & -1172938  ($\pm4$\%) &     475324  ($\pm 9$\%) \\
    & 15Y Swap  & ITM   & 1     & -1681190  ($\pm4$\%) & 240662 ($\pm 12$\%)  \\
    & 30Y Swap  & OTM   & 1     & -2630312  ($\pm5$\%) & 2218071 ($\pm 7$\%) \\    
    & 30Y Swap  & ATM   & 1     & -3756886  ($\pm4$\%)     & 1233129 ($\pm 10$\%)   \\
    & 30Y Swap  & ITM   & 1     & -5173086  ($\pm3$\%)       & 650447 ($\pm 14$\%)   \\
    & 5x10Y Fwd Swap  & OTM   & -1    & -387131  ($\pm9$\%)     & 1373951 ($\pm 4$\%)  \\    
    & 5x10Y Fwd Swap  & ATM   & 1     & -821210  ($\pm5$\%)     & 714338 ($\pm 7$\%)   \\
    & 5x10Y Fwd Swap  & OTM   & 1     & -530527  ($\pm7$\%)     & 1134273 ($\pm 5$\%)  \\
    & 5x10Y Swaption  & OTM   & -1    & -390339  ($\pm9$\%)     & 43011 ($\pm 17$\%)   \\
    & 5x10Y Swaption  & ATM   & 1     & -818308  ($\pm5$\%)     & 36418 ($\pm 17$\%)  \\
    & 5x10Y Swaption  & OTM   & 1     & -519394  ($\pm7$\%)     & 34825 ($\pm 18$\%)   \\
	\midrule
	\multirow{12}[2]{*}{VM} & 15Y Swap & OTM   & 1     & -14124 ($\pm6$\%)   & 17427 ($\pm6$\%) \\
    & 15Y Swap & ATM   & 1     & -13826 ($\pm6$\%) & 17468 ($\pm6$\%) \\
    & 15Y Swap & ITM   & 1     & -13581 ($\pm6$\%) & 17566 ($\pm6$\%) \\
	& 30Y Swap  & OTM   & 1     & -49435 ($\pm6$\%) & 55953 ($\pm7$\%)\\    
    & 30Y Swap  & ATM   & 1     & -47369 ($\pm6$\%) & 53993 ($\pm7$\%) \\   
    & 30Y Swap & ITM   & 1     & -45363 ($\pm6$\%) & 52094 ($\pm6$\%) \\
    & 5x10Y Fwd Swap & OTM   & -1    & -15584 ($\pm6$\%) & 19819 ($\pm6$\%) \\    
    & 5x10Y Fwd Swap & ATM   & 1     & -15685 ($\pm6$\%) & 19949 ($\pm6$\%) \\
    & 5x10Y Fwd Swap & OTM   & 1     & -15940 ($\pm6$\%) & 20014 ($\pm6$\%) \\
    & 5x10Y Swaption & OTM   & -1    & -5477 ($\pm11$\%) & 7672 ($\pm10$\%) \\
    & 5x10Y Swaption & ATM   & 1     & -8359 ($\pm8$\%) & 11508 ($\pm8$\%) \\
    & 5x10Y Swaption & OTM   & 1     & -6247 ($\pm10$\%) & 8383 ($\pm9$\%) \\	
	\midrule
	\multirow{12}[2]{*}{VM and IM} & 15Y Swap & OTM   & 1     & -44 ($\pm11$\%) & 100 ($\pm9$\%) \\
    & 15Y Swap & ATM   & 1     & -25 ($\pm15$\%) & 114 ($\pm9$\%) \\
    & 15Y Swap & ITM   & 1     & -14 ($\pm19$\%) & 137 ($\pm8$\%) \\
    & 30Y Swap  & OTM   & 1     & -14 ($\pm10$\%) & 61 ($\pm12$\%) \\
    & 30Y Swap  & ATM   & 1     & -9 ($\pm12$\%) & 70 ($\pm13$\%) \\
    & 30Y Swap  & ITM   & 1     & -6 ($\pm15$\%) & 84 ($\pm12$\%) \\
    & 5x10Y Fwd Swap & OTM   & -1  & -113 ($\pm9$\%) & 26 ($\pm14$\%) \\
    & 5x10Y Fwd Swap & ATM   & 1     & -45 ($\pm11$\%) & 100 ($\pm9$\%) \\
    & 5x10Y Fwd Swap & OTM   & 1     & -76 ($\pm9$\%) & 92 ($\pm8$\%) \\
    & 5x10Y Swaption & OTM   & -1    & -49 ($\pm19$\%) & 71 ($\pm18$\%) \\
    & 5x10Y Swaption & ATM   & 1     & -43 ($\pm19$\%) & 328 ($\pm10$\%) \\
    & 5x10Y Swaption & OTM   & 1     & -50 ($\pm19$\%) & 259 ($\pm11$\%) \\
    \bottomrule	
	\end{tabular}
\caption{XVA figures for instruments in tab.~\ref{tab:instruments} for the three collateralization schemes considered. 3$\sigma$ confidence intervals in parentheses. Model setup as in tab.~\ref{tab:model_setup}.}
\label{tab:XVA_results}
\end{table}

\clearpage

\section{XVA Model Validation}
\label{sec:XVA model validation}

The previous sections \ref{sec:XVA Pricing Framework} and \ref{sec:XVA numerical calculations} suggest that XVA  calculation depends on a number of assumptions which affect the results in different ways. 
We can look at these assumptions as sources of model risk, which need to be properly addressed. 

The purpose of this section is threefold: i) we want to identify and analyze in detail the most significant sources of model risk; ii) we want to validate our XVA framework by assessing its robustness and tuning the corresponding calculation parameters; iii) we look for a strategy to set the acceptable compromise between accuracy and performance, a very important feature for practical applications. 
These analyses will also lead to a distribution of XVA values which will be the basis to compute a model risk measure in the following sec. \ref{sec:Model Risk}.

In order to ease the presentation we report the results only for a subset of the instruments listed in tab.~\ref{tab:instruments}, mainly the 15Y ATM payer Swap and the 5x10Y ATM physically settled European payer Swaption. Analogous results are obtained for the other instruments and are reported in the corresponding sections of app. \ref{app:additional results}.

\subsection{G2++ Model Calibration}
\label{sec:G2++ Model Calibration}
Our XVA framework is based on the G2++ model, whose parameters $p$ are calibrated on market ATM Swaption prices using the procedure described in app.~\ref{app:G2++ calibration procedure}. 
We will refer to this calibration as the \emph{baseline calibration}. 

Calibrating the model to ATM swaption quotes is a standard approach for at least two reasons: i) ATM quotes are the most liquid, in particular for the most frequently traded expiries/tenors, ii) there is less interest into the smile risk when one has to deal with many counterparties and large netting sets dominated by linear interest rate derivatives, a typical situation for XVA trading desks. 
Nevertheless, this choice is not unique, and different approaches can be adopted according to market conditions, specific trades and the relevance of smile risk. 
For example, a specific calibration approach could be adopted when structuring a new trade for a client, especially in the case of a competitive auctions. 
\par
Since XVA figures depend on the G2++ model parameters, the G2++ model calibration is a source of model risk, and we address it by considering different alternative calibrations, also including the Swaption smile risk. 
In particular, we compare the baseline calibration (parameters denoted with $p$) with six alternative calibrations (parameters denoted with $p_i$, with $i=1,\dots,6$), which we define by tuning the following features: i) the maximum expiry of market points used for calibration, ii) the maximum/minimum strikes of market points used for calibration, and iii) imposing flat volatility G2++ parameters. 
\par 
We report in app.~\ref{app:G2++ model calibration} all the details about the 7 different calibrations.
We show here in tab.~\ref{tab:XVA_calibrations} the XVA figures obtained with the seven different calibrations for the 15Y ATM payer Swap and the 5x10Y ATM physically settled European payer Swaption, both without collateral. Similar results are obtained for the other collateralization schemes. 
We observe that the range of values obtained using different model calibrations is always lower than the $3\sigma$ statistical uncertainty due to MC simulation. Therefore we can conclude that the XVA model risk stemming from different G2++ model calibrations is limited, and that our choice of the baseline calibration on market ATM Swaptions is sufficiently robust for our purposes.
\begin{table}[!t]
	\centering
	\begin{tabular}{l c r r}
		\toprule
		\multicolumn{1}{c}{Instrument} & \multicolumn{1}{c}{G2++ parameters} &\multicolumn{1}{c}{CVA (\euro)} & \multicolumn{1}{c}{DVA (\euro)} \\
		\midrule
		& $p$ 	& -1172938  ($\pm4$\%)     &  475324  ($\pm9$\%)  \\
		& $p_1$ & -1174733 ($\pm4$\%)     &  473529  ($\pm9$\%)  \\
		& $p_2$ & -1178799  ($\pm4$\%)     &  492086  ($\pm9$\%)  \\
		& $p_3$ & -1160878  ($\pm4$\%)     &  460594  ($\pm9$\%)  \\
		Swap 	& $p_4$ & -1169436  ($\pm4$\%)     &  462020 ($\pm9$\%)  \\
		& $p_5$ & -1174460  ($\pm4$\%)     &  493422  ($\pm9$\%)  \\
		& $p_6$ & -1170109  ($\pm4$\%)     &  465102  ($\pm9$\%)  \\
		\cline{2-4}
		& \multicolumn{1}{c}{Avg} & \multicolumn{1}{c}{-1171622 ($\pm4$\%)} & \multicolumn{1}{c}{474582 ($\pm9$\%)}\\ 
		& \multicolumn{1}{c}{Range} & \multicolumn{1}{c}{17921} & \multicolumn{1}{c}{32828}\\ 
		& \multicolumn{1}{c}{Range/Avg} & \multicolumn{1}{c}{1.5\%} & \multicolumn{1}{c}{6.9\%}\\ 
		\midrule     
		& $p$ 	& -818308  ($\pm5$\%)     &  36418  ($\pm17$\%)  \\
		& $p_1$ & -822656 ($\pm5$\%)     &  36530  ($\pm17$\%)  \\
		& $p_2$ & -823830  ($\pm5$\%)     &  37473  ($\pm17$\%)  \\
		& $p_3$ & -809796  ($\pm5$\%)     &  34339  ($\pm17$\%)  \\
		Swaption & $p_4$ & -817334  ($\pm5$\%)     &  33847 ($\pm17$\%)  \\
		& $p_5$ & -821480  ($\pm5$\%)     &  35823  ($\pm17$\%)  \\
		& $p_6$ & -818475  ($\pm5$\%)     &  33896  ($\pm17$\%)  \\  
		\cline{2-4}
		& \multicolumn{1}{c}{Avg} & \multicolumn{1}{c}{-818840 ($\pm5$\%)} & \multicolumn{1}{c}{35475 ($\pm17$\%)}\\ 
		& \multicolumn{1}{c}{Range} & \multicolumn{1}{c}{14034} & \multicolumn{1}{c}{3626}\\ 
		& \multicolumn{1}{c}{Range/Avg} & \multicolumn{1}{c}{1.7\%} & \multicolumn{1}{c}{10.2\%}\\                                        
		\bottomrule
	\end{tabular}
	\caption{XVA figures and 3$\sigma$ confidence intervals (in parentheses) obtained with 7 different G2++ model calibrations for 15Y ATM payer Swap and 5x10Y ATM physically settled European payer Swaption, EUR 100 Mio nominal amount, no collateral. Other model parameters as in tab.~\ref{tab:model_setup}. Since the MC runs are not independent (see sec.~\ref{sec:XVAs MC convergence}), the total MC error is obtained as the average of single MC errors.}
	\label{tab:XVA_calibrations}
\end{table}

\subsection{Time Simulation Grid}
\label{sec:Time Simulation Grid}

The numerical calculation of XVA in eqs. \ref{eq:CVA2} and \ref{eq:DVA2} requires the discretization of the integral on a time grid in correspondence of which the exposure is computed using the Monte Carlo simulation, as shown in app. \ref{app:XVA numerical formulas}. 

The construction of this MC time simulation grid is a crucial step to find an acceptable compromise between accuracy and performance for at least two reasons: i) an high granularity reduces the discretization error but increases the computational time, leading to poor performances, and ii) the presence of the margin period of risk requires a careful distribution of the time simulation points in order to capture the spikes in collateralized exposures, which have material impact on XVA.
As a consequence, both the granularity and the distribution of the time simulation grid are an important model risk factor in XVA calculation.

In order to identify the optimal construction of the time simulation grid we proceed as follows: i) we analyse the spikes in collateralized exposure using the most accurate choice, i.e. a daily grid; ii) we propose a workaround which allows to capture all the spikes using lower granularities and iii) we perform a convergence analysis looking for the granularity which ensures an acceptable compromise between accuracy and performance.

\subsubsection{Spikes Analysis}
\label{sec:Spikes analysis} 

Spikes arising in collateralized exposure are due to the MPoR, since it implies that the collateral available at time step $t_i$ depends on instrument's simulated mark-to-market values at time step $\hat{t}_i = t_i - l$, assumed to be the last date at which VM and IM are fully exchanged (see app.~\ref{app:collateral management}). 
Clearly, the best possible choice in terms of accuracy is a daily time simulation grid, that both reduces the discretization error in the integrals in eqs.~\ref{eq:CVA2} and \ref{eq:DVA2}, and automatically captures all the details of the exposure, including the spikes. 
\par
\begin{figure}[!htbp]
	\begin{subfigure}{0.49\textwidth}
		\caption{EPE/ENE Swap, no collateral}
		\centering
		\includegraphics[width=0.9\linewidth]{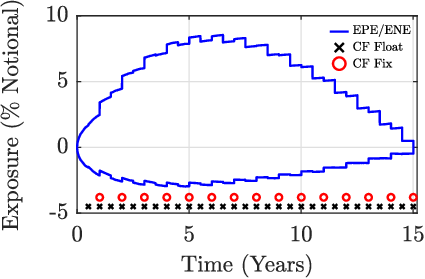}
		\label{fig:irs_15Y_ATM_DailyGrid_noMargins}
	\end{subfigure}
\hfill
\begin{subfigure}{0.49\textwidth}
		\caption{EPE/ENE Swaption, no collateral}
		\centering
		\includegraphics[width=0.9\linewidth]{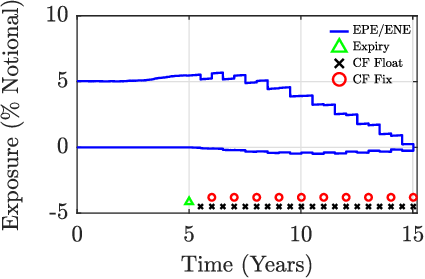}
		\label{fig:swpt_5x10Y_ATM_DailyGrid_noMargins}
	\end{subfigure} \\
	\begin{subfigure}{0.49\linewidth}
		\caption{EPE/ENE Swap, VM.}
		\centering
		\includegraphics[width=0.9\linewidth]{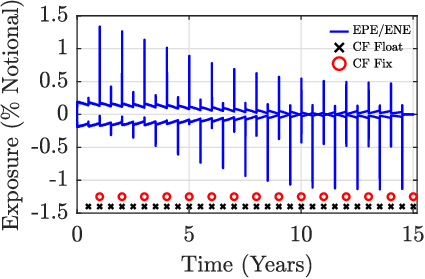}
		\label{fig:irs_15Y_ATM_DailyGrid_VM}
	\end{subfigure}
\hfill
	\begin{subfigure}{0.49\linewidth}
		\caption{EPE/ENE Swaption, VM.}				
		\centering
		\includegraphics[width=0.9\linewidth]{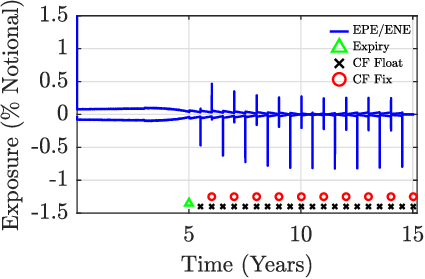}
		\label{fig:swpt_5x10Y_ATM_DailyGrid_VM}
	\end{subfigure} \\
	\begin{subfigure}{0.49\linewidth}
	\caption{EPE/ENE Swap, VM and IM.}	
		\centering
		\includegraphics[width=0.9\linewidth]{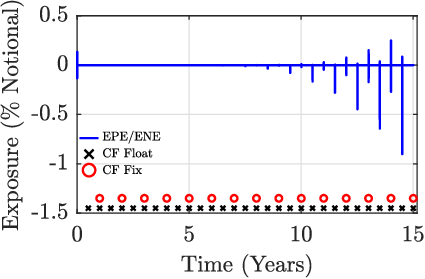}
		\label{fig:irs_15Y_ATM_DailyGrid_VMandIM}
	\end{subfigure}
\hfill
	\begin{subfigure}{0.49\linewidth}			
		\caption{EPE/ENE Swaption, VM and IM.}
		\centering
		\includegraphics[width=0.9\linewidth]{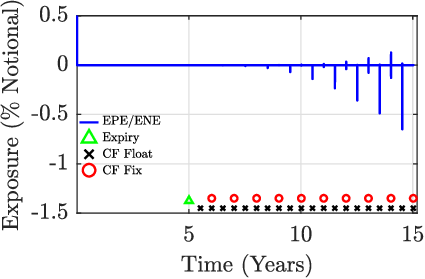}	
		\label{fig:swpt_5x10Y_ATM_DailyGrid_VMandIM}
	\end{subfigure} \\
\caption{EPE and ENE profiles (blue solid lines) for 15Y ATM payer Swap (left-hand side) and 5x10Y ATM physically settled European payer Swaption (right-hand side), EUR 100 Mio nominal amount, on a daily grid for the three collateralization schemes considered (top: no collateral, mid: VM, bottom: VM and IM). Black crosses: floating leg cash flow dates (semi-annual frequency); red circles: fixed cash flow dates (annual frequency); green triangles: Swaption's expiry. To enhance Swaption's plots readability, we omit the collateralized EPE at time steps $t_0$ and $t_1$ where the exposure spikes since collateral is exchanged MPoR days later (mid and bottom right-hand panels). In particular EPE($t_0$) is equal to present Swaption's price (5030423 EUR, approx.~$5\%$ of the nominal amount). Other model parameters as in tab.~\ref{tab:model_setup}. Quantities expressed as a percentage of the nominal amount.}  
\label{fig:irs_swpt_EPE_ENE_DailyGrid}
\end{figure}

We show in fig. \ref{fig:irs_swpt_EPE_ENE_DailyGrid} the EPE/ENE profiles obtained with a daily grid for the 15Y Swap (left-hand side panel) and the 5x10Y Swaption (right-hand side panel) for the three collateralization schemes considered. 
As can be seen, when only VM is considered, spikes emerge at inception as no collateral is posted, and at cash flow dates as sudden changes in future simulated mark-to-market values are captured by VM with a delay due to MPoR. When also IM is considered, spikes closest to maturity persist due to the downward profile of IM.
Further investigations on the nature of the exposure's spikes are reported in app. \ref{app:Time Simulation Grid}.
\par 
The impact of these spikes on XVA figures is significant: with VM only the contribution is respectively of $+7\%$ and $+6\%$ for the 15Y Swap and of $+3\%$ and $+6\%$ for the 5x10Y Swaption. With also IM the exposure between spikes is suppressed, therefore CVA and DVA are completely attributable to spikes. In other words, neglecting the spikes would significantly underestimate the (absolute) XVA figures. 
On the other hand, since using a daily grid is unfeasible in practice, in the next section we look for a possible solution.

\subsubsection{Parsimonious Time Grid}
\label{sec:Parsimonious Time Grid} 

Although a daily grid, as discussed in the previous section \ref{sec:Spikes analysis}, clearly represents the best discrete approximation to compute the XVA integrals in eqs.~\ref{eq:CVA2} and \ref{eq:DVA2}, this choice is often unfeasible in practice because of the poor computational performance, as can be observed in the last column of tab.~\ref{tab:irs_swpt_GridConvergence}. 
Notice that the most time consuming component is the IM, which involves the calculation of several forward sensitivities for each path (see app.~\ref{app:collateral IM calculation methodology}). 
Another bottleneck is the numerical integration of the semi-analytical G2++ pricing formula for Swaptions (see eq.~\ref{eq:g2++_swpt_multi_curve}). 
On the other hand, the adoption of simple, less granular, evenly spaced time grids would be inadequate to capture the spikes in collateralised exposure and could produce biased XVA figures.
\par 
In order to overcome these issues we build a \emph{parsimonious time simulation grid} $\left\{t_i\right\}_{i=0}^N$, which is obtained by joining an \emph{initial time grid} $\left\{\bar{t}_i\right\}$, evenly-spaced with time step $\Delta t$, a \emph{cash flow grid}, including the trade (or portfolio) cash flow dates, and a \emph{collateral time grid}, where each previous date is shifted by the MPoR.
The resulting final \emph{joint time grid} depends on the initial time step $\Delta t$ but is no longer evenly spaced. See  app. \ref{app:Time Simulation Grid} for more details. 
\par
\begin{figure}[!htbp]			
	\begin{subfigure}{0.49\linewidth}
		\centering
		\caption{Std grid, EPE/ENE no collat.}
		\includegraphics[width=0.9\linewidth]{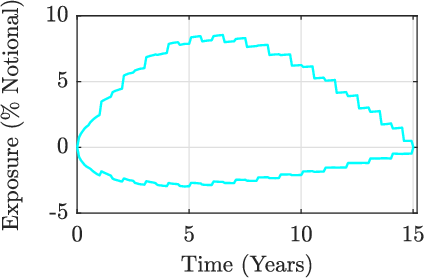}	
		\label{fig:irs_15Y_ATM_standard_grid_noColl}
	\end{subfigure}
\hfill
	\begin{subfigure}{0.49\linewidth}
		\centering
		\caption{Joint grid, EPE/ENE no collat.}
		\includegraphics[width=0.9\linewidth]{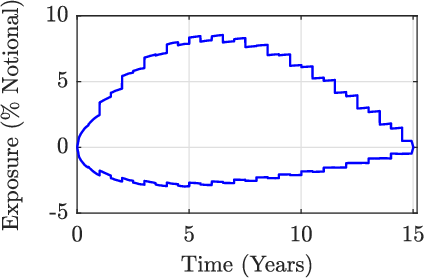}	
		\label{fig:irs_15Y_ATM_joint_grid_noColl}
	\end{subfigure} \\
	\begin{subfigure}{0.49\linewidth}
		\centering
		\caption{Std grid, EPE/ENE VM.}
		\includegraphics[width=0.9\linewidth]{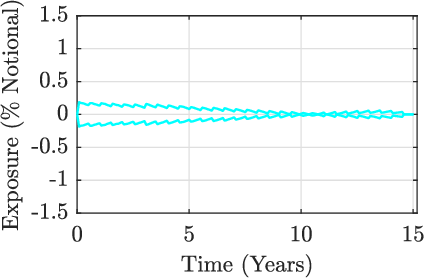}	
		\label{fig:irs_15Y_ATM_standard_grid_VM}
	\end{subfigure}
\hfill
	\begin{subfigure}{0.49\linewidth}
		\centering
		\caption{Joint grid, EPE/ENE VM.}
		\includegraphics[width=0.9\linewidth]{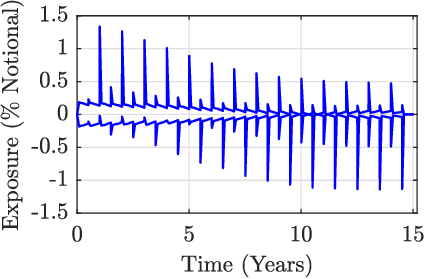}	
		\label{fig:irs_15Y_ATM_joint_grid_VM}
	\end{subfigure} \\
	\begin{subfigure}{0.49\linewidth}
		\centering
		\caption{Std grid, EPE/ENE VM and IM.}
		\includegraphics[width=0.9\linewidth]{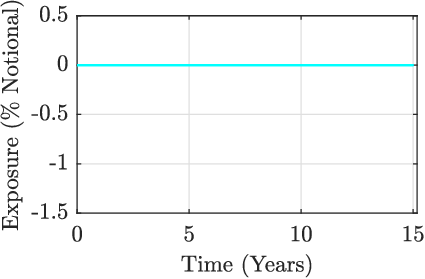}	
		\label{fig:irs_15Y_ATM_standard_grid_IM}
	\end{subfigure}
\hfill
	\begin{subfigure}{0.49\linewidth}
		\centering
		\caption{Joint grid, EPE/ENE VM and IM.}
		\includegraphics[width=0.9\linewidth]{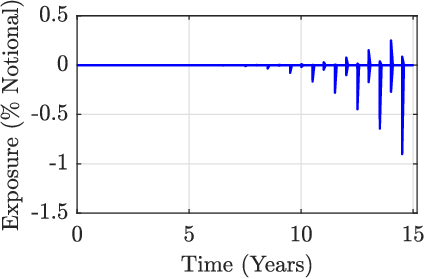}	
		\label{fig:irs_15Y_ATM_joint_grid_IM}
	\end{subfigure} \\
\caption{EPE/ENE profiles for 15Y ATM payer Swap, EUR 100 Mio nominal amount, obtained with standard grid (left-hand side) and joint grid (right-hand side) with monthly granularity. The \emph{standard time grid} is built by primary + collateral time grids (does not include the cash flow time grid). Other parameters as in fig. \ref{fig:irs_swpt_EPE_ENE_DailyGrid}. The joint grid $\Delta t =1M$ exposures are very similar to $\Delta t =1D$ exposures in fig. \ref{fig:irs_swpt_EPE_ENE_DailyGrid} (left-hand side).}
\label{fig:irs_15Y_ATM_joint_vs_standard_grid}
\end{figure}

We show in fig.~\ref{fig:irs_15Y_ATM_joint_vs_standard_grid} the exposure profiles for the 15Y Swap obtained with different grids for the three collateralization schemes considered. For testing purposes we compare the joint time grid with a \emph{standard time grid}, obtained adding the primary time grid and its corresponding collateral time grid, which does not include the cash flow time grid.
\begin{table}[!t]
	\centering
	\resizebox{\linewidth}{!}{
	\begin{tabular}{l l l r r r r r r r r}
		\toprule
		\multirow{2}*{Inst.} & \multirow{2}*{Coll.} & \multirow{2}*{$\Delta t$} & \multirow{1}*{CVA} & \multirow{1}*{Grid err.} & MC err. & \multirow{1}*{DVA} & \multirow{1}*{Grid err.}  & MC & \multirow{1}*{Time} & \multirow{1}*{Time}\\
		& & & (\euro) & (\%) & (\%) & (\euro) & (\%) & err. &  (s) & (\%) \\ 
		\midrule
		&  & 12M & -1119629 & -4.6 & 4.5    &  485660    & 2.4 & 8.4      & 0.45 & 0.8 \\          
		&  & 6M & -1170271  & -0.3 & 4.4      & 478016    & 0.7 & 8.5      & 0.60 & 1.1 \\
		&  None & 3M & -1172139  & -0.1 & 4.4    & 476789 & 0.5   & 8.6  & 0.90 & 1.6 \\
		& & 1M & -1172938 & 0.0 &   4.4      & 475324 & 0.2 &  8.6  &  2.10 &  3.8 \\
		&  & 1D & -1173296 & 0.0 &   4.4      &        474469 & 0.0 &  8.6 & 54.8 & 100 \\
		\cline{2-11}
		& & 12M & -24503 & 68 &  8.1  & 87496  & 377 & 4.5     &  0.90 & 0.8 \\
		& &  6M & -11182 & -23 &  6.2  & 14175 & -23 & 6.2    & 1.20 & 1.1 \\
		Swap & VM &  3M & -12783 & -12 &  6.2  & 16186 & -12 &  6.2     & 1.80 & 1.6 \\
		& &  1M & -13826 & -5.1 & 6.2  & 17468 & -4.8 &  6.2     & 4.20 &  3.8 \\
		& & 1D & -14569  & 0.0 & 6.1  & 18340  & 0.0 & 6.2    &  109.6 & 100 \\
		\cline{2-11}
		& & 12M & -637 & 1817 & 20  & 11953 &  7176 & 7.3   & 12.4 & 0.8 \\
		& &  6M & -25 &-25 &  15  & 114 &  -31 & 8.6   &   16.5 &  1.1 \\
		& VM \& IM &  3M & -25 & -25  & 15  & 114 & -31  & 8.6    & 24.8 & 1.6 \\
		& &  1M & -25 & -25 & 15  & 114 & -31  & 8.6  &  57.8 &  3.8 \\
		& & 1D  & -33 & 0.0  & 14  & 164 & 0.0 & 8.3   &  1507 & 100 \\
		\midrule 
		& &  12M & -775964  & -5.3 &   5.5      &  41357  & 14 &  16     &  4.32 & 0.7 \\
		& &  6M & -815745  & -0.5&  5.4      &  36705  & 0.9 &  17     &  6.07 &  0.9 \\
		& None  &  3M & -817396 & -0.3&   5.4      &  36544  & 0.5 &  17     &  9.57 & 1.5 \\
		& &  1M & -818308  &-0.2 &  5.4      &  36418  & 0.1 &  17     &  23.6 &  3.7 \\
		&& 1D & -819616  & 0.0 & 5.4      &       36367  & 0.0 &  17     &  640.0 &  100 \\
		\cline{2-11}
		& &  12M & -7851  & -4.0& 12  & 54319  & 410 & 6.0     &  8.02 &  0.7 \\
		& &  6M & -7461& -8.7&   8.5  & 10580 & -0.6 &  7.9    & 11.3 & 0.9 \\
		Swaption & VM &  3M & -8002 & -2.1 &  8.5  & 11144 & 4.7 &  7.9    & 17.8 & 1.5 \\
		& &  1M & -8359  & 2.3 & 8.5  & 11508 & 8.1  &  7.9   & 43.8 & 3.7 \\
		& & 1D & -8175 & 0.0 &  8.4  & 10644  &  0.0 & 8.0    &  1189 & 100 \\
		\cline{2-11}
		& &  12M & -136 & 190&  34  & 9301 & 8007 &  9.0    & 119.6 & 0.7 \\
		& &  6M & -43 & -7.8&  19  & 328 & 186  &  10    & 168.1 & 0.9 \\
		& VM \& IM &  3M & -43  & -7.9& 19  & 328 & 186 &  10   & 265.1 & 1.5 \\
		& &  1M & -43 & -7.9&  19  & 328 & 186  &  10     &  653.1 & 3.7 \\
		&  & 1D & -47 & 0.0&  5.5  & 115  & 0.0 & 10     &  17738 & 100 \\
		\bottomrule
	\end{tabular}}
	\caption{XVA convergence analysis for 15Y ATM payer Swap and 5x10Y ATM physically settled European payer Swaption, EUR 100 Mio nominal amount. We used different joint grid granularities $\Delta t$ plus the daily time simulation grid of sec.~\ref{sec:Spikes analysis}. We report the relative grid error, the MC $3\sigma$ error, and the computational time both in absolute (seconds) and relative terms w.r.t.~the daily grid (in parenthesis). Other model parameters as in tab.~\ref{tab:model_setup}.}
	\label{tab:irs_swpt_GridConvergence}
\end{table}
We observe that uncollateralized exposures are similar for both grids (top panels), but the standard grid fails to capture the spikes in exposures with VM (panel \emph{c} vs \emph{d}), because of the lack of the cash flow time grid. Adding the IM with standard grid completely suppresses the residual exposure (panel \emph{e}), thus leading to null XVA figures. Instead, the joint time grid allows to correctly model all spikes in the collateralized exposure (panel \emph{f}) with considerable computational benefits with respect to the daily grid discussed in the previous section. In fact, the joint grid $\Delta t =1M$ exposures in fig. \ref{fig:irs_15Y_ATM_joint_vs_standard_grid} (right-hand side) are very similar to the corresponding $\Delta t =1D$ exposures in fig. \ref{fig:irs_swpt_EPE_ENE_DailyGrid} (left-hand side).
Similar results are obtained for the 5x10Y Swaption (see app. \ref{app:Time Simulation Grid}).
\par 
The parsimonious time simulation grid discussed above is governed by two parameters: i.e. the constant granularity $\Delta t$ used in the initial time grid and the number $n$ of cash flows in the cash flow grid. Since the number of cash flows is fixed exhogenously according to the trade or portfolio under analysis, the other parameter $\Delta t$ can be used to tune the compromise between accuracy and computational performance in the XVA calculation. 
\par 
We show in tab.~\ref{tab:irs_swpt_GridConvergence} the XVA results obtained for the 15Y Swap and the 5x10Y Swaption using the joint time grid with different granularities $\Delta t$, taking the results obtained with the daily time grid as benchmark. 
We observe that uncollateralized XVA, without spikes, show a good convergence already for low granularities, i.e.~$\Delta t = 6$M. In fact, the relative grid error is smaller than the MC $3\sigma$ error. 
Instead, collateralized XVA require higher granularities, up to $\Delta t = 1$M, due to the exposure spikes. In particular, XVA with VM are dominated by the grid error for the Swap (except for DVA with $\Delta t = 1M$), and by the MC error for the Swaption (except for DVA with $\Delta t = 12M,1M$). 
XVA with both VM and IM, very small and highly spike dependent, are mainly dominated by the grid error, except for Swaption's CVA. 
The differences between collateralized Swaps and Swaptions are not surprising, since in the collateralized exposure for physical Swaptions i) cash flows and spikes appear only after the Swaptions' expiry and ii) many MC paths after the Swaptions' expiry date go OTM and give zero prices (e.g. the 5x10Y Swaption goes OTM for 45.2\% of the MC paths). This is clearly visible in fig. \ref{fig:swaption_results}, where the spikes for the 5x10Y collateralized Swaptions (panels \emph{d-i}) are much smaller w.r.t. the corresponding collateralized Swaps in fig. \ref{fig:irs_fwd_results}.
\par 
Looking at the computational performance (last column), we observe that, overall, the computational time is roughly proportional to the number of time simulation steps $N_S$. In particular, the monthly grid is approx. 26-27 times faster than the daily grid and 2.3-2.5 times slower than the quarterly grid. Regarding the instruments, the Swaption is approx. 10-12 times slower than the Swap. Regarding the collateral, adding VM costs approx. a factor of 2, and adding also IM costs another factor of 14-15, in total approx. 28-30 times slower than the uncollateralized case, both for the Swap and the Swaption. 
\par 
In light of this analysis we may confirm that the construction of the time simulation grid is a relevant source of model risk. For the purposes of the present work, we identify $\Delta t = 1M$ as an acceptable compromise between accuracy and performance.

\subsection{Monte Carlo Convergence}
\label{sec:XVAs MC convergence}

The Monte Carlo simulation used in this work for XVA calculation (eq.~\ref{eq:epe_ene}), although computational intensive, allows to manage the complexities inherent XVA calculation, such as collateralization. 
Obviously, the most important parameter for MC is the number of MC scenarios, which has to be tuned to find an acceptable compromise between precision and computational effort. 
\par 
Accordingly, we investigate the XVA convergence with respect to the number of Monte Carlo scenarios $N_{MC}$. In order to do so, we assume the XVA figures calculated with a large number of MC scenarios (i.e. $N_{MC} = 10^6$) as proxies for the ``exact'' XVA figures, and we use them as benchmarks to assess the XVA convergence for smaller numbers of scenarios (always using the same seed in the pseudo-random number generator). Furthermore, in order to investigate the XVA Monte Carlo error, we use the upper and lower bounds on EPE/ENE in eq. \ref{eq:XVA LB UB}.
\par
\begin{figure}[!htbp]
	\begin{subfigure}{0.49\linewidth}
		\centering
		\caption{CVA no collateral.}
		\includegraphics[width=0.9\linewidth]{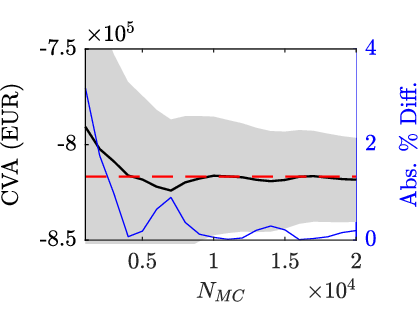}
		\label{fig:swpt_5x10Y_ATM_CVAconvergence_noMargins}
	\end{subfigure}
\hfill
	\begin{subfigure}{0.49\linewidth}
		\centering
		\caption{DVA no collateral.}
		\includegraphics[width=0.9\linewidth]{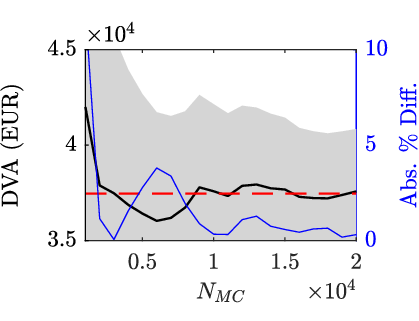}
		\label{fig:swpt_5x10Y_ATM_DVAconvergence_noMargins}
	\end{subfigure}\\
	\begin{subfigure}{0.49\linewidth}
		\centering
		\caption{CVA with VM.}		
		\includegraphics[width=0.9\linewidth]{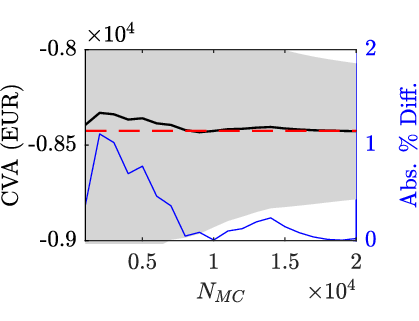}
		\label{fig:swpt_5x10Y_ATM_CVAconvergence_VM}
	\end{subfigure}
\hfill
	\begin{subfigure}{0.49\linewidth}
		\centering	
		\caption{DVA with VM.}
		\includegraphics[width=0.9\linewidth]{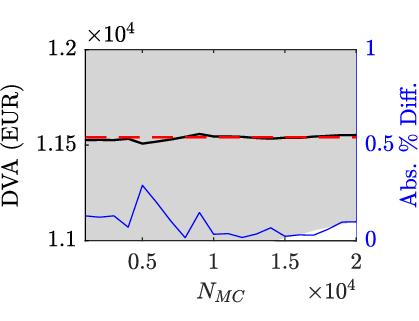}
		\label{fig:swpt_5x10Y_ATM_DVAconvergence_VM}
	\end{subfigure}\\
	\begin{subfigure}{0.49\linewidth}
		\centering
		\caption{CVA with VM and IM.}
		\includegraphics[width=0.9\linewidth]{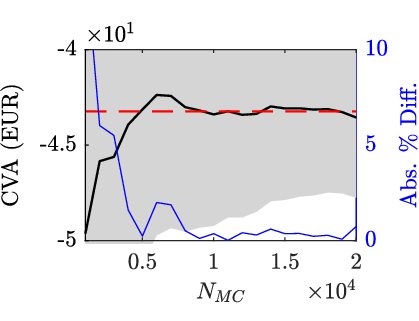}
		\label{fig:swpt_5x10Y_ATM_CVAconvergence_VMandIM}
	\end{subfigure}
\hfill
	\begin{subfigure}{0.49\linewidth}
		\centering
		\caption{DVA with VM and IM.}
		\includegraphics[width=0.9\linewidth]{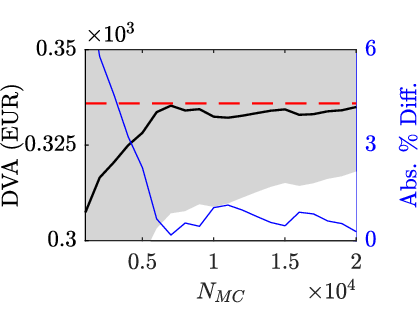}
		\label{fig:swpt_5x10Y_ATM_DVAconvergence_VMandIM}
	\end{subfigure}\\
\caption{CVA (l.h.s) and DVA (r.h.s) convergence diagrams versus number of MC scenarios for the 5x10Y ATM physically settled European payer Swaption, EUR 100 Mio nominal amount, and the three collateralization schemes considered (top: no collateral, mid: VM, bottom: VM and IM). Left-hand scale, black line: simulated XVA; grey area: $3\sigma$ confidence interval; dashed red line: ``exact'' value proxies (we omit their small confidence interval). Right-hand scale, blue line: convergence rate in terms of absolute percentage difference w.r.t. ``exact'' values. Model parameters other than $N_{MC}$ as in tab.~\ref{tab:model_setup}.}
\label{fig:swpt_5x10Y_ATM_convergenceXVA}
\end{figure}

In order to clarify the MC convergence, we show in fig.~\ref{fig:swpt_5x10Y_ATM_convergenceXVA} the XVA convergence diagrams for the 5x10Y Swaption, for the three collateralization schemes considered. 
We observe that XVA converge, for all collateralization schemes, to ``exact'' values with small absolute percentage differences already for few paths (i.e.~$N_{MC} = 1000$). As expected, higher differences can be observed for IM (bottom panels) due to small XVA values; nevertheless, $N_{MC} \geq 5000$ ensures an absolute percentage difference below 5\%. Similar results are obtained for the 15Y Swap (see app. \ref{app:XVAs MC convergence}).
Regarding the computational effort, since it scales linearly with the number of simulated paths, we observe that beyond $N_{MC} = 5000$ the benefits in terms of accuracy would be exceeded by the computational costs, particularly for IM. 
\par 
In light of this analysis we may identify $N_{MC} = 5000$ as an acceptable compromise between accuracy and performance for the purposes of the present work.

\subsection{Forward Vega Sensitivity Calculation}
\label{sec:Vega Sensitivity time simulation}

In this section we report the analyses conducted to validate the approach adopted to calculate the Vega sensitivity when simulating ISDA-SIMM dynamic IM (see app. \ref{app:collateral VM}), which is a source of model risk for the Swaption's XVA.

\subsubsection{Sensitivity to G2++ Parameters}
\label{sec:Sensy to G2++ parameters}

ISDA-SIMM defines Vega sensitivity as the price change with respect to a 1\% shift up in ATM shifted-Black implied volatility. 
Since the G2++ pricing formula for European Swaption does not depend explicitly on the Black implied volatility (see eq.~\ref{eq:g2++_swpt_multi_curve}), in our framework Vega for Swaptions cannot be calculated at future time steps according to ISDA prescriptions. 
In app.~\ref{app:collateral IM calculation methodology} we propose an approximation scheme to calculate forward Vega by shifting up the G2++ model parameters governing the underlying process volatility.  
In order to validate this approach, we compare the Vega obtained at valuation date $t_0$ through eq.~\ref{eq:vega_workaround} with a ``market'' Vega and a ``model'' Vega, both consistent with ISDA prescriptions. 
Specifically, for a given combination of expiry and tenor, we computed the following three Vega sensitivities,
\begin{align}
\nu_{1}(t_0) &=  \frac{V^{\text{Blk}} \bigl( t_0; \sigma^{\text{Blk}}(t_0) + 0.01 \bigr)-V^{\text{Mkt}} (t_0) }{0.01} , \label{eq:vega_black} \\
\nu_{2}(t_0) &=  \frac{V^{\text{G2++}} \bigl(t_0; \hat{p} \bigr)-V^{G2++} \bigl(t_0; p \bigr) }{0.01} , \label{eq:vega_g2++} \\
\nu_{3}(t_0) &=  \frac{V^{\text{G2++}} \bigl( t_0;\sigma + \epsilon_{\sigma},\eta + \epsilon_{\eta} \bigr) - V^{\text{G2++}} \bigl( t_0; \sigma, \eta \bigr) }{\hat{\sigma}^{\text{Blk/G2++}}(t_0) - \sigma^{\text{Blk/G2++}}(t_0)}, \label{eq:vega_workaround_swpt5x10} 
\end{align}
where $\nu_{1}$ denotes the ``market'' Vega obtained by shifting the ATM Black implied volatility by $+1\%$ and re-pricing the Swaption via Black pricing formula; 
$\nu_{2}$ denotes the ``model'' Vega obtained by shifting the ATM Black implied volatility matrix by $+1\%$, re-pricing market Swaptions via Black pricing formula, re-calibrating the G2++ parameters $p$ on these prices, and computing the Swaption price using the re-calibrated parameters $\hat{p}$;
$\nu_{3}$ denotes the Vega obtained according to the approximation outlined in app.~\ref{app:collateral IM calculation methodology}, i.e.~by applying the shocks $\epsilon_{\sigma}$ and $\epsilon_{\eta}$ on the G2++ parameters $\sigma$ and $\eta$ governing the underlying process volatility, recomputing the  G2++ Swaptions' prices and the corresponding Black implied volatilities.
\par
In addition, we also tested eq.~\ref{eq:vega_workaround_swpt5x10} against different values of the shocks $\epsilon_{\sigma}, \epsilon_{\eta}$, considering both $\epsilon_{\sigma} = \epsilon_{\eta}$ and $\epsilon_{\sigma} \neq \epsilon_{\eta}$. In the latter case, we recovered the values for the shocks from the re-calibrated parameters $\hat{\sigma}$ and $\hat{\eta}$ of eq.~\ref{eq:vega_g2++}, i.e. $\epsilon_{\sigma} = 1\%$ and $ \epsilon_{\eta} = 4\%$. The results of the comparison are reported in tab.~\ref{tab:vega_shock_parameters}.
\begin{table}[!htbp]
  \centering
    \begin{tabular}{l l c c c c}
    \toprule
    \multicolumn{1}{c}{Approach} & \multicolumn{1}{c}{G2++ parameters} & \multicolumn{1}{c}{$V$} & \multicolumn{1}{c}{$\hat{V}$} & \multicolumn{1}{c}{$\nu$} & \multicolumn{1}{c}{VR} \\
    \midrule
    eq.~\ref{eq:vega_black} & None & 4968574 & 5183980 & 21540601 & 4857191 \\
    eq.~\ref{eq:vega_g2++} & $\hat{p}$ &5030423 & 5248595 & 21817192 & 4982139 \\
    \multirow{5}[0]{*}{eq.~\ref{eq:vega_workaround_swpt5x10}} & $p, \epsilon_{\sigma} = 0.01, \epsilon_{\eta} = 0.01$ & 5030423 & 5080649 & 21546611 & 4920350 \\
     & $p, \epsilon_{\sigma} = 0.02, \epsilon_{\eta} = 0.02$ & 5030423 & 5155984 & 21535711 & 4917861 \\
     & $p, \epsilon_{\sigma} = 0.04, \epsilon_{\eta} = 0.04$ & 5030423 & 5231313 & 21524697 & 4915345 \\
     & $p, \epsilon_{\sigma} = 0.10, \epsilon_{\eta} = 0.10$ & 5030423 & 5532576 & 21479486 & 4905021 \\
     & $p, \epsilon_{\sigma} = 0.01, \epsilon_{\eta} = 0.04$ & 5030423 & 5249476 & 21522024 & 4914735 \\
    \bottomrule
    \end{tabular}
    \caption{5x10Y ATM physically settled European payer Swaption, EUR 100 Mio nominal amount, comparison between prices $V(t_0)$ and $\hat{V}(t_0)$, Vega sensitivity $\nu(t_0)$ and Vega Risk $VR(t_0)$ at time step $t_0$ for the different calculation approaches and parameterizations considered.$\hat{V}(t_0)$ denotes prices obtained with shifted inputs. Vega Risk is the product between Vega sensitivity and Black implied volatility (see eq.~\ref{eq:Vega_Risk}).}
  \label{tab:vega_shock_parameters}
\end{table}
We observe that Vega sensitivities are fairly aligned among the three approaches and the different shocks values examined. 
Hence, at the initial time step $t_0$ our approximation produces Vega sensitivity and Vega Risk values consistent with those obtained by applying the ISDA definition. Therefore, we assume that this approach can be adopted also for future time steps. As regards the choice of shocks sizes, in order to avoid any arbitrary element, we decide to compute forward Vega by using the re-calibrated $\hat{\sigma}$ and $\hat{\eta}$, corresponding to $\epsilon_{\sigma} = 1\%$ and $ \epsilon_{\eta} = 4\%$.

\subsubsection{Implied Volatility Calculation}
\label{sec:Implied Volatility Calculation}

\begin{figure}[!htbp]
	\centering		
	\includegraphics[width=0.55\linewidth]{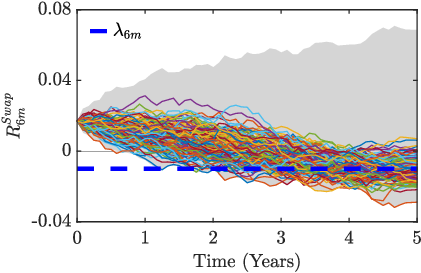}		
\caption{MC time simulation of 5x10Y forward swap rate. The grey area represents the cloud for the stochastic trajectories. Coloured paths are the ones for which the swap rate falls below the Black shift $\lambda_{6\text{m}}(t_0)=1\%$ for at least one time step.}
\label{fig:swpt_ATM_swap_rates}
\end{figure}

Looking closely at the Monte Carlo simulation of forward swap rates, we find that some paths exhibit deeply negative rates, exceeding (in absolute terms) the value of the Black shift $\lambda_{x}(t_0)$ used at the initial time step $t_0$ in the calibration of the model parameters. This feature prevents the calculation of Black implied volatilities at future time steps, needed to compute Vega sensitivity according to eq.~\ref{eq:vega_workaround_swpt5x10}.
In fig.~\ref{fig:swpt_ATM_swap_rates} we show the MC simulation of the 5x10Y forward swap rate, where in 2538 paths out of 5000 (51\% of the total) the rate falls below $\lambda_{6\text{m}}(t_0)=1\%$ for at least one time step.
\begin{table}[!htbp]
	\centering
	\begin{tabular}{c c c c}
		\toprule
		\multicolumn{1}{c}{$\lambda_{6\text{m}}$} & \multicolumn{1}{c}{$\sigma_{6\text{m}}^{\text{Blk}}$} & \multicolumn{1}{c}{$\nu$} & \multicolumn{1}{c}{VR}  \\
		\midrule
		0.01  & 0.2284 & 21522024 & 4914735 \\
		0.04  & 0.1302 & 38342983  & 4992724 \\
		0.06  & 0.0793 & 63259702 & 5016378 \\
		0.08  & 0.0629 & 79814066 & 5021578 \\
		0.10  & 0.0521 & 96350056 & 5024345 \\
		\bottomrule
	\end{tabular}%
	\caption{5x10Y ATM physically settled European payer Swaption, EUR 100 Mio nominal amount. Comparison between shifted-Black implied volatility $\sigma_{6\text{m}}^{\text{Blk}}(t_0)$, Vega sensitivity $\nu(t_0)$ and Vega Risk VR$(t_0)$ at time $t_0$ for different Black shifts $\lambda_{6\text{m}}$. Calculation as in eq.~\ref{eq:vega_workaround_swpt5x10}, with $\epsilon_{\sigma} = 0.01$ and $\epsilon_{\eta} = 0.04$.}
	\label{tab:implied_volatility_displacements}
\end{table}
In light of this fact, in order to ensure Vega sensitivity calculation for each time step and path, we are forced to use Black shift values larger than those necessary and sufficient at time step $t_0$. To this end, we analysed the impact of different Black shifts on shifted-Black implied volatility, Vega sensitivity and Vega Risk at $t_0$. The results for the 5x10Y ATM Swaption are reported in tab.~\ref{tab:implied_volatility_displacements}. 
We observe that shifted-Black implied volatility and Vega sensitivity are highly impacted by the different black shift but the Vega Risk, given by the product of the two quantities (see eq.~\ref{eq:Vega_Risk}), is fairly stable, differing up to a maximum of 2\% w.r.t. the case $\lambda_{6\text{m}}=1\%$. 
\par 
We conclude that Black shift values larger than those typically used at $t_0$ ensure the inversion of the Black formula for each path with an acceptable accuracy in Vega Risk. For this reason we set $\lambda_{6\text{m}} = 6\%$ in our calculations.

\subsection{XVA Sensitivities to CSA Parameters}
\label{sec:XVA Sensitivities to CSA Parameters}

In order to establish the most relevant CSA parameters driving the exposure and the XVA, we analysed the corresponding sensitivities with respect to the most important CSA parameters, i.e. the margin threshold K, the minimum transfer amount MTA, and the length of margin period of risk (MPoR), keeping the other model parameters as in tab.~\ref{tab:model_setup}. 
In carrying out this analysis we distinguished between the following three collateralization schemes for both Swaps and Swaptions: i) XVA with VM only; ii) XVA with VM and IM, with K and MTA applied on VM only; iii) XVA with VM and IM, with K and MTA applied on both VM and IM.
\par 
We report all the results in app.~\ref{app:XVA Sensitivities to CSA Parameters}.
Overall, we found that MPoR does not contribute significantly with respect to K and MTA parameters. In particular, the threshold K is the most important parameter. As expected, for increasing values of K and MTA, the collateralized XVA converges to the uncollateralized value. 
We notice that this analysis does not identifies XVA model risk factors, but it is very useful in practical situations, in particular when collateral agreements are negotiated, as widely happened during the financial benchmarks reform.

\subsection{Monte Carlo vs Analytical XVA}
\label{sec:Monte Carlo Versus Analytical XVAs}

As shown in app.~\ref{app:XVA analytical formulas}, in the case of uncollateralized Swaps, there exist analytical XVA formulas in terms of an integral over the values of co-terminal European Swaptions (see eqs. \ref{eq:CVA_exact_formula_int} and \ref{eq:DVA_exact_formula_int}).
\begin{table}[!htbp]
  \centering
    \begin{tabular}{c c c c r}
    \toprule
    Item &  \multicolumn{1}{c}{$\Delta t$} & {An. G2++ (\euro)} & {An. Black (\euro)} & \multicolumn{1}{c}{MC (\euro)}  \\
    \midrule
    \multirow{5}[1]{*}{CVA ATM Swap} & 12M   & -1091604* & -1016082 & -1119629   ($\pm4$\%)\\
    & 6M    & -1157012* & -1077346 & -1170271  ($\pm4$\%) \\
    & 3M    & -1147871* & -1067726 & -1172139   ($\pm4$\%)\\
    & 1M    & -1152567* & -1072231 & -1172938   ($\pm4$\%)\\
    & 1D    & -1156083* & -1075398 & -1173296   ($\pm4$\%)\\
	\midrule   
    \multirow{5}[1]{*}{DVA ATM Swap} & 12M   & 454023* & 432373 & 485660  ($\pm8$\%)\\
    & 6M    & 471340* & 449890* & 478016   ($\pm9$\%)\\
    & 3M    & 468884* & 447358* & 476789   ($\pm9$\%)\\
    & 1M    & 468721* & 447369* & 475324  ($\pm9$\%)\\
    & 1D    & 469562* & 448170* & 474469   ($\pm9$\%)\\
    \midrule 
    \multirow{5}[1]{*}{CVA OTM Swap} & 12M   & -744144* & -698128* & -732977  ($\pm6$\%) \\
    & 6M    & -783485* & -735481 & -778892   ($\pm6$\%)\\
    & 3M    & -773913* & -725498 & -779280   ($\pm6$\%)\\
    & 1M    & -774432* & -726025 & -779085   ($\pm6$\%)\\
    & 1D    & -775955* & -727392 & -778991  ($\pm6$\%) \\
	\midrule 
    \multirow{5}[1]{*}{DVA OTM Swap}  & 12M   & 788254 & 793430 & 892169   ($\pm6$\%)\\
    & 6M    & 833226* & 840902* & 880117  ($\pm6$\%)\\
    & 3M    & 839022* & 847065* & 882972  ($\pm6$\%)\\
    & 1M    & 846586* & 855381* & 884382  ($\pm6$\%)\\
    & 1D    & 851479* & 860467* & 885227  ($\pm6$\%) \\
    \bottomrule
    \end{tabular}
\caption{XVA figures obtained for 15Y ATM/OTM payer Swap, EUR 100 Mio nominal amount. $\Delta t$  denotes the time step adopted in the XVA analytical formulas \ref{eq:CVA_exact_formula_int} and \ref{eq:DVA_exact_formula_int}. The strip of co-terminal Swaptions has been computed with the G2++ pricing formula in the third column and with Black formula in the fourth column. Results consistent with the $3\sigma$ MC error are marked with an asterisk.}
\label{tab: IRS_15Y_analyticXVA}
\end{table}
The corresponding numerical solution requires the discretization of the integrals on a time grid (see eqs.~\ref{eq:CVA_exact_formula_disc} and \ref{eq:DVA_exact_formula_disc}), whose granularity clearly introduces a model risk in the XVA figures. 
Therefore, we tested these formulas for different time grids with different frequencies. Moreover, given the model independent nature of this approach, we calculated co-terminal Swaptions' prices according to two different approaches:
\begin{enumerate}
\item using our G2++ model, eq.~\ref{eq:g2++_swpt_multi_curve}, with G2++ parameters in tab.~\ref{tab:model_setup};
\item using the shifted-SABR model, using shifted-Black formulas and shifted-lognormal SABR volatilities (see \cite{HagKum02,Obl07}) calibrated on the market swaption cube for each available smile section. We stress that this approach is not straightforward, since typically only a few co-terminal Swaptions entering into the XVA analytical formula correspond to quoted smile sections, where the SABR formula can be directly used. All the remaining Swaptions insisting on non-quoted smile sections require delicate interpolation/extrapolation of the calibrated SABR parameters (see app.\ref{app:XVA analytical formulas} for further details).
\end{enumerate}
The purpose of this analysis is twofold: one the one hand, we want to test the results of the Monte Carlo approach against analytical formulas, on the other hand, we want to quantify the model risk stemming from the use of  alternative pricing models. 
\par 
The results for the 15Y ATM and OTM payer Swaps are shown in tab.~\ref{tab: IRS_15Y_analyticXVA},	 
We observe that, with respect to the Monte Carlo approach (last column), analytical formulas generally underestimate CVA and DVA values. 
The G2++ results (third column) are always consistent with the $3\sigma$ Monte Carlo error (marked with an asterisk), with only one exception (the DVA of the OTM Swap with annual time grid granularity). This evidence  confirms the robustness of the Monte Carlo simulation parameterized as discussed in the previous sec. \ref{sec:XVAs MC convergence}. 
Instead, the SABR results (fourth column) show considerable differences, particularly for the CVA, which is consistent with $3\sigma$ Monte Carlo error only in one case. This is not surprising, since we are using two completely different dynamics of the underlying risk factors (G2++ for MC vs SABR for anaytical).
In terms of computational performance, the analytical approach is obviously much faster than that of Monte Carlo approach: even with a daily time grid robust results can be obtained almost immediately, meaning that, for an uncollateralized Swap, the analytical approach can replace the Monte Carlo approach whereas performance is critical.

\subsection{Tuning Accuracy vs Performance}
\label{sec:accuracy vs performance}

The model validations performed in the previous sections allowed to identify the most important model risk factors and the corresponding calculation parameters governing the XVA framework and affecting the XVA figures, both in terms of accuracy and computational performance, which we summarize in tab. \ref{tab:model_setup} below. 
\begin{table}[h]
	\centering
	\begin{tabular}{c c c c}
		\toprule
		\multicolumn{1}{c}{Parameters Class} & \multicolumn{1}{c}{Parameter} & Sec. & \multicolumn{1}{c}{Value}  \\
		\midrule
		G2++ model parameters & $p$ & \ref{sec:G2++ Model Calibration} & tab.~\ref{tab:calibrations_parameters} \\
		\midrule
		\multirow{5}{*}{Calculation parameters}&
		$\Delta t$ & \ref{sec:Time Simulation Grid} & $1$ month \\
		& $N_{MC}$ & \ref{sec:XVAs MC convergence} & $5000$ \\
		& $\epsilon_{\sigma}$ & \multirow{3}{*}{\ref{sec:Vega Sensitivity time simulation}} & $0.01$\\
		& $\epsilon_{\eta}$ & & $0.04$\\
		& $\lambda_{x}$ & & $0.06$ \\
		\midrule
		\multirow{3}{*}{CSA parameters}&
		$l$   & \multirow{3}{*}{\ref{sec:XVA Sensitivities to CSA Parameters}} & $2$ days \\
		& K   & & $0$ EUR\\
		& MTA & & $0$ EUR \\
		\bottomrule
	\end{tabular}
	\caption{XVA framework parameters and respective values adopted according to the model validations performed in the previous sections.}
	\label{tab:model_setup}
\end{table}
\par
In the last column we report the parameter values identified in our model validation analyses which set our acceptable compromise between accuracy and performance. 
Regarding the CSA parameters, we considered bilateral CSA with $\text{K}=\text{MTA}=0$ both for VM and IM, with $l=2$ days, which is a common practice and also get close to the perfect collateralization case.
\par 
Essentially, the most important parameters are the number of time steps $N_S$ in the time simulation grid and the number $N_{MC}$ of Monte Carlo scenarios. Their product $N_C = N_S \times N_{MC}$ is proportional to the computational time $T_C$ required for the XVA calculation, i.e. $T_C = \alpha N_C$, where the proportionality coefficient $\alpha$ depends on the hardware available, and all the rest being the same. 
Hence, given a computational budget  $\Delta T$, i.e. the maximum time that one is willing to wait to compute the XVA figures, tuning precision vs performance roughly amounts to set $N_S$ and $N_{MC}$ such that $T_C \leq \Delta T$.
\par 
We stress that this choice is not unique, since it depends on the specific context, in particular: 
i) the trades or portfolio under analysis, 
ii) the presence of collateral, in particular the IM,
iii) the hardware available, 
iv) the calculation time constraints,
v) the desired level of accuracy, 
vi) the purpose of the XVA calculation, e.g. either structuring a single trade for a client, or end of day XVA revaluation, or end of quarter accounting fair value measurement,
vii) the purpose of the model validation, e.g. either for the Front Office quants developing the XVA engine for the XVA trading desk, or for the Model Validation quants challenging the Front Office framework.
\par
The considerations above answer to our first and second research questions reported in sec. \ref{sec:introduction}.

\section{XVA Model Risk}
\label{sec:Model Risk}

According to the EU regulation (see \cite{EUParlPrd13,EUComPrd16}) financial institutions are required to apply prudent valuation to fair-valued positions in order to mitigate their valuation risk, i.e. the risk of losses deriving from the valuation uncertainty in the exit price of financial instruments.
The prudent value has to be computed on the top of the fair value, including possible fair valuation adjustments accounted in the income statement, considering 9 different valuation risk factors at the $90\%$ confidence level from a distribution of exit prices. The corresponding 9 differences between the prudent value and the fair value, called Additional Valuation Adjustments are aggregated and finally deducted from the Common Equity Tier 1 (CET1) capital
In particular, the Model Risk (MoRi) AVA, envisaged in art.~11 of \cite{EUComPrd16}, comprises the valuation uncertainty linked to the ``\emph{potential existence of a range of different models or model calibrations used by market participants}''. Accordingly, for MoRi AVA the prudent value at a 90\% confidence level corresponds to the 10\textsuperscript{th} percentile of the distribution of the plausible prices obtained from different models/parameterizations\footnote{Notice that we conventionally adopt positive/negative prices for assets/liabilities.}. 
\par 
Hence, we compute a MoRi AVA based on the analyses described in the previous sec.~\ref{sec:XVA model validation}. In particular, we build the distribution of XVA exit prices by considering the following four sources of model risk: i) G2++ model calibration approach (see sec.~\ref{sec:G2++ Model Calibration}), ii) time grid construction approach and related granularity $\Delta t$ (see sec.~\ref{sec:Time Simulation Grid}), iii) number of MC scenarios $N_{MC}$ (see sec.~\ref{sec:XVAs MC convergence}), and iv) the fast analytical XVA formulas for uncollateralized Swaps with different time grid granularities and SABR pricing formulas for the strip of co-terminal Swaptions (see sec.~\ref{sec:Monte Carlo Versus Analytical XVAs}). 
\par
\begin{figure}[!htbp]			
	\begin{subfigure}{0.49\linewidth}
		\centering
		\caption{CVA distribution Swap no collat.}
		\includegraphics[width=0.9\linewidth]{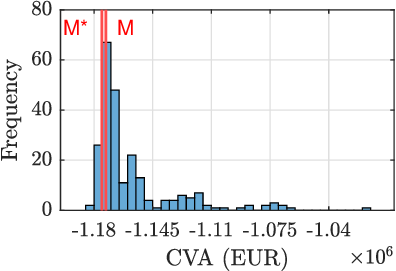}	
		\label{fig:irs_15Y_ATM_CVA_hist}
	\end{subfigure}
	\hfill
	\begin{subfigure}{0.49\linewidth}
		\centering
		\caption{CVA distribution Swaption no collat.}
		\includegraphics[width=0.9\linewidth]{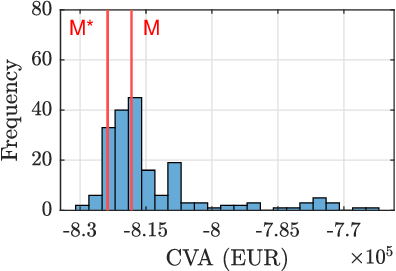}	
		\label{fig:swpt_5x10Y_ATM_CVA_hist}
	\end{subfigure} \\
	\caption{CVA distributions for the 15 years ATM payer Swap (left-hand side) and for the 5x10 years ATM physically settled European payer Swaption (right-hand side) without collateral. Red lines denote the XVA framework ($M$) and the prudent XVA framework ($M^{*}$).}
	\label{fig:irs_swpt_CVA_histogram}
\end{figure}

 
According to the ranges of parameter values examined in sec.~\ref{sec:XVA model validation} for the four sources of model risk above, we would obtain a distribution of XVA exit prices including 1440 points for the uncollateralized Swaps\footnote{1440 = 7 G2++ calibrations x 2 time grids x 5 values of $\Delta t$ x 20 values of $N_{MC}$ + 40 analytical XVA, where 40 = 7 G2++ calibrations x 5 $\Delta t$ for G2++ model + 5 $\Delta t$ for SABR model.} and 1400 points for the uncollateralized Swaption\footnote{1400 = 1440 - 40, since for the Swaption there are no XVA analytical formulas.}.
Since the production of such an high number of XVA exit prices is computationally prohibitive, we restrict our analysis by considering 236 points\footnote{236 = 7 G2++ calibrations x (2 time grids + 5 $\Delta_t$ + 20 $N_{MC}$ + 1 baseline) + 40 analytical XVA.} for the Swap, and 196 points\footnote{196 = 236 - 40.} for the Swaption. 
Therefore, we compute MoRi AVA at time $t_0$ as\footnote{\cite{EUComPrd16} prescribes an aggregation coefficient equal to 0.5 to take into account diversification benefit, which we do not consider here.}
\begin{align}
\begin{split}
\text{AVA}^{\text{MoRi}}(t_0) 
&= V(t_0;M) - \text{PV}(t_0;M^{*})\\
&= \text{CVA}(t_0;M) - \text{CVA}(t_0;M^{*}),
\end{split}
\label{eq:AVA}
\end{align}
where:
\begin{itemize}
	\item $V(t_0;M) = V_0(t_0) + \text{XVA} \left(t_0; M \right)$ is the fair-value of the instrument, intended as the price obtained from our XVA framework, denoted here with $M$ (see tab.~\ref{tab:model_setup});
	\item $\text{PV}(t_0;M^{*}) = V_0(t_0) + \text{XVA}(t_0;M^{*})$ is the prudent value obtained from the prudent XVA framework, denoted by $M^{*}$, determined as the 10\textsuperscript{th} percentile of the XVA exit price distribution. In other words, $M^{*}$ ensures that one can exit the XVA at a price equal to or larger than $\text{PV}(t_0;M^{*})$ with a degree of certainty equal to or larger than 90\%\footnote{Notice that, according to our conventions, $\text{PV}(t_0) \leq V(t_0)$ and $\text{AVA}(t_0)\geq 0$};
	\item the final formula reduces to the CVA only since the EU regulation expressly excludes any own credit risk component, as the DVA, which is filtered out from the CET1 capital; furthermore, we are not considering the valuation uncertainty related to the base value $V_0(t_0)$.
\end{itemize}
We report in figs.~\ref{fig:irs_15Y_ATM_CVA_hist} and \ref{fig:swpt_5x10Y_ATM_CVA_hist} the CVA distributions for the Swap and the Swaption, respectively. We observe that both distributions have a positive skew, with many points concentrated around the left tail. The less conservative points falling in the right tail are attributable to the analytical formulas and to MC simulations with a low number of scenarios and/or less granular time grids for both instruments, as visible in tabs.~\ref{tab:swap_AVA_MoRi_noCollateral_app} and \ref{tab:swpt_AVA_MoRi_noCollateral_app} which detail these distributions.
In the following tabs.~\ref{tab:swap_AVA_MoRi_noCollateral} and \ref{tab:swpt_AVA_MoRi_noCollateral} we show a summary of the full tabs.~\ref{tab:swap_AVA_MoRi_noCollateral_app} and \ref{tab:swpt_AVA_MoRi_noCollateral_app}. 
Looking at the 15Y Swap, the MoRi AVA, corresponding to the prudent XVA framework ${M_{24}}$ (calibration $p_2$, joint grid, $\Delta t = 1$M and $N_{MC} = 18000$), is equal to $0.20\%$ of the CVA obtained from our XVA framework. 
For the 5x10Y Swaption, the MoRi AVA, corresponding to ${M_{20}}$ (calibration $p_1$, joint grid, $\Delta t = 1$M and $N_{MC} = 14000$), is equal to $0.66\%$ of the CVA.
\begin{table}[!htbp]
	\centering
	\begin{tabular}{l l l l l l l l l}
		\toprule
		& \multicolumn{6}{c}{XVA framework} &       &       \\
		\cline{2-7}
		\multirow{3}*{Item} &	\multirow{3}*{$M_j$} & \multirow{3}*{Model} & \multicolumn{4}{c}{Parameters} & \multicolumn{1}{c}{\multirow{3}*{CVA (\euro)}}   & \multicolumn{1}{c}{\multirow{3}*{DVA (\euro)}}    \\
		\cline{4-7} 
		& &       & G2++ & Time & $\Delta t$ & $N_{MC}$  &       &        \\
		& & & calib. & grid & & & &  \\
		\midrule 
		Min & $M_{1}$ & MC    & $p_{2}$ & Joint & 1M    &  6000 &  -1180939 & 495052  \\
		10\textsuperscript{th} pct & $M_{24}$ & MC    & $p_{2}$ & Joint & 1M    &  18000 &  -1175336** & 486628  \\
		 & $M$ & MC    & $p$ & Joint & 1M    &  5000 &  -1172938* & 475324  \\
		50\textsuperscript{th} pct & $M_{117}$ & MC    & $p_{4}$ & Joint & 1M    &  8000 &  -1167879 & 459438  \\
		90\textsuperscript{th} pct & $M_{212}$ & MC    & $p_{1}$ & Joint & 12M   &  5000 &  -1121235 & 483889  \\
		Max & $M_{235}$ & An Blk & NA    & Std.  & 12M   & NA    & -1016082 & 432373 \\
		\midrule
		\multicolumn{6}{l}{XVA$(t_0;M)$} & & -1172938 & 475324  \\
		\multicolumn{6}{l}{XVA$(t_0;M_{24})$} & & -1175336 & 486628 \\
		\multicolumn{6}{l}{AVA$^{\text{MoRi}}(t_0)$} & & \multicolumn{1}{c}{2398} & \\
		\bottomrule
	\end{tabular}
	\caption{15 years ATM payer Swap, no collateral. Summary of the distribution of XVA exit prices reported in tab.~\ref{tab:swap_AVA_MoRi_noCollateral_app}, sorted by increasing CVA. Each row represents a different combination of model and parameters $M_j$, $j = 1,...,235$. $M$ labels CVA obtained from XVA framework denoted by an asterisk, CVA obtained from prudent XVA framework $M_{24}$ denoted by double asterisk.} 
	\label{tab:swap_AVA_MoRi_noCollateral}
\end{table}
\begin{table}[!htbp]
	\centering
	\begin{tabular}{l l l l l l l l l}
		\toprule
		& \multicolumn{6}{c}{XVA framework} &       &       \\
		\cline{2-7}
		\multirow{3}*{Item} &	\multirow{3}*{$M_j$} & \multirow{3}*{Model} & \multicolumn{4}{c}{Parameters} & \multicolumn{1}{c}{\multirow{3}*{CVA (\euro)}}   & \multicolumn{1}{c}{\multirow{3}*{DVA (\euro)}}    \\
		\cline{4-7} 
		& &       & G2++ & Time & $\Delta t$ & $N_{MC}$  &       &        \\
		& & & calib. & grid & & & &  \\
		\midrule 
		Min & $M_{1}$ & MC    & $p_{2}$ & Joint & 1M    & 7000  & -829878 & 37102 \\
		10\textsuperscript{th} pct & $M_{20}$ & MC    & $p_{1}$ & Joint & 1M    & 14000 & -823724** & 37859\\
		 & $M$ & MC    & $p$ & Joint & 1M    & 5000  & -818308* & 36418\\
		50\textsuperscript{th} pct & $M_{97}$ & MC    & $p_{6}$ & Joint & 3M    & 5000  & -817586 & 33997\\
		90\textsuperscript{th} pct & $M_{176}$ & MC    & $p_{3}$ & Joint & 1M    & 2000  & -794334 & 35981\\
		Max & $M_{195}$ & MC    & $p_{3}$ & Std.  & 12M   & 5000  & -764292 & 39784\\
		\midrule
		\multicolumn{6}{l}{XVA$(t_0;M)$} & & -818308 & 36418\\
		\multicolumn{6}{l}{XVA$(t_0;M_{20})$} & & -823724 & 37859\\
		\multicolumn{6}{l}{AVA$^{\text{MoRi}}(t_0)$} & & \multicolumn{1}{c}{5416} &\\
		\bottomrule
	\end{tabular}
	\caption{5x10 years ATM physically settled European payer Swaption, no collateral. Summary of the distribution of XVA exit prices reported in tab.~\ref{tab:swpt_AVA_MoRi_noCollateral_app}. The structure of the table is the same of tab.~\ref{tab:swap_AVA_MoRi_noCollateral}.}
	\label{tab:swpt_AVA_MoRi_noCollateral}
\end{table}
\par 
In conclusion, we observe that the small relative AVA values are due to the fact that our XVA framework produces already conservative CVA figures, and most of the mass of the XVA distribution is concentrated in the left tail. 
A different compromise between accuracy and performance, e.g. faster but less accurate, may produce more significant relative AVA values, leading to non-negligible CET1 reductions in the case of large financial institutions with important XVA figures.
\par 
The considerations above answer to our third research question reported in sec. \ref{sec:introduction}.

\section{Conclusions}
\label{sec:conclusions}

In this work we investigated the XVA model risk. 
To this scope we focused on an industry-standard realistic and complete XVA modelling framework, typically used by XVA trading desks, based on multi-curve time-dependent volatility G2++ stochastic dynamics calibrated on real market data, i.e.~distinct discounting and forwarding yield curves, CDS spread curves, and swaption volatility cube. The numerical XVA calculation is based on a multi-step Monte Carlo simulation including both dynamic variation margin and initial margins under the ISDA Standard Initial Margin Model. 
We applied this framework to the most common linear and non-linear interest rates derivatives, i.e. Swaps and European Swaptions with different maturities and strikes. 
Within this context, we formulated  in sec. \ref{sec:introduction} three research questions, to which we report the corresponding answers below.
\begin{itemize}
	
	\item[A1] Within the XVA modelling framework above, we were able to identify and investigate the most important model risk factors to which XVA exposure modelling and thus XVA are most sensitive, and to measure the associated computational effort. 
	In particular, we showed that a crucial model risk factor is the construction of a MC time simulation grid able to capture the spikes arising in collateralized exposure during the margin period of risk, which have a material impact on XVA figures. To this end, we proposed a strategy to build a parsimonious and efficient grid which ensures to capture all the spikes and reduce the computational effort. 
	Regarding the MC simulation, we observed a convergence of XVA figures even for a limited number of MC scenarios, leaving room for further saving of computational time if necessary. 
	Regarding the simulation of the initial margin, further assumptions are required to compute G2++ forward vega sensitivities according to the ISDA-SIMM prescriptions. The related model risk has been addressed by ensuring that our calculation strategy is aligned with two alternative approaches, both consistent with ISDA-SIMM definition at time $t_0$ (i.e.~valuation date) and by avoiding any arbitrary elements in the choice of the associated parameters. 
	Finally, we showed that XVA analytical formulas for uncollateralized Swaps represent a useful tool for validating the MC results and to speed up XVA calculations. In this case we found that model risk arises from the discrete time grid used in the XVA analytical formula and from the model used to price the corresponding strip of co-terminal European Swaptions. XVA figures obtained with the G2++ pricing formulas, consistent with the G2++ dynamics of the underlying risk factors, resulted to be superior with respect to those obtained with the SABR model, which assumes a different dynamics.
	
	\item[A2] The model risk analyses above allowed to identify a parametrization of the XVA modelling framework allowing a compromise between accuracy and performance, i.e. leading to sufficiently robust XVA figures in a reasonable time, a very important feature for practical applications. Obviously, this choice is not unique, and our analyses allow to adapt the parameters to different contexts and purposes of the XVA calculation. 
	
	\item[A3] Finally, based on the large number of different parameterizations considered in the analyses above, we were able to estimate the XVA model risk using the Additional Valuation Adjustment (AVA) envisaged by the EU regulation as the 10th percentile of the XVA distribution, corresponding to the 90\% confidence level for XVA.
\end{itemize}
\par 
Our framework is general and could be extended to include other valuation adjustments, e.g.~Funding Valuation Adjustment (FVA) and Margin Valuation Adjustment (MVA), other financial instruments, and XVA calculation at portfolio level. The computational performance could be enhanced by using last generation high dimensional scrambled Sobol sequences generators, which allow to reduce the number of scenarios while keeping the MC error under control (see. e.g.~\cite{AtaKuc20,ScoBia21}). 
Adjoint algorithmic differentiation (\cite{CapGil12,HugSav20}) or Chebyshev decomposition (\cite{MarPal21}) could be be used to speed up and stabilize sensitivities calculation for initial margin modelling.


\begin{appendices}

\section{Theoretical Framework}
\label{app:theoretical framework}

In this appendix we detail the theoretical framework used in this work.

\subsection{Pricing with Collateral and XVA}
\label{app:pricing approach}

We describe here the general no-arbitrage, additive pricing formulas for financial instruments subject to XVA.

Assuming no arbitrage and the usual probabilistic framework ($\Omega,\mathcal{F},\mathcal{F}_t,Q$) with market filtration $\mathcal{F}_{t}$ and risk-neutral probability measure $Q$, the general pricing formula of a financial instrument with payoff $V(T)$ paid at time $T>t$ is
\begin{align}
V(t) &= V_{0}(t) + \text{XVA}(t)\label{eq:fair value},\\
V_{0}(t) &=\mathbb{E}^{Q}\Brack{D(t;T)V(T)\vert\mathcal{F}_t}
=P(t;T) \mathbb{E}^{Q^T}\Brack{V(T)\vert\mathcal{F}_t},\label{eq:base value}\\
D(t;T) &= \dfrac{B(t)}{B(T)}=e^{-\int_{t}^{T}r(u)du},\label{eq:stochastic discount factor}\\
P(t;T) &=\mathbb{E}^{Q}\Brack{D(t;T)\vert\mathcal{F}_t},\label{eq:ZCBrf}
\end{align}
where the base value\footnote{In order to ease the notation, in the following sections we omit subscript 0 unless clearly necessary, denoting the base value simply with $V$.}, or mark to market), $V_{0}(t)$ in eq.~\ref{eq:base value} is interpreted as the price of the financial instrument under perfect collateralization\footnote{An ideal Credit Support Annex (CSA) ensuring a perfect match between the price $V_{0}(t)$ and the corresponding collateral at any time $t$. This condition is realised in practice with a real CSA minimizing any friction between the price and the collateral, i.e. with daily margination, cash collateral in the same currency of the trade, flat overnight collateral rate, zero threshold and minimum transfer amount.}, the discount (short) rate $r(t)$ in eq.~\ref{eq:stochastic discount factor} is the corresponding collateral rate, $B(t)$ is the collateral bank account growing at rate $r(t)$, $D(t;T)$ is the stochastic collateral discount factor, $P(t;T)$ is the perfectly collateralized Zero Coupon Bond (ZCB) price, and $Q^T$ is the $T$-forward probability measure associated to the numeraire $P(t;T)$. 

Valuation adjustments in eq.~\ref{eq:fair value}, collectively named XVA, represent a crucial and consolidated component in modern derivatives pricing which takes into account additional risk factors not included among the risk factors considered in the base value $V_0$ in eq.~\ref{eq:base value}. 
These risk factors are typically related to counterparties default, funding, and capital, leading, respectively, to Credit/Debt Valuation Adjustment (CVA/DVA), Funding Valuation Adjustment (FVA), often split into Funding Cost/Benefit Adjustment (FCA/FBA), Margin Valuation Adjustment (MVA), Capital Valuation Adjustment (KVA). A complete discussion on XVA may be found e.g. in \cite{Gre20}. 
For XVA pricing we must consider the enlarged filtration 
$ \mathcal{G}_{t}= \mathcal{F}_{t}\vee\mathcal{H}_{t}\supseteq\mathcal{F}_{t}$ 
where $ \mathcal{H}_{t}=\sigma(\{\tau\leq u\}:u\leq t)$ 
is the filtration generated by default events. More
details can be found in a number of papers, see e.g.~\cite{BriMor13,Bri18,BriFraPal19} and references therein.

\subsection{Financial Instruments}
\label{app:financial instruments and pricing formulas}

We describe here the detailed list of financial instruments considered in this work and their corresponding pricing formulas.

\subsubsection{Instruments' List}
\label{app:financial instruments}

According to the discussion in sec. \ref{sec:XVA numerical calculations}, we show in the following tab.~\ref{tab:instruments} the complete list of financial instruments considered in this work.
\begin{table}[h]
	\centering 
	\begin{tabular}{l r c l}
		\toprule
		\multicolumn{1}{c}{Instrument} & \multicolumn{1}{c}{$\omega$} & $K$ & \multicolumn{1}{c}{Moneyness} \\ 
		\midrule
		\multirow{3}{*}{15Y Swap } & 1  & $ 0.0167 $& out-of-the-money \\
		& 1 & $0.0117$ & at-the-money \\
		& 1 & $0.0067$ & in-the-money   \\
		\midrule
		\multirow{3}{*}{30Y Swap } & 1  & $ 0.0188 $ & out-of-the-money \\
		& 1 & $0.0138$ & at-the-money  \\
		& 1 & $0.0088$ & in-the-money   \\
		\midrule
		\multirow{3}{*}{5x10Y forward Swap} & -1 & $ 0.0120 $ & out-of-the-money \\
		& 1 & $0.0170$ & at-the-money  \\
		& 1 & $0.0220$ & out-of-the-money \\
		\midrule
		\multirow{3}{*}{5x10Y Swaption} & -1 & $0.0120$& out-of-the-money \\
		& 1 & $0.0170$ & at-the-money  \\
		& 1 & $0.0220$ & out-of-the-money \\
		\bottomrule
	\end{tabular}
	\caption{Details on the financial instruments analysed in this work. All instruments are denominated in EUR, with nominal amount $N=100$ Mio, semi-annual floating leg tied to EURIBOR 6M, annual fixed leg. $\omega = +/- 1$ stands for payer/receiver (referred to the fixed leg of the underlying Swap) and $K$ denotes the Swaps' fixed rate or the Swaption's strike.}
	\label{tab:instruments}
\end{table}

\subsubsection{Interest Rate Swap}
\label{app:Interest Rate Swap}

A Swap is a contract which allows the exchange of a fixed rate $K$ against a floating rate, characterised by the following time schedules 
\begin{equation}
	\begin{split}
		& \textbf{S}=\{S_{0},\dots,T_{i},\dots,S_{m}\}\text{, fixed leg schedule,} \\
		& \textbf{T}=\{T_{0},\dots,T_{j},\dots,T_{n}\}\text{, floating leg schedule,} \\
		& S_0=T_{0} \text{, } S_m=T_{n},
	\end{split}
	\label{eq:swap_schedules}
\end{equation}
and by the following payoffs for the fixed and floating cash flows, respectively,
\begin{equation}
\begin{split}
&\textbf{Swaplet}_{\text{fix}}(S_i;S_{i-1},S_i,K) = NK\tau_{K}(S_{i-1},S_i), \\
&\textbf{Swaplet}_{\text{float}}(T_j;T_{j-1},T_j) = NR_x(T_{j-1},T_j)\tau_{R}(T_{j-1},T_j),
\end{split}
\label{eq:swaplet_payoffs}
\end{equation}
where $\tau_{K}$ and $\tau_{R}$ are the year fractions for fixed and floating rate conventions, respectively, and $R_x(T_{j-1},T_j)$ is the underlying spot floating rate with tenor $x$, consistent with the time interval $\left[ T_{j-1}, T_{j} \right]$ (e.g. $x=6M$ for EURIBOR 6M and semi-annual coupons). 

The price of the Swap at time $t\leq T_n = S_m$ is given by the sum of the prices of fixed and floating cash flows occurring after $t$,
\begin{equation}
\begin{split}
\textbf{Swap}(t;\textbf{T},\textbf{S},K,\omega) & =  N \omega  \left[ \sum_{j=\eta_L(t)}^{n} P(t;T_j) F_{x,j}(t)\tau_{R}(T_{j-1},T_j) - K A(t;\textbf{S}) \right] \\
A(t;\textbf{S}) & = \sum_{i=\eta_K(t)}^{m}P(t;S_i)\tau_{K}(S_{i-1},S_i),
\end{split}
\label{eq:swap_price}
\end{equation}
where $N$ is the nominal amount, $\omega = +/- 1$ denotes a payer/receiver Swap (referred to the fixed leg), $\eta_L(t)=\min\{j\in\{1,...,n\}\text{ s.t. } T_j\geq t\}$ and $\eta_K(t)=\min\{i\in\{1,...,m\}\text{ s.t. } S_j\geq t\}$ are the first future cash flows in the Swap's schedules, $A(t;\textbf{S})$ is the Swap annuity, and $F_{x,j}(t)$ is the forward rate observed at time $t$, fixing at future time $T_{j-1}$ and spanning the future time interval $\Brack{T_{j-1},T_{j}}$, given by
\begin{equation}
\begin{split}
&F_{x,j} (t) = F_x(t;T_{j-1},T_j) = \mathbb{E}^{Q^{T_j}} \left[ R_x(T_{j-1},T_j) \vert\mathcal{F}_t \right],\\
&F_{x,j} (t) = R_x(T_{j-1},T_j)\;\textit{for } t\geq T_{j-1}.
\end{split}
\label{eq:forward_rate}
\end{equation}
By construction, the forward rate $F_{x,j}(t)$ is a martingale under the forward measure $Q^{T_j}$ associated to the discounting numeraire $P(t;T_j)$.
The par Swap rate $R_{x}^{\text{Swap}}(t;\textbf{T},\textbf{S})$, i.e.~the fixed rate $K$ such that the Swap is worth zero, is given by
\begin{equation}
R_{x}^{\text{Swap}}(t;\textbf{T},\textbf{S})= \dfrac{\sum_{j=\eta_L(t)}^{n} P(t;T_j) F_{x,j}(t)\tau_{R}(T_{j-1},T_j)}{A(t;\textbf{S})}.
\label{eq:Swap_Rate_R}
\end{equation}
The Swap's price in terms of the par Swap rate can be expressed as 
\begin{equation}
\textbf{Swap}\left(t;\textbf{T},\textbf{S},K,\omega\right) = N\omega \left[ R_{x}^{\text{Swap}}(t;\textbf{T},\textbf{S})-K \right] A(t;\textbf{S}).
\label{eq:Swap_price_R}
\end{equation}

IBOR forward rates $F_{x,j}(t)$ in eqs. \ref{eq:swap_price}-\ref{eq:Swap_Rate_R} are computed from IBOR ZCB curves 
$\mathcal{C}_x(t) = \left\{T\rightarrow P_x(t;T)\right\}$, built from homogeneous market Swap quotes (i.e. with the same underlying IBOR tenor $x$) using the usual expression of forward rates
\begin{equation}
F_{x,j}(t) = \dfrac{1}{\tau_F(T_{j-1},T_j)}\left[\dfrac{P_x(t;T_{j-1})}{P_x(t;T_{j})}-1\right],
\label{eq:forward_rate_ZCB}
\end{equation}
where $\tau_{F}$ is the year fraction with the forward rate convention and $P_x(t;T_{j})$ can be interpreted as the price of a risky ZCB issued by an average IBOR counterparty\footnote{Namely an issuer with a credit risk equal to the average credit risk of the IBOR panel, see e.g.~\cite{Mor09}.}. 
Expressions \ref{eq:Swap_Rate_R} and \ref{eq:forward_rate_ZCB} are consistently used during the bootstrapping procedure. 

Discounting ZCBs $P(t;T_{j})$ in eq. \ref{eq:swap_price} are computed from discounting ZCB curve 
$\mathcal{C}(t) = \left\{T\rightarrow P(t;T)\right\}$, built from market quotes of Overnight Indexed Swaps (OIS\footnote{OIS are fixed vs floating Swaps where the floating cash flows are based on a daily compounded overnight rate, e.g. \euro STR. Their pricing formula is similar to eq. \ref{eq:swap_price}, see e.g. \cite{ScaBia20} for a detailed derivation.}). 
We notice that the discounting curve $\mathcal{C}(t)$ is also required to build IBOR curves $\mathcal{C}_x(t)$ using recursively eq. \ref{eq:swap_price}. 
Overall, this procedure is commonly called multi-curve bootstrapping, since OIS and IBOR curves with different tenors are involved.

The market quotes OIS and IBOR Swaps with different tenors and maturities, which, along with other similar quoted instruments (i.e. Forward Rate Agreements, Futures, Basis Swaps) can be used to build OIS and multiple IBOR yield curves for each rate tenor $x$.
see e.g.~\cite{AmeBia13} for a detailed discussion of market quotes and multi-curve bootstrapping. 
In App. \ref{app:Market Data} we report the yield curves used in this paper.

\subsubsection{European Swaption}
\label{app:european swaption}

In this paper we consider physically-settled European Swaptions, i.e. contracts which give to the holder the right to enter, at a given expiry date $T_{e}$, into a Swap contract starting at $T_{0} \geq T_{e}$ as described in app. \ref{app:Interest Rate Swap}. 

The payoff can be written as
\begin{equation}
\begin{split}
\textbf{Swaption}&(T_{e};\textbf{T},\textbf{S},K,\omega) 
 = \max \left[ \textbf{Swap}(T_{e};\textbf{T},\textbf{S},K,\omega);0 \right], \\
& = N A(T_{e}; \textbf{S}) \max \left\{ \omega \left[ R_{x}^{\text{Swap}}(T_{e};\textbf{T},\textbf{S})-K \right];0\right\},
\end{split}
\label{eq:swaption_payoff}
\end{equation} 
where $N$ is the nominal amount, $A(T_{e}; \textbf{S})$ is the Swap annuity in eq. \ref{eq:swap_price} and $R_{x}^{\text{Swap}}(T_{e};\textbf{T},\textbf{S})$ is the Swap rate in eq. \ref{eq:Swap_Rate_R}, both evaluated at expiry date $T_e$.

The market practice is to value such Swaptions through the shifted-Black formula, assuming a shifted log-normal driftless dynamics with instantaneous volatility $\sigma_{x} (t; \textbf{T}, \textbf{S})$ for the evolution of the Swap rate $R_{x}^{\text{Swap}}(T_{e};\textbf{T},\textbf{S})$ under its corresponding discounting Swap measure\footnote{Since $ \mathbb{E}^{Q} \left[ D(t;T)  V(T) \vert\mathcal{F}_t\right] = A(t;\textbf{S}) \mathbb{E}^{Q_{S}} \left[ \frac{V(T)}{A(T;\textbf{S})} \vert\mathcal{F}_t \right]$.} $Q_{S}$ associated to the numeraire $A(t,\textbf{S})$. 
The shifted-Black price at time $t$ is thus given by
\begin{equation}
\begin{split}
&\textbf{Swaption}(t;\textbf{T},\textbf{S},K,\omega) \\
&\quad = N \mathbb{E}^{Q} \left\{ D(t;T_e) A(T_e;\textbf{S}) \max \left[ \omega \left[ R_{x}^{\text{Swap}}(T_{e};\textbf{T},\textbf{S})-K \right];0 \right] \vert\mathcal{F}_t \right\}, \\ 
&\quad = N A(t; \textbf{S}) \mathbb{E}^{Q_{S}} \left\{ \max \left[ \omega \left[ R_{x}^{\text{Swap}}(T_{e};\textbf{T},\textbf{S})-K \right];0 \right] \vert\mathcal{F}_t \right\}, \\
&\quad = N A(t; \textbf{S}) \text{Black} \left[ R_{x}^{\text{Swap}}(t;\textbf{T},\textbf{S})+\lambda_x, K+\lambda_x, v_x(t;\textbf{T},\textbf{S}), \omega \right],\\
&v_x(t;\textbf{T},\textbf{S}) = \int_{t}^{T_{e}} \sigma_{x} (u; \textbf{T}, \textbf{S})^2 du,\quad
\sigma_{x} (t; T_{e}, \textbf{T}, \textbf{S}) = \sqrt{\frac{v_x(t; \textbf{T}, \textbf{S})}{\tau_x(t,T_{e})}},
\end{split}
\label{eq:Swaption_price_black}
\end{equation} 
where $\lambda_x$ is the constant log-normal shift, $v_x(t;\textbf{T},\textbf{S})$ is the shifted log-normal implied forward variance, and $\sigma_{x} (t; T_{e}, \textbf{T}, \textbf{S})$ is the shifted log-normal implied forward volatility.
Actually, eq. \ref{eq:Swaption_price_black} is not used ``as is" for pricing purposes\footnote{more sophisticated pricing models are used, typically based on SABR stochastic volatility model, see \cite{HagKum16}.}, but, rather, as a standard tool to imply shifted-Black volatilities from market prices.

The market quotes physically-settled Swaptions with different expiry dates, underlying Swap lenghts, and strikes, i.e. a 3D structure called Swaption price cube. In App. \ref{app:Market Data} we report the price cube used in this paper.

\subsection{G2++ Model}
\label{app:Multi-curve G2++ model}

We describe here the multi-curve time-dependent volatility two-factor Shifted-Vasicek gaussian model, also known as G2++, adopted in this work, including the corresponding pricing formulas for Swaps and European Swaptions, the methodologies used to calibrate the model parameters to market data, and the Monte Carlo simulation. 
We refer to e.g. \cite{BriMer06} for the standard single-curve constant-volatility G2++ version and to \cite{Ken10a} for the multi-curve G1++ version.

\subsubsection{Short-Rates Dynamics}
\label{app:short rate dynamics}

The dynamics of the instantaneous short rate processes under the risk-neutral measure $Q$ are given by
\begin{align}
\begin{split}
&r^{c}(t) = x(t) + y(t) + \varphi^{c}(t), \qquad  r^{c}(0) = r_0^{c},\\
&dx(t) = -ax(t)dt + \sigma(t) dW_1(t), \qquad x(0) = 0, \\
&dy(t) = -by(t)dt + \eta(t) dW_2(t), \qquad y(0) = 0, \\
&d \left\langle W_1,W_2\right\rangle\left(t\right) = \rho dt,\\
&\sigma(t)= \sigma\Gamma(t),\; \eta(t)=\eta\Gamma(t),\;\Gamma(t) = \sum_i \Gamma_i \mathds{1}_{[T_i,T_{i+1}]},
\end{split}
\label{eq:short_rate_dynamics}
\end{align}
where $c\in\{d,x\}$, with $d$ and $x$ denoting the discount curve $\mathcal{C}_d$ and the forward curve $\mathcal{C}_x$, respectively, $-1\leqslant \rho \leqslant 1$, $\Gamma: \mathbb{R} \rightarrow \mathbb{R}^{+}$ is a piece-wise constant function on time intervals $[T_i,T_{i+1}]$, $\varphi^{c}: \mathbb{R}^+ \rightarrow \mathbb{R}$, and $r^{c}_{0}, a, b, \sigma, \eta \in \mathbb{R}$. 

By integrating eqs.~\ref{eq:short_rate_dynamics} we obtain, for each $s<t$,
\begin{multline}
r^c(t) =  x(s)e^{-a(t-s)} + y(s)e^{-b(t-s)} \\
\quad + \sigma \int_s^t \Gamma(u) e^{-a(t-u)} dW_1(u) + \eta \int_s^t \Gamma(u) e^{-b(t-u)} dW_2(u) + \varphi^c(t),
\end{multline}
from which we can see that $r^c(t)$, conditional on the sigma-field $\mathcal{F}_s$ generated by the pair $(x,y)$ up to time $s$, is normally distributed with mean ad variance, respectively,
\begin{align}
\mathbb E \left(r^c(t)\vert\mathcal F_s \right) &= x(s)e^{-a(t-s)} + y(s) e^{-b(t-s)} + \varphi^c(t), \\
\mathrm{Var}(r^c(t)\vert\mathcal F_s) &= \frac{\sigma^2}{2a} \sum_{i=u}^v \Gamma_i^2 e^{-2a t} e^{2a(T_{i+1}-T_i)} + \frac{\eta^2}{2b} \sum_{i=u}^v \Gamma_i^2 e^{-2b t} e^{2b(T_{i+1}-T_i)} \nonumber \\
&\quad+ 2 \frac{\rho \eta \sigma}{a+b}(1-e^{-(a+b)t}) \sum_{i=u}^v \Gamma_i^2 e^{(a+b)(T_{i+1}-T_i)},
\end{align}
where for the variance we assumed, without loosing generality, that $T_u = s$ and $T_{v+1} = t$ for a certain $u$ and $v$ with $u<v$.

\subsubsection{Pricing}
\label{app:G2++pricing}

\paragraph{Interest Rate Swap}

In our G2++ framework, the price at time $t$ of a Swap's floating coupon is given by
\begin{multline}
\textbf{Swaplet}_{\text{float}}(t;T_{j-1},T_j) 
= N \mathbb E^Q_t \left[ D_d(t;T_j) R_x(T_{j-1},T_j) \tau_{R}(T_{j-1},T_j) \right] \\
= N \left[ \frac{\phi^d(T_{j-1};T_j)}{\phi^x(T_{j-1};T_j)} P_d(t;T_{j-1}) - P_d(t;T_j) \right], 
\label{eq:swap_sc}
\end{multline}
where $\phi^c(t;T) = \exp \left\{-\int_t^T \varphi^c(u)du \right\}$, $c\in\{d,x\}$ with $d$ and $x$ denoting respectively the discount $\mathcal{C}_d$ and the forward $\mathcal{C}_x$ curves. 
This result comes from the fact that, setting $z(s)=x(s)+y(s)$,
\begin{align}
\begin{split}
&D_c(t;T) = e^{-\int_{t}^{T}r^c(s)ds} = \phi^{c}(t;T)e^{-\int_{t}^{T}z(s)ds},\\
&P_c(t;T) = \mathbb E^{Q}_t\left[D_c(t;T)\right] = \phi^{c}(t;T)\mathbb E^{Q}_t\left[e^{-\int_{t}^{T}z(s)ds}\right],\\
&\mathbb E^{Q}_t\left[D_d(t;T_j) R_x(T_{j-1},T_j) \tau_R(T_{j-1},T_j)\right] \\
&\quad=\mathbb E^{Q}_t \left[ \phi^{d}(t;T_{j})e^{-\int_{t}^{T_{j}}z(s)ds}\left(\dfrac{1}{P_x(T_{j-1};T_{j})}-1\right)\right] \\
&\quad=\mathbb E^{Q}_t \left[ \phi^{d}(t;T_{j})e^{-\int_{t}^{T_{j-1}}z(s)ds}
\dfrac
{\mathbb E_{T_{j-1}}^{Q} \left[e^{-\int_{T_{j-1}}^{T_{j}}z(s)ds}\right]}
{\mathbb \phi^{x}(T_{j-1};T_{j})E_{T_{j-1}}^{Q} \left[e^{-\int_{T_{j-1}}^{T_{j}}z(s)ds}\right]}\right] - P_d(t;T_{j})\\
&\quad=\mathbb E^{Q}_t \left[ \dfrac{\phi^{d}(t;T_{j-1})\phi^{d}(T_{j-1};T_{j})}{\phi^{x}(T_{j-1};T_{j})}e^{-\int_{t}^{T_{j-1}}z(s)ds} \right]-P_d(t;T_{j})\\
&\quad=\frac{\phi^d(T_{j-1};T_j)}{\phi^x(T_{i-1};T_j)} P_d(t;T_{j-1}) - P_d(t;T_{j},), \qquad \forall j.
\end{split}
\label{eq:demo_swaplet_float}
\end{align}
Therefore, the G2++ Swap price is given by
\begin{align}
\textbf{Swap}&(t;\textbf{T},\textbf{S},K,\omega) \nonumber\\
& = N \omega \left[ \sum_{j=1}^n \left[\psi(T_{j-1};T_j) P_d(t;T_{j-1}) - P_d(t;T_j) \right]- K A(t;\textbf{S})\right],
\label{g2++_swap}
\end{align}
with
\begin{equation}
\psi(T_{j-1};T_j) = \frac{\phi^d(T_{j-1};T_j)}{\phi^x(T_{j-1};T_j)} = \frac{P_d^{M}(0;T_j)}{P_d^{M}(0;T_{j-1})} \frac{P_{x}^{M}(0;T_{j-1})}{P_{x}^{M}(0;T_j)},  \label{eq:psi}
\end{equation}
where $P_c^{M}(0;T)$ is the ZCB price observed on the market at time $t=0$ for maturity $T$, and the second equality follows from
\begin{align}
\phi^c(T_{j-1};T_j) &= \exp \left\{-\int_{T_{j-1}}^{T_j} \varphi^c(u)du \right\} \nonumber \\
&= \frac{P_{c}^{M}(0;T_j)}{P_{c}^{M}(0;T_{j-1})} \exp \left\{ -\frac{1}{2} \left[ V(0;T_j)  - V(0;T_{j-1}) \right]\right\},
\end{align}
where $V(t;T_j)$ is the variance of the random variable $I(t;T_j) = \int_t^{T_j} [x(s) + y(s)] ds$, which is normally distributed with mean $M(t;T_j)$ (see \cite{BriMer06} for the proof). 
By considering the piece-wise constant volatility parameters introduced in app.~\ref{app:short rate dynamics} we obtain the following expression
\begin{align}
\begin{split}
& V(s;t) \\
&= \frac{\sigma^2}{a^2} \sum_{i=u}^v \Gamma_i^2 \left[ (T_{i+1}-T_i) - \frac{2e^{-at}}{a} \left(e^{aT_{i+1}}-e^{aT_i} \right) - \frac{e^{-2at}}{2a} \left(e^{2aT_{i+1}}-e^{2aT_i} \right) \right] \\
&+ \frac{\eta^2}{b^2} \sum_{i=u}^v \Gamma_i^2 \left[ (T_{i+1}-T_i) - \frac{2e^{-bt} }{b}\left(e^{bT_{i+1}}-e^{bT_i} \right) - \frac{e^{-2bt}}{2b}\left(e^{2bT_{i+1}}-e^{2bT_i} \right) \right] \\
&+ \frac{2\rho \eta \sigma}{ab} \sum_{i=u}^v \Gamma_i^2 \left[ (T_{i+1}-T_i) - \frac{e^{-at}}{a} \left(e^{aT_{i+1}}-e^{aT_i} \right) - \frac{e^{-bt}}{b} \left(e^{bT_{i+1}}-e^{bT_i} \right) \right.\\
&\left.\hspace{2.5cm} +\frac{e^{-(a+b)t}}{a+b} \left( e^{(a+b)T_{i+1}}-e^{(a+b)T_i}\right)\right],
\end{split}
\end{align}
where here we assumed that $T_u = s$ and $T_{v+1} = t$. 

The multi-curve G2++ Swap price can be written in terms of the single-curve G2++ Swap price assuming the existence of a vector of $n+1$ coefficients $c=\left\{c_0,\cdots,c_n\right\}$ such that
\begin{align}
\textbf{Swap}(t;\textbf{T},\textbf{S},K,\omega) 
&= c_0\ \omega \left( P(t;T_0) - \sum_{j=1}^{n} c_j P(t;T_j) \right),
\label{eq:swap_mc1}
\end{align}
where
\begin{equation}
\begin{split}
c_0 &= \psi(T_0;T_1), \\
c_j &= c_{0}^{-1} \left[ 1-\psi(T_{j};T_{j+1})  \mathds{1}_{ \left\{T_j \neq T_n \right\}} + K \tau_K(S_{i},S_{i+1})  \mathds{1}_{ \left\{T_j = S_i \right\}} \right], \quad j = 1,\dots,n,
\end{split}
\label{eq:hat_c_vector}
\end{equation}
where we assumed that the the schedule $\textbf{S}$ is a subset of schedule $\textbf{T}$.
As a practical example, we consider a payer Swap with unitary nominal amount and maturity 2 years with semi-annual floating leg and annual fixed leg, characterized by the schedules $\textbf{T}=\{T_{0},T_{1},T_{2},T_{3},T_{4}\}$ and $\textbf{S}=\{S_{0},S_{1},S_{2}\}$, where $T_0 = S_0$, $T_2 = S_1$ and $T_4 = S_2$. 
The G2++ price is given by
\begin{equation}
\begin{split}
\textbf{Swap}(t;\textbf{T},\textbf{S},K,1) 
&= P_d(t;T_0) \psi(T_0;T_1) \\
&- P_d(t;T_1) \left[ 1 - \psi(T_1;T_2) \right] \\
&- P_d(t;T_2) \left[ 1 - \psi(T_2;T_3) + K \tau_K(S_0,S_1) \right] \\
&- P_d(t;T_3) \left[ 1 - \psi(T_3;T_4) \right] \\
&- P_d(t;T_4) \left[ 1+K \tau_K(S_1,S_2) \right],
\end{split}
\end{equation} 
from which it is possible to recognize the following coefficients
\begin{equation}
\begin{split}
c_0 &= \psi(T_0;T_1) \\
c_1 &= c_0^{-1} \left[ 1 - \psi(T_1;T_2) \right] \\
c_2 &= c_0^{-1} \left[ 1 - \psi(T_2;T_3) + K \tau_K(S_0,S_1) \right] \\
c_3 &= c_0^{-1} \left[ 1 - \psi(T_3;T_4) \right] \\
c_4 &= c_0^{-1} \left[ 1+K \tau_K(S_1;S_2) \right].
\end{split}
\end{equation}

\paragraph{European Swaption}

The price at time $t$ of an European Swaption expiring at $T_e$ under the forward measure $Q^{T_e}$ can be written, using the single-curve G2++ Swap pricing formula in eq.~\ref{eq:swap_mc1} as 
\begin{equation}
\begin{split}
& \textbf{Swaption}(t;\textbf{T},\textbf{S},K,\omega) \\ 
& = N P(t;T_e) \mathbb{E}_t^{Q^{T_e}} \left\{ \max \left[ \textbf{Swap}(T_e;\textbf{T},\textbf{S},K,\omega);0 \right] \right\} \\
& = N P(t,T_e) \mathbb{E}_t^{Q^{T_e}} \left\{ \max \left[ c_0\,\omega \left( P(T_e;T_0) - \sum_{j=1}^{n} c_j P(T_e;T_j) \right) ; 0 \right] \right\}, 
\end{split}
\label{eq:app_swpt_price}
\end{equation}
where the coefficients $c_i, i=0,...,m$ are given by eq. \ref{eq:hat_c_vector}. 
Since the G2++ Swaption price in eq.~\ref{eq:app_swpt_price} above is reduced to a single-curve expression, the semi-analytical single-curve G2++ pricing formula derived in \cite{BriMer06} (Theorem 4.2.3) is preserved and can be generalised to piece-wise constant volatility parameters as follows,
\begin{multline}
\textbf{Swaption}(0;\textbf{T},\textbf{S},K,\omega) = N \omega P_d(0;T_e) \\
\int_{-\infty}^{+\infty}\frac{e^{-\frac{1}{2}(\frac{s-\mu_x}{\sigma_x})^2}}{\sigma_x\sqrt{2\pi}}\left[ \Phi\left[-\omega h_1(s)\right]-\sum_{j=1}^{n}\lambda_j(s)e^{\mathcal{K}_j(s)} \Phi\left[-\omega h_2(s)\right]\right] ds,
\label{eq:g2++_swpt_multi_curve}
\end{multline}
where
\begin{align}
&h_1(s) = \frac{\bar{q}-\mu_y}{\sigma_y\sqrt{1-\rho_{xy}^2}}-\frac{\rho_{xy}(s-\mu_x)}{\sigma_x\sqrt{1-\rho_{xy}^2}}, \\
&h_2(s) = h_1(s)+\mathcal{B}(b,T_e,T_j)\sigma_y\sqrt{1-\rho_{xy}^2},\\
&\lambda_j(s) = c_j \mathcal{A}_d(T_e;T_j)e^{-\mathcal{B}(a,T_e,T_j)s}, \\
&\mathcal{K}_j(s) = -\mathcal{B}(b,T_e,T_j)\left[\mu_y-\frac{1}{2}(1-\rho_{xy}^2)\sigma_y^2 \mathcal{B}(b,T_e,T_j)+\rho_{xy}\sigma_y\frac{s-\mu_x}{\sigma_x}\right],\\
&\mathcal{A}(t_1;t_2) = \frac{P_{c}(0;t_2)}{P_{c}(0;t_1)} 
\exp \left\{\frac{1}{2} \left[ V(t_1;t_2) - V(0;t_2) + V(0;t_1) \right] \right\}, \\
&\mathcal{B}(z,t_1;t_2) = \frac{1-e^{-z(t_2-t_1)}}{z}, \quad z\in\{a,b\},
\end{align}
and $\bar{q}=\bar{q}(s)$ is the solution of the following equation
\begin{equation}
\sum_{j=1}^{n} c_j\mathcal{A}(T_e;T_j)e^{-\mathcal{B}(a,T_e,T_j)s-\mathcal{B}(b,T_e,T_j)\bar{q}}=1,
\end{equation}
and, by setting $T_u = 0$ and $T_{v+1} = T_e$,
\begin{align}
\mu_x & = - M^{T_e}_x(0;T_e) ,\\
\mu_y & = - M^{T_e}_y(0;T_e),\\	
\sigma_x & = \sqrt{ \frac{\sigma^2}{2a} \sum_{i=u}^v \Gamma_i^2 e^{-2a T_e} e^{2a(T_{i+1}-T_i)}},	\\
\sigma_y & = \sqrt{\frac{\eta^2}{2b} \sum_{i=u}^v \Gamma_i^2 e^{-2b T_e} e^{2b(T_{i+1}-T_i)}}, \\
\rho_{xy} & =\frac{\rho\sigma\eta}{(a+b)\sigma_x\sigma_y} \left[1-e^{-(a+b)T_e} \right] \sum_{i=u}^v \Gamma_i^2 e^{(a+b)(T_{i+1}-T_i)},
\end{align}
where $M^{T_e}_x(0;T_e)$ and $M^{T_e}_y(0;T_e)$ are drift components stemming from the dynamics of the processes $x$ and $y$ under the forward measure $T_e$ given, setting $T_u = 0$ and $T_{v+1} = T_e$, by
\begin{multline}
M^{T}_x(0;T_e) = \frac{\sigma^2}{a} \sum_{i=u}^v \Gamma_i^2 \left[ \frac{e^{-aT_e}}{a}\left(e^{a T_{i+1}} - e^{a T_{i}}\right) - \frac{e^{-a(T_e+T)}}{2a} \left(e^{2a T_{i+1}} - e^{2a T_{i}}\right) \right] \\
 \quad - \frac{\rho\sigma\eta}{b(a+b)} \sum_{i=u}^v \Gamma_i^2 \left[ \frac{e^{-aT_e}}{a} \left(e^{a T_{i+1}} - e^{a T_{i}} \right) -\frac{e^{-(aT_e+bT)}}{a+b}\left(e^{(a+b)T_{i+1}} - e^{(a+b)T_{i}} \right) \right], 
\label{eq:MT_x}
\end{multline}
\begin{multline}
M^{T}_y(0;T_e) = \frac{\eta^2}{b} \sum_{i=u}^v \Gamma_i^2 \left[ \frac{1}{b}e^{-bT_e}\left(e^{b T_{i+1}} - e^{b T_{i}}\right) - \frac{e^{-b(T_e+T)}}{2b} \left(e^{2b T_{i+1}} - e^{2b T_{i}}\right) \right] \\
 \quad - \frac{\rho\sigma\eta}{a(a+b)} \sum_{i=u}^v \Gamma_i^2 \left[ \frac{e^{-bT_e}}{b} \left(e^{b T_{i+1}} - e^{b T_{i}} \right) -\frac{e^{-(aT_e+bT)}}{a+b}\left(e^{(a+b)T_{i+1}} - e^{(a+b)T_{i}} \right) \right]. 
\label{eq:MT_y}	
\end{multline}

\subsubsection{Calibration}
\label{app:G2++ calibration procedure}
We calibrate the G2++ model parameters on ATM Swaption prices following a two-steps procedure.
\begin{enumerate}
\item \textbf{Constant volatility calibration}: firstly, we obtain the G2++ parameters $a,b,\sigma,\eta,\rho$ in eq. \ref{eq:short_rate_dynamics} by minimizing the following objective function, which represents the Mean Squared Relative Error (MSRE) between market prices $V^{\text{mkt}}$ and G2++ model prices $V^{\text{G2++}}$ for each combination of expiry $ \{ \xi_i \}_{i=1}^{M} $ and tenor $ \{ \mathfrak{T}_j \}_{j=1}^{N}$,
\begin{equation}
\left\{ \hat{a},\hat{b},\hat{\sigma},\hat{\eta},\hat{\rho} \right\} = \argminA_{ \left\{ a,b,\sigma,\eta,\rho \right\} } \frac{1}{M N}\sum_{i=1}^{M}\sum_{j=1}^{N}\left[ \frac{V^{\text{G2++}}(a,b,\sigma,\eta,\rho;\xi_i,\mathfrak{T}_j)+\epsilon}{V^{\text{mkt}}(\xi_i,\mathfrak{T}_j)+\epsilon} -1\right] ^2,
\label{A1}	
\end{equation}
where $\epsilon$ is a regularization parameter set to 10 bps in order to avoid divergence in case of small prices.
\item \textbf{Time-dependent volatility calibration}: secondly, given the G2++ parameters $\hat{a},\hat{b},\hat{\sigma},\hat{\eta},\hat{\rho}$ obtained above, we calibrate the piece-wise function $\Gamma(t)$ in eq. \ref{eq:short_rate_dynamics} using an iterative forward procedure to obtain at step $i=1,\cdots,M$ the piece $\Gamma_i = \Gamma(\xi_i)$ using the previous pieces $\{ \hat{\Gamma}_k \}_{k=1}^{i-1} $ by minimizing the same MSRE objective function 
\begin{equation}
\hat{\Gamma}_i = \argminA_{\Gamma_i} \frac{1}{N} \sum_{j=1}^{N}\left[ \frac{V^{\text{G2++}}(\hat{a},\hat{b},\hat{\sigma} \Gamma_i, \hat{\eta} \Gamma_i,\hat{\rho};\xi_i,\mathfrak{T}_j)+\epsilon}{V^{\text{mkt}}(\xi_i,\mathfrak{T}_j)+\epsilon} -1 \right]^2,
\label{A2}	
\end{equation}
\end{enumerate}
Notice that, at first order, the price calibration using the MSRE objective function above is equivalent to a Vega-weighted volatility calibration. In fact, if we consider the following minimization problem
\begin{equation}
\hat{p} = \argminA_{p} \frac{1}{M N} \sum_{i=1}^{M} \sum_{j=1}^{N} \left\{ w_{i,j}  \left[ \sigma(p;\xi_i,\mathfrak{T}_j) - \sigma^{\text{mkt}}(\xi_i,\mathfrak{T}_j) \right]^2 \right\},
\end{equation}
where $\sigma$ is the (normal or log-normal) volatility implied in G2++ option prices and where $\{w_{1,1}, \dots, w_{i,j}, \dots, w_{M,N} \} $ are calibration weights set equal to normalized Vega sensitivities,
\begin{equation}
w_{i,j} = \frac{\nu_{i,j}}{\sum_{i=1}^{M} \sum_{j=1}^{N} \nu_{i,j}}, \quad \sum_{i=1}^{M} \sum_{j=1}^{N} w_{i,j} = 1,
\end{equation}
where $\nu_{i,j}$ is the corresponding (normal or log-normal) Vega sensitivty. At first order,
\begin{equation}
\nu_{i,j} \left[ \sigma(p;\xi_i,\mathfrak{T}_j) - \sigma^{\text{mkt}}(\xi_i,\mathfrak{T}_j) \right] \approx V(p;\xi_i,\mathfrak{T}_j) - V^{\text{mkt}}(\xi_i,\mathfrak{T}_j).
\end{equation}
The formulas above are restricted to the ATM case but can be easily extended to the more
general case of ITM and OTM Swaptions prices with strikes $k_h$, with $h = 1,\dots, K_i$ (since the number of strikes depends on the given expiry).

Full details about the different G2++ model calibrations are reported in app. \ref{app:G2++ model calibration}.

\subsubsection{Monte Carlo Simulation}
\label{app:G2++ MC simulation}

In \cite{BriMer06} the dynamics of $x(t)$ and $y(t)$ in eqs. \ref{eq:short_rate_dynamics} are rewritten under the $T$-forward measure $Q^T$, which is convenient for Monte Carlo simuation.  
Here we generalize that MC simulation scheme for time-dependent volatility parameters as follows
\begin{align}
x(t) &= x(s)e^{-a(t-s) } - M^{T}_{x}(s;t) + N_{1}(s,t), \label{eq:g2++_mc_scheme_x} \\
y(t) &= y(s)e^{-b(t-s) } - M^{T}_{y}(s;t)+ N_{2}(s,t), \label{eq:g2++_mc_scheme_y}
\end{align}
where $s \leq t \leq T$, $M^{T}_{x}(s;t)$ and $M^{T}_{y}(s;t)$ are the drift components defined in app.~\ref{app:G2++pricing}, and $N(t-s)$ is a two-dimensional normal random vector with zero mean and $2\times 2$ covariance matrix given by
\begin{align}
\begin{split}
\Sigma(s,t)_{1,1} &= \frac{\sigma^2}{2a} \sum_{i=u}^v \Gamma_i^2 e^{-2a t} e^{2a(T_{i+1}-T_i)},\\
\Sigma(s,t)_{1,2} &= \frac{\rho \eta \sigma}{a+b} \left[1-e^{-(a+b)t} \right] \sum_{i=u}^v \Gamma_i^2 e^{(a+b)(T_{i+1}-T_i)},\\
\Sigma(s,t)_{2,1} &= \Sigma(s,t)_{1,2},\\
\Sigma(s,t)_{2,2} &= \frac{\eta^2}{2b} \sum_{i=u}^v \Gamma_i^2 e^{-2b t} e^{2b(T_{i+1}-T_i)},
\end{split}
\end{align}
where $T_u = s$ and $T_{v+1} = t$.

The MC simulation is performed using a discrete time grid $\left\{t=t_0,t_1,\cdots,t_{N_T}=T\right\}$. The mark-to-market of each instrument at future time step $t_i$ is calculated by simulating its future market risk factors using the pair $\left\{x(t_i), y(t_i)\right\}$ computed from  eqs.~\ref{eq:g2++_mc_scheme_x} and \ref{eq:g2++_mc_scheme_y}, for $i = 1,\dots,N_T$.
It should be noticed that, for our purposes, we do not need to simulate $r(t)$ since the price of a Swap or an European Swaptions can be obtained directly through $x(t)$ and $y(t)$ (see app.~\ref{app:G2++pricing}).

\subsection{XVA Pricing} 
\label{app:XVA pricing}

We describe here the Credit and Debt Valuation Adjustments (CVA and DVA), which take into account the risks related to counterparties default on derivative transactions. 
Below we give their definitions, we derive their pricing formulas, we show how they can be computed in practice and we also consider the analytical expressions available for linear derivatives. 

\subsubsection{XVA Definitions and Formulas}
\label{app:XVA formulas}

Let us consider a contract between a bank (B) and a counterparty (C) engaged at time $t$ in a derivative contract with final maturity at time $T>t$, which can default at future times $\tau_B$ and $\tau_C$. 
Hence, we can distinguish six cases, according to the six possible orderings of the three time instants $T,\tau_B,\tau_C$ as follows.
\begin{itemize}
\item In the two cases, either $T<\tau_B<\tau_C$ or $T<\tau_C<\tau_B$, i.e. when both counterparties default after the contract maturity, the present value of the contract $V(t)$ is unchanged, since all the due cash flows will be regularly exchanged. Hence, in this case the fair value of the trade is given, according to eq. \ref{eq:fair value}, by the mark to market without any adjustment, i.e. $V(t) = V_0(t)$. Since counterparties' default is not effective, we may associate this value to a perfect collateralization, even if the contract is not subject to a collateral agreement (CSA).
\item In the two cases, either $\tau_C < \tau_B < T$ or $\tau_C < T < \tau_B$, i.e. when the counterparty defaults before the bank and contract maturity $T$, and the bank has positive exposure with respect to the counterparty, the bank suffers a loss equal to the replacement cost of the position. The CVA is defined as the discounted value of the expected future loss suffered by the bank due to the default of the counterparty, and it is a negative quantity from the bank's perspective. Hence, in this case the fair value of the contract is given, according to eq. \ref{eq:fair value}, by the mark to market plus the (negative) CVA, i.e. $V(t) = V_0(t)+\text{CVA}(t)$.
\item In the two last cases, either $\tau_B < \tau_C < T$ or $\tau_B < T < \tau_C$, i.e. when the bank defaults before the counterparty and contract maturity $T$, and the counterparty has a positive exposure with respect to the bank (i.e. the bank has negative exposure with respect to the counterparty), the counterparty suffers a loss equal to the replacement cost of the position, and therefore computes a (negative) CVA. This is by definition the DVA of the bank, i.e. the discounted value of the expected future gain suffered by the bank due to the default of the counterparty, and it is a positive quantity from the bank's perspective. Hence, in this case the fair value of the contracy is given, according to eq. \ref{eq:fair value}, by the mark to market plus the (positive) DVA, i.e. $V(t) = V_0(t)+\text{DVA}(t)$.
\end{itemize}
Overall, considering all the six possible cases, we have the total fair value $V(t) = V_0(t)+\text{CVA}(t)+\text{DVA}(t)$, where the CVA/DVA definitions above translate into the following pricing expressions,
\begin{align}
\text{CVA}(t) &= -\mathbb{E}^Q\left[D(t;\tau_\text{C}) LGD_\text{C}(\tau_\text{C}) \left[ H(\tau_\text{C}) \right]^+  \mathds{1}_{\{ t < \tau_C < \tau_B < T \}} \vert\mathcal{G}_t \right] \label{eq:CVA1} ,\\
\text{DVA}(t) &= - \mathbb{E}^Q \left[ D(t;\tau_B) LGD_\text{B}(\tau_\text{B}) \left[ H(\tau_B) \right]^- \mathds{1}_{\{ t < \tau_B < \tau_C < T \}} \vert\mathcal{G}_t \right] \label{eq:DVA1} , \\
H(\tau_\text{X}) &= \mathbb{E}_{\tau_\text{X}}^Q \left[ V_0 (\tau_\text{X}) \right] - C(\tau_\text{X}) 
\label{eq:H1} ,
\end{align}
where $H(\tau_\text{X})$ denotes the bank's exposure at time $t<\tau_\text{X} \leq T$ in the event of default of X, for $X\in\{B,C\}$, $V_0(\tau_\text{X})$ is the mark to market\footnote{Here we assume ``risk free'' close-out at the mark to market, without any further adjustment.} of the instrument at time $\tau_\text{X}$, $C(\tau_\text{X})$ denotes generically the collateral available at time $\tau_\text{X}$ (see app.~\ref{app:collateral modelling} for collateral modelling), and $LGD_\text{X}(\tau_\text{X}) = 1 - R_\text{X}(\tau_\text{X})$ is the Loss Given Default of $X$ at time $\tau_\text{X}$, which represents the percentage amount of the exposure expected to be lost in case of X's default, with $R_\text{X}$ denoting the corresponding Recovery Rate. 
We stress that the exposure may refer to single contracts between the bank and the counterparty, or, more generally, to groups of contracts subject to netting and/or collateral agreements.

Assuming independent interest rate and default processes\footnote{We neglect here the wrong way risk arising when the exposure with the counterparty is inversely related to the creditworthiness of the counterparty itself.} and constant $LGD_\text{X}$, the expectations in eqs.~\ref{eq:CVA1} and \ref{eq:DVA1} simplify to
\begin{align}
\text{CVA}(t) 
&  = -  \int_{t}^{T} \mathbb{E}^Q\left[D(t;u) LGD_\text{C} \left[ H(u) \right]^{+} \vert\mathcal{F}_t \right] \mathbb{E}^Q \left[  \mathds{1}_{\{ t < \tau_C < \tau_B < T \}} \vert\mathcal{F}_t \right] du , \nonumber \\
& = - LGD_\text{C} \int_{t}^{T} \mathbb{E}^Q\left[ D(t;u) \left[ H(u) \right]^{+} \vert\mathcal{F}_t \right] S_B(t,u) dQ_C(t,u) \label{eq:CVA2} , \\
\text{DVA}(t) 
& = -  \int_{t}^{T} \mathbb{E}^Q\left[ D(t;u)LGD_\text{B} \left[ H(u) \right]^{-} \vert\mathcal{F}_t \right] \mathbb{E}^Q \left[  \mathds{1}_{\{ t < \tau_B < \tau_C < T \}} \vert\mathcal{F}_t \right] du , \nonumber \\
& = - LGD_\text{B} \int_{t}^{T} \mathbb{E}^Q\left[ D(t;u) \left[ H(u) \right]^{-} \vert\mathcal{F}_t \right] S_C(t,u) dQ_B(t,u) \label{eq:DVA2} , \\
S_X(t,u) 
&= \mathbb{E}^Q \left[\mathds{1}_{\left\{ \tau_X > u \right\} } \vert\mathcal{F}_t \right] = \mathbb{E}^Q \left[ e^{-\int_t^u \gamma_X(s)ds} \vert\mathcal{F}_t \right] = 1-Q_X(t,u),
\label{eq:survival probability X}
\end{align}
where $S_X(t,u)$ is the survival probability of X until time $u$ valued at $t$, $\gamma_X$ denotes the stochastic hazard rate of X, $dQ_X(t,u) = Q_X(t,u,u+du)$ is the marginal default probability of X referred to the infinitesimal time interval $[u,u+du]$ valued at $t$, and $\mathcal{F}_t$ denotes the market filtration (see e.g. \cite{BieRut04,BriMor13,Bri18,BriFraPal19}).

Similar formulas for other XVA can be found in e.g.~\cite{Gre20}. 
The survival probabilities in eq. \ref{eq:survival probability X} are computed from default curves built from market CDS quotes through standard bootstrapping procedure.

\subsubsection{XVA Numerical Formulas}
\label{app:XVA numerical formulas}
The numerical XVA calculation requires the discretization of the integrals in eqs.~\ref{eq:CVA2},~\ref{eq:DVA2} using a time grid $\{t = t_0,\cdots,t_i,\cdots, t_{N_T} = T\}$ as follows (see e.g.~\cite{Gre20,BriMor13}),
\begin{align}
\text{CVA}(t) 
&\simeq - LGD_C \sum_{i=1}^{N_T} \mathcal{H}^{+} (t;t_i) \mathcal{S}_B(t,t_i) \Delta \mathcal{Q}_C(t,t_i),
\label{eq:CVA_disc}\\
\text{DVA}(t) 
&\simeq - LGD_B \sum_{i=1}^{N_T} \mathcal{H}^{-} (t;t_i) \mathcal{S}_C(t,t_i) \Delta \mathcal{Q}_B(t,t_i),
\label{eq:DVA_disc}\\
\mathcal{H}^{\pm}(t;t_i) 
&= \mathbb{E}^Q\left[ D(t;t_i) \left[ H(t_i) \right]^{\pm} \vert \mathcal{F}_t  \right] 
= P(t;t_i)  \mathbb{E}^{Q^{t_i}} \left[ \left[ H(t_i) \right]^{\pm} \vert \mathcal{F}_t  \right],
\label{eq:epe_ene}
\end{align}
where $\mathcal{H}^{+} (t;t_i)$ is the Expected Positive Exposure (EPE) at time $t$, discretized on interval $(t_{i-1},t_i ]$, $ \mathcal{H}^{-} (t;t_i)$ is the Expected Negative Exposure (ENE), $\mathcal{S}_X(t,t_i) = 1 - \mathcal{Q}_X(t,t_i)$ is the survival probability of X at time $t$, referred to time interval $[t,t_i]$ and $\Delta \mathcal{Q}_X(t,t_i) = \mathcal{Q}_X(t,t_i) - \mathcal{Q}_X(t,t_{i-1})$ is the marginal default probability of X at time $t$, referred to interval $(t_{i-1},t_i]$. 
\par 
In general, the exposure $\mathcal{H}^{\pm}(t;t_i)$ in eq. \ref{eq:epe_ene} cannot be computed analytically, except in particular cases as discussed in app. \ref{app:XVA analytical formulas} below, and one has to resort to Monte Carlo simulation,
\begin{equation}
\mathcal{H}^{\pm}(t;t_i) 
\simeq P(t;t_i) \frac{1}{N_{MC}} \sum_{m=1}^{N_{MC}} \left[ H_{m}(t_i)\right]^{\pm},
\label{eq:epe_ene_MC}
\end{equation}
where $H_m(t_i)$ is the exposure at time step $t_i$ for MC path $m$, which may include both Variation and Initial margins.      
\par 
In order to investigate the XVA Monte Carlo error, we build, for each time step $t_i$, the following $n\sigma$ upper and lower bounds on EPE/ENE,
\begin{align}
\begin{split}
\mathcal{H}^{\pm}_{\text{UB}}(t_0;t_i)  & = \mathcal{H}(t_0;t_i)^{\pm} + n \frac{\sigma_{\mathcal{H}^{\pm}(t_0;t_i)}}{\sqrt{N_{MC}}}, \\
\mathcal{H}^{\pm}_{\text{LB}}(t_0;t_i)   & = \mathcal{H}(t_0;t_i)^{\pm} - n \frac{\sigma_{\mathcal{H}^{\pm}(t_0;t_i)}}{\sqrt{N_{MC}}},
\end{split}
\end{align}
where 
\begin{equation}
\sigma_{\mathcal{H}^{\pm}(t_0;t_i)} = \sqrt{\frac{1}{N_{MC}-1} \sum_{m=1}^{N_{MC}} \left[ P(t_0;t_i) \left[ H_{m}(t_i) \right]^{\pm} - \mathcal{H}^{\pm}(t_0;t_i) \right]^{2}},
\end{equation}
is the MC standard deviation of the EPE/ENE in eq. \ref{eq:epe_ene}.
Then, we may use the quantities above to get the following confidence interval for CVA/DVA 
\begin{align}
\begin{split}
	\text{CVA}_x(t_0) &=  -LGD_C\sum_{i = 1}^{N_T} \mathcal{H}^{+}_x(t_0;t_i) \mathcal{S}_B(t_0,t_i) \Delta \mathcal{Q}_C(t_0,t_i), \\
	\text{DVA}_x(t_0) &=  -LGD_B\sum_{i = 1}^{N_T} \mathcal{H}^{-}_x(t_0;t_i) \mathcal{S}_C(t_0,t_i) \Delta \mathcal{Q}_B(t_0,t_i),
\end{split}
\label{eq:XVA LB UB}
\end{align}
where $x\in\left\{UB,LB\right\}$, using, e.g., $n=3 \sigma$.

\subsubsection{XVA Analytical Formulas}
\label{app:XVA analytical formulas}
In the special case of single linear derivative without collateral it is possible to solve analytically the eqs.~\ref{eq:CVA2} and \ref{eq:DVA2}, with considerable benefits in terms of computational time (see e.g.~\cite{BriMor13} and references therein). 
Nevertheless, this approach can be used to validate the results obtained through the Monte Carlo simulation, as discussed in sec.~\ref{sec:Monte Carlo Versus Analytical XVAs}.

In the case of an uncollateralized interest rate Swap, the CVA at time $t$ can be written, from eq. \ref{eq:CVA2}, as
\begin{equation}
\begin{split}
&\text{CVA}(t) \\
&= - LGD_C \int_{t}^{T} \mathbb{E}^Q \left\{ D(t;u) \left[ \textbf{Swap}(u;\textbf{T},\textbf{S},K,\omega)\right]^{+} \vert \mathcal{F}_t  \right\} S_B(t,u) dQ_C(t,u)\\
&= - LGD_C \int_{t}^{T}  \textbf{Swaption}(t;u,\textbf{T},\textbf{S},K,\omega) S_B(t,u) dQ_C(t,u) ,
\end{split}
\label{eq:CVA_exact_formula_int}
\end{equation}
where $\textbf{Swaption}(t;u,\textbf{T},\textbf{S},K,\omega)$ denotes the price at time $t$ of an European Swaption expiring at time $u$ on a Swap having tenor equal to $\mathcal{T} =T-u$.

Since $[x]^{-} = [-x]^{+}$, the companion analytical DVA can be obtained as
\begin{equation}
\begin{split}
& \text{DVA}(t) \\
&= - LGD_B \int_{t}^{T}  \mathbb{E}^Q \left\{ D(t;u)\left[\textbf{Swap}(u;\textbf{T},\textbf{S},K,\omega)\right]^{-} \vert \mathcal{F}_t  \right\} S_C(t,u) dQ_B(t,u), \\
& = - LGD_B \int_{t}^{T}  \mathbb{E}^Q \left\{D(t;u)\left[\textbf{Swap}(u;\textbf{T},\textbf{S},K,-\omega)\right]^{+} \vert\mathcal{F}_t\right\} S_C(t,u) dQ_B(t,u), \\					 
& = - LGD_B \int_{t}^{T}  \textbf{Swaption}(t;u,\textbf{T},\textbf{S},K,-\omega) S_C(t,u) dQ_B(t,u).\\
\end{split}
\label{eq:DVA_exact_formula_int}
\end{equation}
By discretizing the above integrals using a time grid $\{t = t_0,\cdots,t_i,\cdots, t_{N_T} = T\}$, with $T$ being the maturity of the Swap, we obtain
\begin{align}
\text{CVA}(t) 
&\simeq -LGD_C \sum_{i=1}^{N_{T}} \textbf{Swaption}(t;t_i,\textbf{T},\textbf{S},K,\omega)\mathcal{S}_B(t,t_i) \Delta \mathcal{Q}_C(t,t_i) 
\label{eq:CVA_exact_formula_disc} , \\
\text{DVA}(t) 
&\simeq -LGD_B \sum_{i=1}^{N_{T}} \textbf{Swaption}(t;t_i,\textbf{T},\textbf{S},K,-\omega)\mathcal{S}_C(t,t_i) \Delta \mathcal{Q}_B(t,t_i) 
\label{eq:DVA_exact_formula_disc} . 
\end{align}
In conclusion, the CVA of a single, uncollateralized payer (receiver) Swap at time $t$ can be written as a weighted sum of co-terminal payer (receiver) European Swaptions expiring at time $t_i$ to enter in a Swap with tenor $\mathcal{T} =T-t_i$. 
Symmetrically, the DVA at time $t$ can be written as a weighted sum of co-terminal receiver (payer) European Swaptions.

Given a discrete time grid to compute CVA and DVA according to equations \ref{eq:CVA_exact_formula_disc} and \ref{eq:DVA_exact_formula_disc} the implementation of the pricing of each co-terminal Swaption involves a number of steps, enumerated below in the case of shifted-SABR framework:
\begin{enumerate}
	\item calculation of the shifted Black implied volatility cube starting form the quoted Swaption price cube (see \ref{tab:mkt_swpt_prices_cube});
	\item calibration of the shifted-SABR model for each combination of expiries $\xi$ and tenors $\mathfrak{T}$;
	\item rearrangement of the calibrated parameters in the corresponding 4 parameters' matrices $\alpha, \beta, \nu, \rho$;
	\item For each time step $t_i$
	\begin{enumerate}
		\item compute the set of SABR parameters $p_i^{SABR}=\{\alpha_i,\beta_i, \nu_i, \rho_i \}$ through a 2-dimensional interpolation and extrapolation\footnote{We interpolated and extrapolated linearly.} on tenor/expiry market grid for expiry $t_i-t$ and tenor $T-t_i$
		\begin{equation}
			\begin{split}
				\alpha_i&=\text{2DInterp}(\alpha;t_i-t,T-t_i)\\
				\beta_i&=\text{2DInterp}(\beta;t_i-t,T-t_i)\\
				\nu_i&=\text{2DInterp}(\nu;t_i-t,T-t_i)\\	
				\rho_i&=\text{2DInterp}(\rho;t_i-t,T-t_i);
			\end{split}
		\end{equation}
		\item compute the underlying forward starting Swap Rate
		\begin{equation}
			\textbf{Swap}(t_i;\textbf{T},\textbf{S},K,\omega);
		\end{equation} 
		\item compute the SABR shifted-lognormal volatility 
		\begin{equation} 		\sigma^{SABR}(\textbf{Swap}(t_i;\textbf{T},\textbf{S},K,\omega),K,p_i^{SABR})
		\end{equation} 
		using the standard closed-form formula (see \cite{HagKum02,Obl07}). It's worth to notice that the the co-terminal swaptions are in general ITM/OTM, even if the underlying swap is ATM;
		\item compute the Swaption price using shifted-Black formula
		\begin{equation}
			\begin{split}
				\textbf{Swaption}&(t;t_i,\textbf{T},\textbf{S},K,\omega)=A_d(t_i,S)\times\\
				&\text{Black}\left[\textbf{Swap}(t_i;\textbf{T},\textbf{S},K,\omega)+\lambda ,K+\lambda,v^{SABR}\right]
			\end{split}
		\end{equation}
		where $v^{SABR}=v^{SABR}(\textbf{Swap}(t_i;\textbf{T},\textbf{S},K,\omega),K,p_i^{SABR})$ is related to the SABR shifted-lognormal volatility as follows
		\begin{equation} 	
			\begin{split}	
				\sigma^{SABR}(\textbf{Swap}(t_i;\textbf{T},&\textbf{S},K,\omega),K,p_i^{SABR})=\\
				&\sqrt{\dfrac{v^{SABR}(\textbf{Swap}(t_i;\textbf{T},\textbf{S},K,\omega),K,p_i^{SABR})}{\tau(t_i,T)}}.
			\end{split}
		\end{equation} 
	\end{enumerate}
\end{enumerate}

\section{Collateral Modelling}
\label{app:collateral modelling}

In this appendix we introduce collateralization and describe the assumptions made for calculating VM and IM, with a particular focus on ISDA-SIMM dynamic IM.        

With the aim to reduce the systemic risk posed by non-cleared OTC derivatives, the BCBS-IOSCO framework (see \cite{BCBSIOSCO15}) requires institutions engaging in these transactions to post bilaterally Variation Margin (VM) and Initial Margin (IM) on a daily basis at a netting set level. 
On the one hand, VM aims at covering the current exposure stemming from changes in instrument's mark to market by reflecting its current size. 
On the other hand, IM aims at covering the potential future exposure that could arise, in the event of default of the counterparty, from changes in instrument's mark to market in the period between the last VM exchange and the close-out of the position, also known as margin period of risk (MPoR).

\subsection{Collateral Management}
\label{app:collateral management}

The ISDA Master Agreement represents the most common legal framework which governs bilateral OTC derivatives transactions. In particular, the Credit Support Annex (CSA) to ISDA Master Agreement provides the terms under which collateral is posted, along with rules for the resolution of collateral disputes. Certain CSA parameters affect the residual exposure, namely: eligible assets (cash, cash equivalent, government bonds), margin call frequency, threshold (K), defined as the maximum amount of allowed unsecured exposure before any margin call is made, and minimum transfer amount (MTA), defined as the minimum amount that can be transferred for each margin call.   

Under perfect collateralization, counterparty risk is suppressed resulting in null XVA. Theoretically, this corresponds to an ideal CSA ensuring a perfect match between the price $V_{0}(t)$ and the corresponding collateral at any time $t$.
This condition is approximated in practice with a CSA minimizing any friction between the mark to market and the collateral, i.e.~cash collateral in the same currency of the trade, daily margination, flat overnight collateral rate, zero threshold and minimum transfer amount. 
Nevertheless, real CSA introduces some frictions causing divergences between the price and the corresponding collateral and hence a non-null counterparty risk. For example, collateral transfers are not instantaneous events and may also take several days to complete in case of disputes; for this reason, in the event of default, the collateral actually available to the non-defaulting party at the close-out date may differ from the prescribed one.

A simple way to capture these divergences is to assume that VM and IM available at time step $t_i$ depend on the instrument price computed at time $\hat{t}_i = t_i-l$. In this way the time interval $\left[ t_i-l,t_i \right]$ represents the MPoR, $\hat{t}$ is the last date for which collateral was fully posted and $t_i$ is the close-out date, and $l$ is the length of the MPoR. This implies that, while both counterparties stop simultaneously to post collateral for the entire MPoR, contractual cash flows are fully paid. Although simplistic, this assumption can be deemed appropriate in relation to the purposes of our work. More advanced models are developed in \cite{AndPyk16,AndPyk17}) and could be easily taken into account in our framework.

In practical terms, the inclusion of MPoR requires a secondary time grid, built by defining for each $t_i$ of the principal time grid a look-back time point $\hat{t}_i= t_i - l$ such that $\hat{t}_i$ is the collateral calculation date. Formally, from bank's perspective the collateralized exposure $H_m(t_i)$ at time step $t_i$ for a generic path $m$, can be written as\footnote{Notice that when VM only is considered the collateralized exposure simply reduces to $H_m(t_i)=V_{0,m} \left(t_i \right)-\text{VM}_m \left(t_i;V_{0,m} (\hat{t}_i),\text{K}_\text{VM},\text{MTA}_\text{VM} \right)$ }
\begin{equation}
\resizebox{\linewidth}{!}{$
H_m(t_i) = \begin{cases}
\left[ V_{0,m} \left(t_i \right)-\text{VM}_m \left(t_i;V_{0,m} (\hat{t}_i),\text{K}_\text{VM},\text{MTA}_\text{VM} \right) - \text{IM}_m^{C}\left( t_i;V_{0,m} (\hat{t}_i),\text{K}_\text{IM},\text{MTA}_\text{IM} \right) \right]^{+} &\text{ if } V_{0,m}-\text{VM}_m \geq 0 \\
\left[ V_{0,m} \left(t_i \right)-\text{VM}_m \left(t_i;V_{0,m} (\hat{t}_i),\text{K}_\text{VM},\text{MTA}_\text{VM} \right) + \text{IM}_m^{B}\left( t_i;V_{0,m} (\hat{t}_i),\text{K}_\text{IM},\text{MTA}_\text{IM} \right) \right]^{-} &\text{ if } V_{0,m}-\text{VM}_m < 0   	
\end{cases}
$}
\label{eq:exposure vm_im}	
\end{equation}

\noindent
Here we assumed that a bilateral and symmetrical CSA is in place, VM is netted, and IM is posted by the Counterparty ($\text{IM}_m^C$) or by the bank ($\text{IM}_m^B$) into a segregated account as required by the regulation (see \cite{BCBSIOSCO15}). Therefore, from bank's perspective, VM can be positive (if posted by the counterparty) or negative (if posted by the bank), while IM is always positive. For the sake of generality we distinguished IM posted by the Bank and the one posted by the Counterparty; in fact, for instruments with optionality, it may happen that $\text{IM}_m^C\neq\text{IM}_m^B$ due to convexity effects introduced by the non-linear dependence on the exposure within Curvature Margin definition (see eq. \eqref{eq:curvature_margin}).

\subsection{Variation Margin}
\label{app:collateral VM}

VM modelling is fairly straightforward as it depends on instrument's mark to market, together with K$_\text{VM}$ and MTA$_\text{VM}$. We calculated VM available at time step $t_i$ for a generic path $m$ through the following formula
\begin{align}
	& \text{VM}_m \left(t_i;V_{0,m} (\hat{t}_i),\text{K}_\text{VM},\text{MTA}_\text{VM} \right) =  \widehat{\text{VM}}_m(\hat{t}_{i-1}) \nonumber \\
	& + \mathds{1}_{ \left\{ \left\lvert \left( V_{0,m}(\hat{t}_i)-\text{K}_\text{VM}\right)^{+} - \ \widehat{\text{VM}}^{+}_m(\hat{t}_{i-1}) \right\rvert > \text{MTA}_\text{VM} \right\} } \bigl[ ( V_{0,m}(\hat{t}_i)-\text{K}_\text{VM})^{+} - \widehat{\text{VM}}^{+}_m(\hat{t}_{i-1}) \bigr] \nonumber \\
	& + \mathds{1}_{ \left\{ \left\lvert \left( V_{0,m}(\hat{t}_i)+\text{K}_\text{VM}\right)^{-} - \ \widehat{\text{VM}}^{-}_m(\hat{t}_{i-1}) \right\rvert > \text{MTA}_\text{VM} \right\} } \bigl[ ( V_{0,m}(\hat{t}_i)+\text{K}_\text{VM})^{-} - \widehat{\text{VM}}^{-}_m(\hat{t}_{i-1}) \bigr]  \label{eq:vm}	\\
	& \widehat{\text{VM}}_m(\hat{t}_{i-1}) = \frac{\text{VM}_m(\hat{t}_{i-1})}{P(\hat{t}_{i-1},\hat{t}_{i})} ,
\end{align}
where $\widehat{\text{VM}}_m(\hat{t}_{i-1})$ is the value of VM just before its update at $\hat{t}_i$, and the second and the third terms of eq.~\ref{eq:vm} correspond to the amount of VM posted at time step $\hat{t}_i$ by the counterparty and the bank respectively. In particular, in the second term, the counterparty will update VM for the amount exceeding the threshold and VM already in place, provided that this amount is greater than the minimum transfer amount; the same holds for the bank in case of negative exposure (third term). We imposed null VM for $t_i = t_0$ and $t_i = t_{N_{T}}$.

\subsection{Initial Margin}
\label{app:collateral IM}

\subsubsection{ISDA Standard Initial Margin Model}
\label{app:collateral IM definitions}

In 2013 ISDA, in cooperation with entities first impacted by bilateral initial margin requirements, started developing the Standard Initial Margin Model (SIMM) with the aim to provide market participant with a uniform risk-sensitive model for calculating bilateral IM (see \cite{ISDA13}), preventing both potential disputes between counterparties related to IM determination with different internal models and the overestimation of margin requirements due to the use of the non-risk-sensitive standard approach (see \cite{BCB13}). The first version of the model was published in 2016. 
On an annual basis the model parameters are recalibrated and the methodology is reviewed in order to ensure that regulatory requirements are met. Since our work relies on market data at 28 December 2018, we considered the ISDA-SIMM Version 2.1 which was effective from 1 December 2018 to 30 November 2019 (see \cite{ISDA18}).
\par 
In general, ISDA-SIMM is a parametric VaR model based on Delta, Vega and Curvature (i.e.~``pseudo'' Gamma) sensitivities, defined across risk factors by asset class, tenor and expiry, computed in line with specific definitions. More in detail, each trade of a portfolio (under a certain CSA agreement) is assigned to a Product Class among Interest Rates \& FX, Credit, Equity and Commodity. Since a given trade may have sensitivity to different risk factors, six Risk Classes are defined among Interest Rate, FX, Credit (Qualifying), Credit (non-Qualifying), Equity and Commodity. The margin contributions stemming from the different Risk Classes are combined by means of an aggregation function taking account of Risk Classes correlations. Formally, IM for a generic instrument can be written as
\begin{align}
\text{IM} &= \sqrt{\sum_{x}\text{M}^2_x + \sum_{x} \sum_{s \neq x} \psi_{x,s} \text{M}_x \text{M}_s},  \\
\text{M}_x &= \text{DeltaMargin}_x + \text{VegaMargin}_x + \text{CurvatureMargin}_x,
\end{align}
where M$_x$ is the margin component for the Risk Class\footnote{For Credit (Qualifying) Risk Class, which includes instruments whose price is sensitive to correlation between the defaults of different credits within an index or basket (e.g.~CDO tranches), an additional margin component, i.e.~the BaseCorrMargin shell be calculated (see \cite{ISDA18}).} $x$, with $x \in  \left\{\text{Interest Rate, FX,}\dots \right\}$,  and $\psi_{x,s}$ is the correlation matrix between Risk Classes. 
IM at portfolio level is obtained by adding together IM contributions from each trade. see \cite{ISDA16,ISDA18} for a complete description of the model and underlying assumptions.

In our case, Swaps and European Swaptions are assigned to the Interest Rates \& FX Product Class and exposed only to Interest Rate (IR) Risk Class, thus $\text{IM} =  \text{M}_{\text{IR}}$. Moreover, for a Swap $\text{IM} = \text{DeltaMargin}_{\text{IR}}$ given the linearity of its payoff. 
As discussed above, we imposed that collateral available at time step $t_i$ is function of instrument's value (sensitivities in IM case) at time $\hat{t}_i = t_i - l$. Therefore, at time step $t_i$ and for a generic path $m$ the ISDA-SIMM dynamic IM is given by 
\begin{align}
\text{IM}^{\text{Swap}}_m(t_i) &	= \text{DeltaMargin}_{\text{IR},m}\left(t_i;V_m(\hat{t}_i)\right), \\
\text{IM}^{\text{Swpt}}_m(t_i) & = \text{DeltaMargin}_{\text{IR},m}\left(t_i;V_m(\hat{t}_i)\right)+ \text{VegaMargin}_{\text{IR},m}\left(t_i;V_m(\hat{t}_i)\right)  \nonumber \\
&\quad + \text{CurvatureMargin}_{\text{IR},m}\left(t_i;V_m(\hat{t}_i)\right).
\end{align} 
Finally, allowing for K$_\text{IM}$ and MTA$_\text{IM}$
\begin{equation}
\begin{split}
& \text{IM}^{p}_m\bigl( t_i;V_m (\hat{t}_i),\text{K}_\text{IM},\text{MTA}_\text{IM} \bigr) \\
& = \mathds{1}_{ \left\{ \left(\text{IM}^p_m (t_i) - \text{K}_\text{IM} \right)^{+} > \text{MTA}_\text{IM} \right\}} \left[ \left(\text{IM}^p_m (t_i) - \text{K}_\text{IM} \right)^{+} \right], \quad p \in \left\{ \text{Swap, Swpt} \right\}.
\end{split}
\label{eq:im}
\end{equation}
Similar to VM, we imposed null IM for $t_i = t_0$ and $t_i = t_{N_{T}}$.

\subsubsection{ISDA-SIMM for Swaps and Swaptions}
\label{app:SIMM Formulas}

In this appendix we describe the implementation of the ISDA Standard Initial Margin Model with respect to interest rate Swaps and Swaptions investigated in this paper (at trade level). In particular, we considered ISDA-SIMM Version 2.1 in order to be consistent with the valuation date considered (28 December 2018) \footnote{Version 2.1 was effective from 1 December 2018 to 30 November 2019 when Version 2.2 was published. In particular, the values reported in this appendix refer to Version 2.1.} used in our analyses. Further details may be found in \cite{ISDA18}. 
As shown in app.~\ref{app:collateral IM definitions}, IM for a Swap and an European Swaption is given by
\begin{align}
	\text{IM}^{\text{Swap}} &	= \text{DeltaMargin}^{\text{IR}},\\
	\text{IM}^{\text{Swpt}}  &	= \text{DeltaMargin}^{\text{IR}} + \text{VegaMargin}^{\text{IR}} + \text{CurvatureMargin}^{\text{IR}}.
\end{align} 
The details on the calculation process for the three IM components are given in sections below.

\paragraph{Delta Margin for Interest Rate Risk Class}
Delta Margin for IR Risk Class is computed through the following step-by-step process.
\begin{enumerate}
	\item Calculation of IR Delta sensitivities vector
	\begin{equation}
	\Delta^{c} = [\Delta^{c}_{2\text{w}},\Delta^{c}_{1\text{m}},\Delta^{c}_{3\text{m}},\Delta^{c}_{6\text{m}},\Delta^{c}_{1\text{y}},\Delta^{c}_{2\text{y}},\Delta^{c}_{3\text{y}},\Delta^{c}_{5\text{y}},\Delta^{c}_{10\text{y}},\Delta^{c}_{15\text{y}},\Delta^{c}_{20\text{y}},\Delta^{c}_{30\text{y}}],
	\end{equation}
	where $c\in\{f,d\}$. See app.~\ref{app:collateral IM calculation methodology} for details on definition and calculation methodology.  
	\item Calculation of Weighted Sensitivity WS$_{j,c}$ for the $j$-th SIMM tenor through the following formula
	\begin{equation}
	\text{WS}_{j,c} = \text{RW}_{j} \Delta^{c}_{j} \text{CR}_b,
	\label{eq:ws_delta}
	\end{equation}
	where:
	\begin{itemize}
		\item RW$_j$ is the Risk Weight for the $j$-th SIMM tenor (tab.~\ref{tab:risk_weights}). ISDA specifies three different vectors based upon the volatility of the currency in which the instrument is denominated\footnote{Low Volatility currencies: JPY; Regular Volatility currencies: USD, EUR, GBP, CHF, AUD, NZD, CAD, SEK, NOK, DKK, HKD, KRW, SGD and TWD; High Volatility currencies: all other currencies.};
		\item $\Delta^c_j$ is Delta sensitivity corresponding to the $j$-th SIMM tenor for the interest rate curve $c$;
		\item CR$_b$ is the Concentration Risk Factor for the Currency Group $b$. In this case ISDA specifies four Currency Groups based upon the volatility of the currency in which the instrument is denominated\footnote{Low Volatility currencies: JPY; Regular Volatility well-traded currencies: USD, EUR, GBP; Regular Volatility less well-traded currencies: CHF, AUD, NZD, CAD, SEK, NOK, DKK, HKD, KRW, SGD, TWD; High Volatility currencies: all other currencies. \label{item:fn_currency_group}}. CR$_b$ is calculated as follows
		\begin{equation}
		\text{CR}_b = \text{max} \left\{1,\left( \frac{\vert \sum_{j,c} \Delta^{c}_{j}\vert}{\text{T}_b} \right)^\frac{1}{2}\right\},
		\label{eq:cr_delta}
		\end{equation}
		where T$_b$ is the Concentration Threshold for the Currency Group $b$. For Regular Volatility well-traded currencies T$_b =$ 210 USD  Mio/bp.
	\end{itemize}       
	\item Calculation of Delta Margin through the aggregation of Weighted Sensitivities
	\begin{equation}
	\text{DeltaMargin}^\text{IR} = \sqrt{\sum_{c,j} \text{WS}_{j,c}^2 + \sum_{c,j} \sum_{(k,l) \neq (c,j)} \phi_{c,k} \rho_{j,l} \text{WS}_{j,c} \text{WS}_{l,k}},
	\label{eq:K_delta}
	\end{equation}
	where:
	\begin{itemize}
		\item $\phi_{c,k} =$ 98\%  is the correlation between interest rate curves of the same currency;
		\item $\rho_{j,l}$ is the correlation matrix between SIMM tenors (tab.~\ref{tab:correlation_tenors}).
	\end{itemize}
\end{enumerate}

\paragraph{Vega Margin for Interest Rate Risk Class}
Vega Margin for IR Risk Class is computed through the following step-by-step process.
\begin{enumerate}
	\item Calculation of Vega Risks vector 
	\begin{equation}
	\resizebox{\linewidth}{!}{$
		\text{VR} = [\text{VR}_{2\text{w}}, \text{VR}_{1\text{m}}, \text{VR}_{3\text{m}}, \text{VR}_{6\text{m}}, \text{VR}_{1\text{y}}, \text{VR}_{2\text{y}}, \text{VR}_{3\text{y}}, \text{VR}_{5\text{y}}, \text{VR}_{10\text{y}}, \text{VR}_{15\text{y}}, \text{VR}_{20\text{y}}, \text{VR}_{30\text{y}}],$}
	\end{equation}
	where:
	\begin{equation}
	\text{VR}_{j} = \nu_{j} \sigma^{\text{Blk}}_{j},
	\label{eq:Vega_Risk}
	\end{equation}
	is the Vega Risk for an European Swaption with expiry equal to tenor $j$, with $\nu_{j}$ and $\sigma^{\text{Blk}}_{j}$ being, respectively, IR Vega sensitivity and Black implied volatility. See app.~\ref{app:collateral IM calculation methodology} for details on definition and calculation methodology.  
	\item Calculation of Vega Risk Exposure VRE$_j$ for the $j$-th SIMM expiry through the following formula
	\begin{equation}
	\text{VRE}_{j} = \text{VRW}^{\text{IR}} \text{VR}_j \text{VCR}_b,
	\label{eq:VR_vega_margin}
	\end{equation}
	where:
	\begin{itemize}
		\item VRW$^{\text{IR}}= 0.16$  is the Vega Risk Weight for IR Risk Class; 
		\item VCR$_b$ is the Vega Concentration Risk Factor for the Currency Group $b$ calculated as\footnote{See footnote \ref{item:fn_currency_group}.}
		\begin{equation}
		\text{VCR}_b = \text{max} \left\{ 1,\left( \frac{\vert \sum_{j} \text{VR}_{j}\vert}{\text{VT}_b} \right)^\frac{1}{2} \right \},
		\end{equation}
		where VT$_b$ is the Concentration Threshold for the Currency Group $b$. For Regular Volatility well-traded currencies T$_b = 2200$ USD  Mio.
	\end{itemize}
	\item Calculation of Vega Margin through the aggregation of the Vega Risk Exposures
	\begin{equation}
	\text{VegaMargin}^\text{IR} = \sqrt{\sum_{j} \text{VRE}_{j}^2 + \sum_{j} \sum_{l \neq j} \rho_{j,l} \text{VRE}_{j} \text{VRE}_{l}},
	\label{eq:K_vega}
	\end{equation}
	where $\rho_{j,l}$ is the correlation matrix between expiries (tab.~\ref{tab:correlation_tenors}).
\end{enumerate}

\paragraph{Curvature Margin for Interest Rate Risk Class}
Curvature Margin for IR Risk Class is computed through the following step-by-step process.
\begin{enumerate}
	\item Calculation of Curvature Risk vector by applying a Scaling Function (tab.~\ref{tab:scaling_function}) to Vega Risk vector calculated for Vega Margin, in particular, for the $j$-th element
	\begin{equation}
	\text{CVR}_{j} = \text{SF}(t_{j}) \text{VR}_{j},
	\label{eq:CVR_curvature_margin}
	\end{equation}
	the value of the scaling function SF$(t_{j})$ is defined as
	\begin{equation}
	\text{SF}(t_{j}) = 0.5 \text{ min} \left\{ 1, \frac{14 \text{ days}}{t_{j} \text{ days}} \right\},
	\label{eq:scaling_function}
	\end{equation} 
	where $t_{j}$ is the time (in calendar days) from the valuation date to the $j$-th SIMM expiry. The values of the Scaling Function for all SIMM expiries are reported in tab.~\ref{tab:scaling_function}.
	\item Aggregation of Curvature Risk vector elements
	\begin{equation}
	K = \sqrt{\sum_{j} \text{CVR}_{j}^2 + \sum_{j} \sum_{l \neq j} \rho^2_{j,l} \text{CVR}_{j} \text{CVR}_{l}},
	\label{eq:K_curvature}
	\end{equation}
	where $\rho_{j,l}$ is the correlation matrix between expiries (tab.~ \ref{tab:correlation_tenors}).
	\item Calculation of Curvature Margin through the following formula
	\begin{equation}
	\text{CurvatureMargin}^\text{IR} = \frac{\text{max} \left\{0, \sum_j \text{CVR}_{j} + \lambda K \right\}} {\text{HVR}_{\text{IR}}^2},
	\label{eq:curvature_margin}
	\end{equation}
	where: 
	\begin{itemize}
		\item HVR$_{\text{IR}} = 0.62$ is the Historical Volatility Ratio for the IR Risk Class;
		\item  $\lambda = (\Phi^{-1}(0.995)^2-1)(1+\theta)-\theta$, with $\Phi^{-1}(0.995)$ is the 99.5$^\text{th}$ percentile of the standard normal distribution, and
		\begin{equation}
		\theta = \text{min}\left\{0,\frac{\sum_j \text{CVR}_j}{\sum_j \vert\text{CVR}_j\vert} \right\}.
		\end{equation}
	\end{itemize}
\end{enumerate}
\begin{table}[h]
	\centering
		\begin{tabular}{lcccccccccccc}
			\toprule
			Tenor & 2w    & 1m    & 3m    & 6m    & 1y    & 2y    & 3y    & 5y    & 10y   & 15y   & 20y   & 30y \\
			\midrule
			Risk Weight    & 114   & 115   & 102   & 71    & 61    & 52    & 50    & 51    & 51    & 51    & 54    & 62 \\
			\bottomrule
	\end{tabular}
	\caption{Risk Weights RW$_j$ parameters which apply to IR Delta tenors for Regular Volatility currencies.}
	\label{tab:risk_weights}
\end{table}
\begin{table}[h]
	\centering
	\resizebox{\linewidth}{!}{
		\begin{tabular}{l|cccccccccccc}
			\toprule
			Tenor/Expiry & 2w    & 1m    & 3m    & 6m    & 1y    & 2y    & 3y    & 5y    & 10y   & 15y   & 20y   & 30y \\
			\midrule
			2w    &       & 63\%  & 59\%  & 47\%  & 31\%  & 22\%  & 18\%  & 14\%  & 9\%   & 6\%   & 4\%   & 5\% \\
			1m    & 63\%  &       & 79\%  & 67\%  & 52\%  & 42\%  & 37\%  & 30\%  & 23\%  & 18\%  & 15\%  & 13\% \\
			3m    & 59\%  & 79\%  &       & 84\%  & 68\%  & 56\%  & 50\%  & 42\%  & 32\%  & 26\%  & 24\%  & 21\% \\
			6m    & 47\%  & 67\%  & 84\%  &       & 86\%  & 76\%  & 69\%  & 60\%  & 48\%  & 42\%  & 38\%  & 33\% \\
			1y    & 31\%  & 52\%  & 68\%  & 86\%  &       & 94\%  & 89\%  & 80\%  & 67\%  & 60\%  & 57\%  & 53\% \\
			2y    & 22\%  & 42\%  & 56\%  & 76\%  & 94\%  &       & 98\%  & 91\%  & 79\%  & 73\%  & 70\%  & 66\% \\
			3y    & 18\%  & 37\%  & 50\%  & 69\%  & 89\%  & 98\%  &       & 96\%  & 87\%  & 81\%  & 78\%  & 74\% \\
			5y    & 14\%  & 30\%  & 42\%  & 60\%  & 80\%  & 91\%  & 96\%  &       & 95\%  & 91\%  & 88\%  & 84\% \\
			10y   & 9\%   & 23\%  & 32\%  & 48\%  & 67\%  & 79\%  & 87\%  & 95\%  &       & 98\%  & 97\%  & 94\% \\
			15y   & 6\%   & 18\%  & 26\%  & 42\%  & 60\%  & 73\%  & 81\%  & 91\%  & 98\%  &       & 99\%  & 97\% \\
			20y   & 4\%   & 15\%  & 24\%  & 38\%  & 57\%  & 70\%  & 78\%  & 88\%  & 97\%  & 99\%  &       & 99\% \\
			30y   & 5\%   & 13\%  & 21\%  & 33\%  & 53\%  & 66\%  & 74\%  & 84\%  & 94\%  & 97\%  & 99\%  &  \\
			\bottomrule
	\end{tabular}}
	\caption{Correlation parameters $\rho_{j,l}$ between tenors/expiries.}
	\label{tab:correlation_tenors}
\end{table}
\begin{table}[h]
	\centering
	\resizebox{\linewidth}{!}{
		\begin{tabular}{lcccccccccccc}
			\toprule
			Expiry  & 2w    & 1m    & 3m    & 6m    & 12m   & 2y    & 3y    & 5y    & 10y   & 15y   & 20y   & 30y \\
			\midrule
			SF    & 50.0\% & 23.0\% & 7.7\% & 3.8\% & 1.9\% & 1.0\% & 0.6\% & 0.4\% & 0.2\% & 0.1\% & 0.1\% & 0.1\% \\
			\bottomrule
	\end{tabular}}
	\caption{Scaling Function values for SIMM expiries calculated through eq.~\ref{eq:scaling_function}, where expiries have been converted to calendar days using the convention that 12m = 365 days, with pro-rata scaling for other tenors, e.g.~1m $= \frac{365}{12}$ days and 5y $= 365 \cdot 5$ days.}
	\label{tab:scaling_function}
\end{table}

\subsubsection{Forward Sensitivity Calculation}
\label{app:collateral IM calculation methodology}

The most challenging task underlying ISDA-SIMM dynamic IM is the simulation of forward sensitivities coherently with ISDA definitions, since the subsequent application of weights and aggregation functions is straightforward if we assume that parameters and aggregation rules do not change during the lifetime of the trade. In this section we report the methodology used to compute forward sensitivities. 

According to ISDA, Delta for the IR Risk Class is defined as price change with respect to a 1 bp shift up in a given tenor\footnote{ISDA defines for both OIS and IBOR curves the following 12 tenors at which Delta shall be computed: 2w, 1m, 3m, 6m, 1y, 2y, 3y, 5y, 10y, 15y, 20y, 30y.}
of the interest rate curve, expressed in monetary terms. Moreover, ISDA specifies that if computed by the internal system at different tenors, Delta shall be linearly re-allocated onto the SIMM tenors. In general, the price of an instrument which depends on an interest rate curve $\mathcal{C}_{c}$ depends explicitly on the zero rates of the curve, which, in turn, depends on the market rates from which the curve is constructed via bootstrapping procedure (see e.g.~\cite{AmeBia13}). Therefore, being $Z^{c} = [Z^{c}_{1}, \dots, Z^{c}_{N_{Z}} ]$ and $R^{c} = [R^{c}_{1}, \dots, R^{c}_{N_{R}} ]$, respectively, the zero rates and the market rates in correspondence of the term structure of the same curve $\mathcal{C}_c$, we calculated the sensitivity with respect to the $j$-th market rate $R^{c}_{j}$ at a generic time step $t_i$ as (to ease the notation we neglect subscripts referring to the path) 
\begin{align}
\Delta^{x}_{j} (t_i) = \frac{\partial V(t_i)}{\partial R^{x}_{j}(t_i)} & 	= \sum^{N^{x}_{Z}}_{k = 1} \frac{\partial V(t_i)}{\partial Z^{x}_{k}(t_i)}  J^{x,x}_{j,k}(t_0), \quad  j = 1,\dots,N^{x}_{R}, 
\label{eq:delta_fwd} \\
\Delta^{d}_{j} (t_i) = \frac{\partial V(t_i)}{\partial R^{d}_{j}(t_i)} & 	= \sum^{N^{d}_{Z}}_{k = 1} \frac{\partial V(t_i)}{\partial Z^{d}_{k}(t_i)} J^{d,d}_{j,k}(t_0) + \sum^{N^{x}_{Z}}_{k = 1} \frac{\partial V(t_i)}{\partial Z^{x}_{k}(t_i)} J^{x,d}_{j,k}(t_0), \, j = 1,\dots,N^{d}_{R}
\label{eq:delta_disc}
\end{align}
where $x$ and $d$ denote, respectively, the forwarding curve $\mathcal{C}_{x}$ and the discounting curve $\mathcal{C}_{d}$,
\begin{equation}
J^{m,n}_{j,k}(t_0) = \frac{\partial Z^{m}_{k}(t_0) }{\partial R^{n}_{j}(t_0)}, 
\end{equation}
is an element of the Jacobian matrix (with $m \neq n $), assumed to be constant for each $t_i > t_0$, and 
\begin{equation}
\frac{\partial V(t_i)}{\partial Z^{c}_{k}(t_i)} \approx \frac{V \bigl( t_i;Z^{c}_{k}(t_i)+h \bigr)-V \bigl( t_i;Z^{c}_{k}(t_i) \bigr)}{h} ,
\end{equation}
is the zero rate Delta sensitivity at $t_i$, with $h=10^{-4}$. The last term of eq.~\ref{eq:delta_disc} takes into account the indirect Delta sensitivity component of the forwarding zero curve to the discounting zero curve, due to the exogenous nature of the bootstrapping procedure.    
In line with ISDA prescriptions, for each tenor $j$ of the curve $\mathcal{C}_c$, we multiplied Delta sensitivity by shock size and linearly allocated this quantity onto the SIMM tenors, obtaining the (1-by-12) Delta vector
\begin{equation}
\Delta^{c}(t_i) = \left[ \Delta^{c}_{2\text{w}}(t_i), \dots, \Delta^{c}_{30\text{y}}(t_i) \right].
\end{equation}
We then calculated and aggregated the Weighted Sensitivities in order to get Delta Margin (see eqs.~\ref{eq:ws_delta} and \ref{eq:K_delta}).      

According to ISDA, Vega for the IR Risk Class is defined as price change with respect to a 1\% shift up in at-the-money (ATM) Black implied volatility $\sigma_{x}^{\text{Blk}}$, formally
\begin{equation}
\begin{split}
\nu(t_i) &= \frac{\partial V (t_i)}{\partial \sigma_{x}^{\text{Blk}}(t_i;T_{e},\textbf{T},\textbf{S})} \\
& \approx \frac{ V \bigl( t_i; \sigma_{x}^{\text{Blk}}(t_i;T_{e},\textbf{T},\textbf{S}) + h \bigr) - V \bigl( t_i; \sigma_{x}^{\text{Blk}}(t_i;T_{e},\textbf{T},\textbf{S}) \bigr) }{h}.
\end{split}
\end{equation}
where we use superscript Blk to distinguish between Black implied volatility $\sigma_{x}^{\text{Blk}}$ and G2++ parameter $\sigma$, with $h=10^{-2}$.
\newline
Vega sensitivity shall be multiplied by implied volatility to obtain the Vega Risk for expiry $T_{e}$ and linearly allocated onto the SIMM expiries, which correspond to the tenors defined for Delta sensitivity.
Since the G2++ European Swaption pricing formula (see eq.~\ref{eq:g2++_swpt_multi_curve}) does not provide for an explicit dependence on Black implied volatility, when performing time simulation Vega cannot be calculated according to the definition above. To overcome this limit we propose the following approximation
\begin{equation}
\frac{\partial V (t_i)}{\partial \sigma_{x}^{\text{Blk}}(t_i;T_{e},\textbf{T},\textbf{S})} \approx \frac{V \bigl( t_i;\sigma + \epsilon_{\sigma},\eta + \epsilon_{\eta} \bigr) - V \bigl( t_i; \sigma, \eta \bigr) }{\hat{\sigma}_{x}^{\text{Blk/G2++}}(t_i;\lambda_x,T_{e},\textbf{T},\textbf{S}) - \sigma_{x}^{\text{Blk/G2++}}(t_i;\lambda_x,T_{e},\textbf{T},\textbf{S})},
\label{eq:vega_workaround}
\end{equation}    
where $\epsilon_{\sigma}$ and $\epsilon_{\eta}$ are shocks applied on G2++ model parameters governing the underlying process volatility, and implied volatilities $\hat{\sigma}_{x}^{\text{Blk}}(t_i)$ and $\sigma_{x}^{\text{Blk}}(t_i)$ are obtained respectively from European Swaption's prices $V (t_i;\sigma + \epsilon_{\sigma},\eta + \epsilon_{\eta})$ and $V( t_i; \sigma, \eta)$ by solving the shifted Black pricing formula (see eq.~\ref{eq:Swaption_price_black}).
In sec.~\ref{sec:Vega Sensitivity time simulation} we report the analyses conducted to validate this approach and to select the values for the shocks $\epsilon_{\sigma}$, $\epsilon_{\eta}$ and the Black shift $\lambda_x$. 
In line with ISDA prescriptions, we multiplied Vega sensitivity by implied volatility $\sigma_{x}^{\text{Blk}}$ and linearly allocated the resulting Vega Risk onto the SIMM expiries, obtaining the (1-by-12) Vega Risk vector
\begin{equation}
\text{VR}(t_i) = \left[ \text{VR}_{2\text{w}}(t_i), \dots , \text{VR}_{30\text{y}}(t_i) \right].
\end{equation}
We then calculated and aggregated the Vega Risk Exposures to get Vega Margin (see eqs.~\ref{eq:VR_vega_margin} and \ref{eq:K_vega}).      

Curvature for the IR Risk Class is calculated by using an approximation of the Vega-Gamma relationship (see \cite{ISDA16}). The (1-by-12) Curvature Risk Vector
\begin{equation}
\text{CVR}(t_i) = \left[ \text{CVR}_{2\text{w}}(t_i), \dots , \text{CVR}_{30\text{y}}(t_i) \right]
\end{equation}     
is obtained multiplying the Vega Risk vector by a Scaling Function. We then aggregated the elements of Curvature Risk vector to get Curvature Margin (see eqs.~\ref{eq:K_curvature} and \ref{eq:curvature_margin}).

\section{Additional Results}
\label{app:additional results}

In this appendix we report additional details and comments with respect to sections~\ref{sec:XVA numerical calculations}, \ref{sec:XVA model validation} and \ref{sec:Model Risk}.

\subsection{XVA Numerical Simulation}
\label{app:exposure}

\begin{figure}[H]
\begin{subfigure}[t]{0.3\textwidth}
\caption{OTM, no collateral.}
    \includegraphics[width=\linewidth]{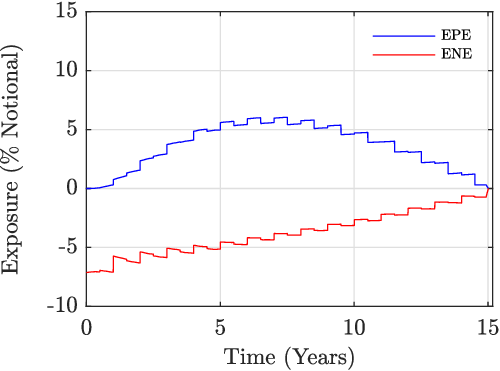}
\label{fig:irs_15Y_OTM_noMargins}
\end{subfigure}
\hfill
\begin{subfigure}[t]{0.3\textwidth}
\caption{ATM, no collateral.}
  \includegraphics[width=\linewidth]{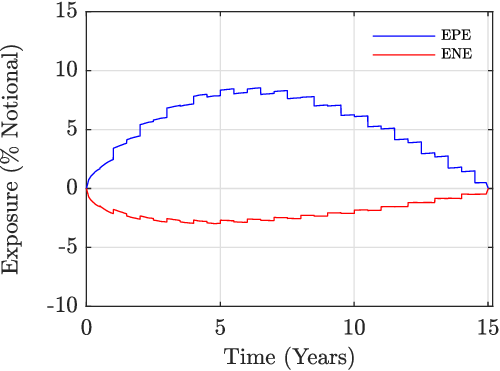}
\label{fig:irs_15Y_ATM_noMargins}
\end{subfigure}
\hfill
\begin{subfigure}[t]{0.3\textwidth}
\caption{ITM, no collateral.}
    \includegraphics[width=\linewidth]{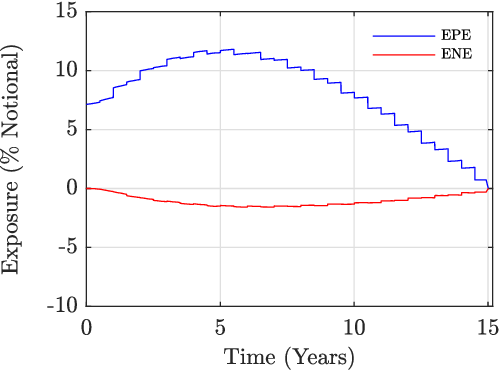}
\label{fig:irs_15Y_ITM_noMargins}
\end{subfigure}\\

\begin{subfigure}[t]{0.3\textwidth}
\caption{OTM, with VM.}
    \includegraphics[width=\linewidth]{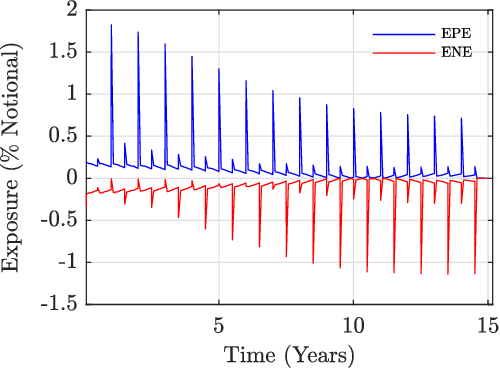}
\label{fig:irs_15Y_OTM_VM_app}
\end{subfigure}
\hfill
\begin{subfigure}[t]{0.3\textwidth}
\caption{ATM, with VM.}
  \includegraphics[width=\linewidth]{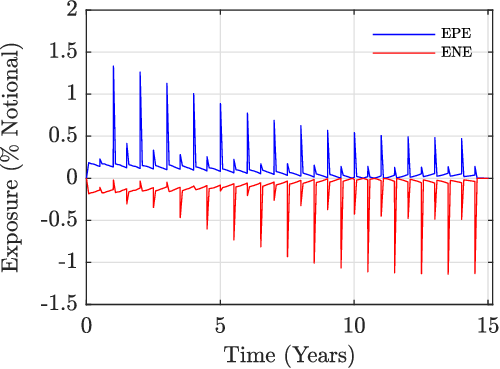}
\label{fig:irs_15Y_ATM_VM_app}
\end{subfigure}\hfill
\begin{subfigure}[t]{0.3\textwidth}
\caption{ITM, with VM.}
    \includegraphics[width=\linewidth]{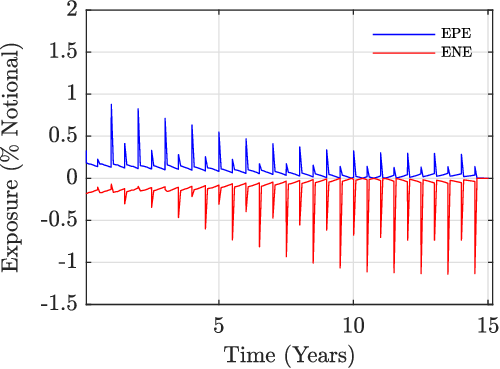}
\label{fig:irs_15Y_ITM_VM_app}
\end{subfigure}\\

\begin{subfigure}[t]{0.3\textwidth}
\caption{OTM, with VM and IM.}
    \includegraphics[width=\linewidth]{Figures/Fig2g.eps}
\label{fig:irs_15Y_OTM_VM_IM_app}
\end{subfigure}
\hfill
\begin{subfigure}[t]{0.3\textwidth}
\caption{ATM, with VM and IM.}
  \includegraphics[width=\linewidth]{Figures/Fig2h.eps}
\label{fig:irs_15Y_ATM_VM_IM_app}
\end{subfigure}
\hfill
\begin{subfigure}[t]{0.3\textwidth}
\caption{ITM, with VM and IM.}
    \includegraphics[width=\linewidth]{Figures/Fig2i.eps}
\label{fig:irs_15Y_ITM_VM_IM_app}
\end{subfigure}
\caption{15Y payer Swaps, EPE and ENE profiles (blue and red solid lines respectively) for the different collateralization schemes considered (top: no collateral, mid: VM, bottom: VM and IM) and moneyness (left: OTM, mid: ATM, right: ITM). To enhance plots readability of collateralized profiles, we excluded the initial time step $t_0$ showing a very high exposure not yet mitigated by the collateral exchanged MPoR days later. Quantities expressed as a percentage of the nominal amount (EUR 100 Mio). Model setup as in tab.~\ref{tab:model_setup}.}
\label{fig:irs_15Y_results}
\end{figure}

\begin{figure}[H]
\begin{subfigure}[t]{0.3\textwidth}
\caption{OTM, no collateral.}
    \includegraphics[width=\linewidth]{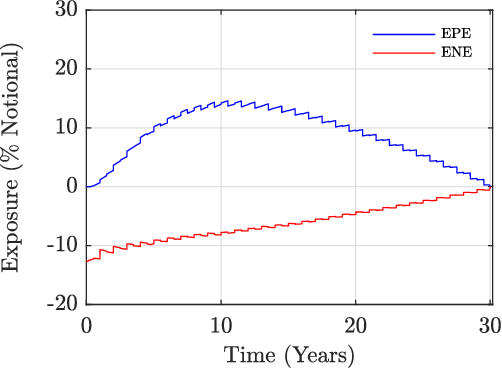}
\label{fig:irs_30Y_OTM_noMargins}
\end{subfigure}
\hfill
\begin{subfigure}[t]{0.3\textwidth}
\caption{ATM, no collateral.}
  \includegraphics[width=\linewidth]{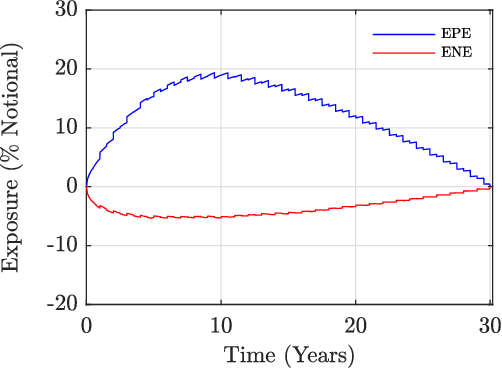}
\label{fig:irs_30Y_ATM_noMargins}
\end{subfigure}
\hfill
\begin{subfigure}[t]{0.3\textwidth}
\caption{ITM, no collateral.}
    \includegraphics[width=\linewidth]{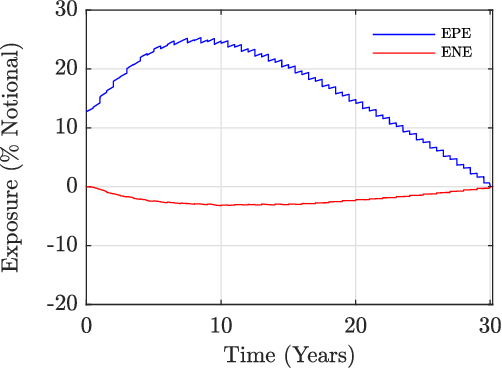}
\label{fig:irs_30Y_ITM_noMargins}
\end{subfigure}\\

\begin{subfigure}[t]{0.3\textwidth}
\caption{OTM, with VM.}
    \includegraphics[width=\linewidth]{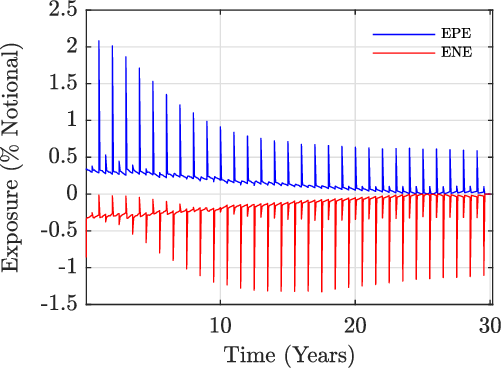}
\label{fig:irs_30Y_OTM_VM_app}
\end{subfigure}
\hfill
\begin{subfigure}[t]{0.3\textwidth}
\caption{ATM, with VM.}
  \includegraphics[width=\linewidth]{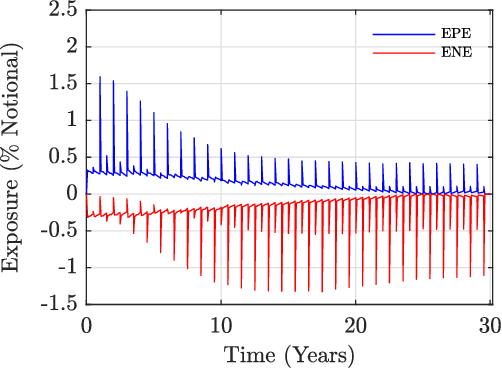}
\label{fig:irs_30Y_ATM_VM_app}
\end{subfigure}
\hfill
\begin{subfigure}[t]{0.3\textwidth}
\caption{ITM, with VM.}
    \includegraphics[width=\linewidth]{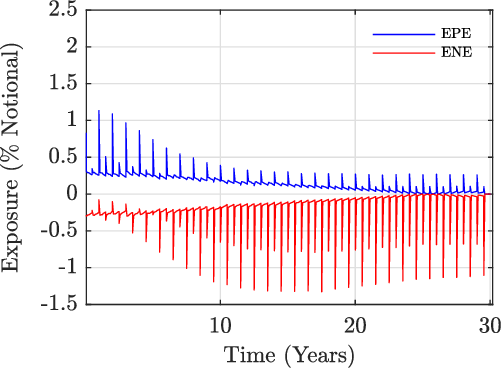}
\label{fig:irs_30Y_ITM_VM_app}
\end{subfigure}\\

\begin{subfigure}[t]{0.3\textwidth}
\caption{OTM, with VM and IM.}
    \includegraphics[width=\linewidth]{Figures/Fig3g.eps}
\label{fig:irs_30Y_OTM_VM_IM_app}
\end{subfigure}
\hfill
\begin{subfigure}[t]{0.3\textwidth}
\caption{ATM, with VM and IM.}
  \includegraphics[width=\linewidth]{Figures/Fig3h.eps}
\label{fig:irs_30Y_ATM_VM_IM_app}
\end{subfigure}
\hfill
\begin{subfigure}[t]{0.3\textwidth}
\caption{ITM, with VM and IM.}
    \includegraphics[width=\linewidth]{Figures/Fig3i.eps}
\label{fig:irs_30Y_ITM_VM_IM_app}
\end{subfigure}
\caption{30 years payer Swaps. Other settings as in fig. \ref{fig:irs_15Y_results}.}
\label{fig:irs_30Y_results}
\end{figure}

\begin{figure}[H]
\begin{subfigure}[t]{0.3\textwidth}
\caption{Receiver OTM, no collat.}
    \includegraphics[width=\linewidth]{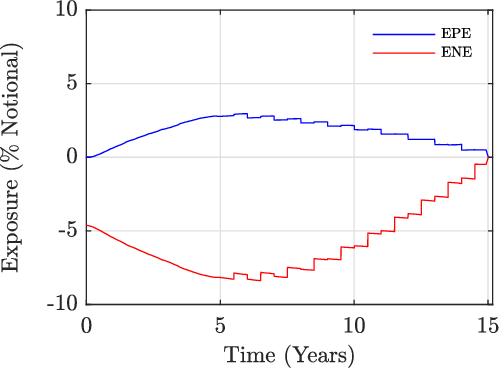}
\label{fig:irs_fwd_receiver_OTM_noMargins}
\end{subfigure}
\hfill
\begin{subfigure}[t]{0.3\textwidth}
\caption{Payer ATM, no collat.}
  \includegraphics[width=\linewidth]{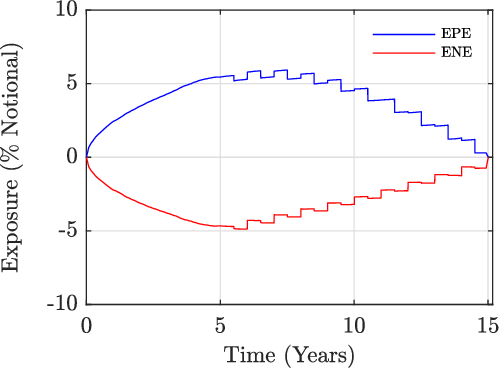}
\label{fig:irs_fwd_ATM_noMargins}
\end{subfigure}
\hfill
\begin{subfigure}[t]{0.3\textwidth}
\caption{Payer OTM, no collat.}
    \includegraphics[width=\linewidth]{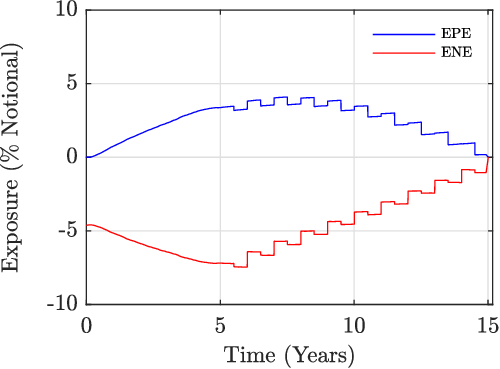}
\label{fig:irs_fwd_payer_OTM_noMargins}
\end{subfigure}\\

\begin{subfigure}[t]{0.3\textwidth}
\caption{Receiver OTM, with VM.}
    \includegraphics[width=\linewidth]{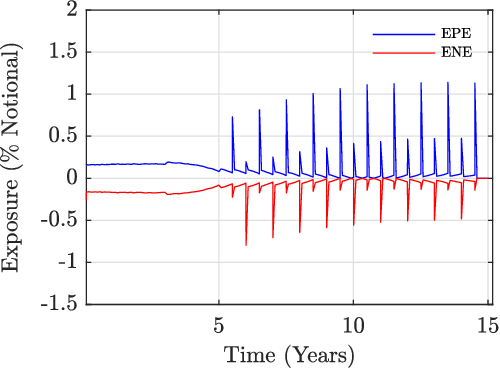}
\label{fig:irs_fwd_receiver_OTM_VM_app}
\end{subfigure}
\hfill
\begin{subfigure}[t]{0.3\textwidth}
\caption{Payer ATM, with VM.}
  \includegraphics[width=\linewidth]{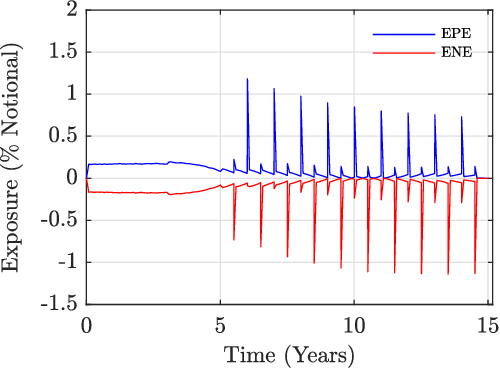}
\label{fig:irs_fwd_ATM_VM_app}
\end{subfigure}
\hfill
\begin{subfigure}[t]{0.3\textwidth}
\caption{Payer OTM, with VM.}
    \includegraphics[width=\linewidth]{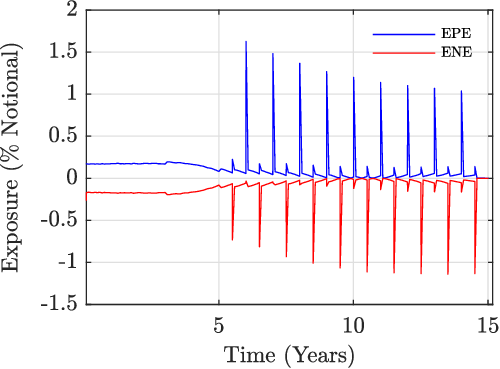}
\label{fig:irs_fwd_payer_OTM_VM_app}
\end{subfigure}\\

\begin{subfigure}[t]{0.3\textwidth}
\caption{Receiver OTM, with VM and IM.}
    \includegraphics[width=\linewidth]{Figures/Fig4g.eps}
\label{fig:irs_fwd_receiver_OTM_VM_IM_app}
\end{subfigure}
\hfill
\begin{subfigure}[t]{0.3\textwidth}
\caption{Payer ATM, with VM and IM.}
  \includegraphics[width=\linewidth]{Figures/Fig4h.eps}
\label{fig:irs_fwd_ATM_VM_IM_app}
\end{subfigure}
\hfill
\begin{subfigure}[t]{0.3\textwidth}
\caption{Payer OTM, with VM and IM.}
    \includegraphics[width=\linewidth]{Figures/Fig4i.eps}
\label{fig:irs_fwd_payer_OTM_VM_IM_app}
\end{subfigure}
\caption{5x10Y forward Swaps. Other settings as in fig. \ref{fig:irs_15Y_results}.}
\label{fig:irs_fwd_results}
\end{figure}

\begin{figure}[H]
\begin{subfigure}[t]{0.3\textwidth}
\caption{Receiver OTM, no collat.}
    \includegraphics[width=\linewidth]{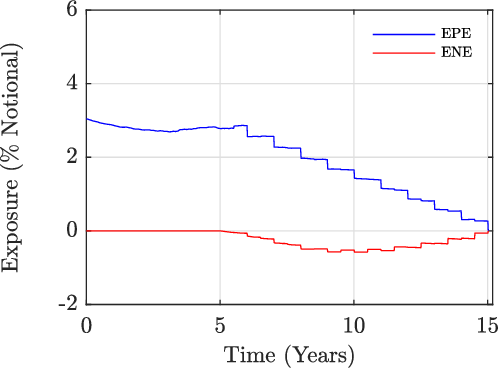}
\label{fig:swpt_receiver_OTM_noMargins}
\end{subfigure}
\hfill
\begin{subfigure}[t]{0.3\textwidth}
\caption{Payer ATM, no collat.}
  \includegraphics[width=\linewidth]{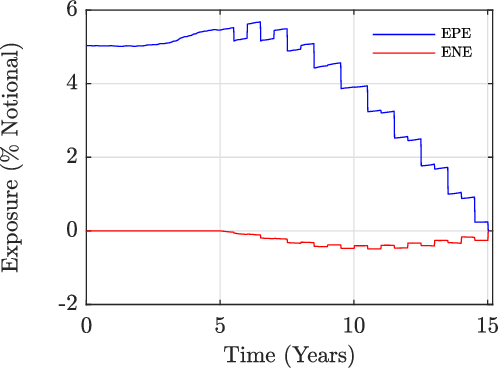}
\label{fig:swpt_ATM_noMargins}
\end{subfigure}
\hfill
\begin{subfigure}[t]{0.3\textwidth}
\caption{Payer OTM, no collat.}
    \includegraphics[width=\linewidth]{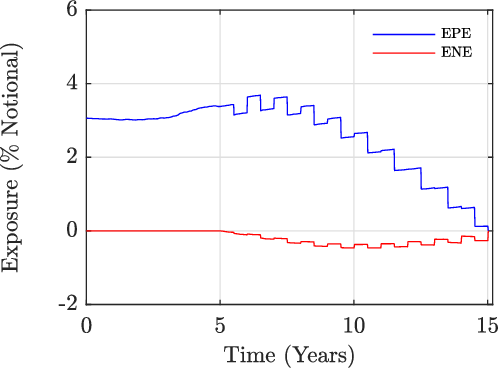}
\label{fig:swpt_payer_OTM_noMargins}
\end{subfigure}\\

\begin{subfigure}[t]{0.3\textwidth}
\caption{Receiver OTM, with VM.}
    \includegraphics[width=\linewidth]{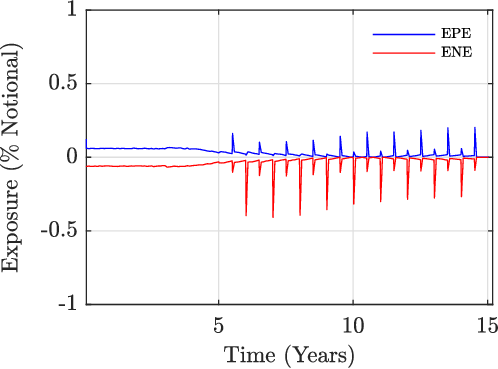}
\label{fig:swpt_receiver_OTM_VM_app}
\end{subfigure}
\hfill
\begin{subfigure}[t]{0.3\textwidth}
\caption{Payer ATM, with VM.}
  \includegraphics[width=\linewidth]{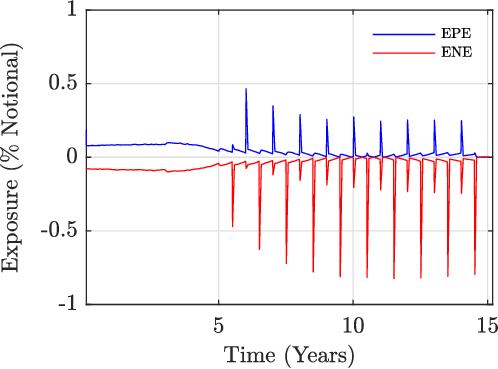}
\label{fig:swpt_ATM_VM_app}
\end{subfigure}
\hfill
\begin{subfigure}[t]{0.3\textwidth}
\caption{Payer OTM, with VM.}
    \includegraphics[width=\linewidth]{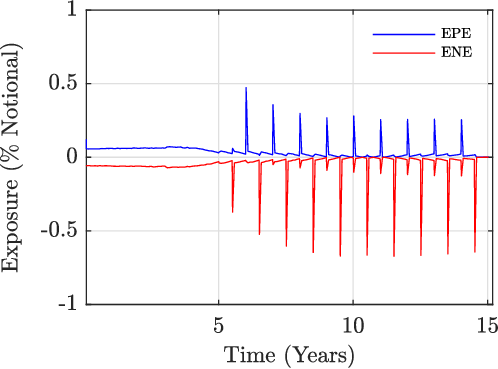}
\label{fig:swpt_payer_OTM_VM_app}
\end{subfigure}\\

\begin{subfigure}[t]{0.3\textwidth}
\caption{Receiver OTM, with VM and IM.}    \includegraphics[width=\linewidth]{Figures/Fig5g.eps}
\label{fig:swpt_receiver_OTM_VM_IM_app}
\end{subfigure}
\hfill
\begin{subfigure}[t]{0.3\textwidth}
\caption{Payer ATM, with VM and IM.}
  \includegraphics[width=\linewidth]{Figures/Fig5h.eps}
\label{fig:swpt_ATM_VM_IM_app}
\end{subfigure}
\hfill
\begin{subfigure}[t]{0.3\textwidth}
\caption{Payer OTM, with VM and IM.}
    \includegraphics[width=\linewidth]{Figures/Fig5i.eps}
\label{fig:swpt_payer_OTM_VM_IM_app}
\end{subfigure}
\caption{5x10Y physically settled European Swaptions (left: reicever OTM, mid: payer ATM, right: payer OTM). Other settings as in fig. \ref{fig:irs_15Y_results}.}
\label{fig:swaption_results}
\end{figure}

In this section we report additional results concerning the exposure calculation discussed in sec. \ref{sec:exposure results}. Figures \ref{fig:irs_15Y_results}-\ref{fig:swaption_results} show the exposure profiles for the four financial instruments (15Y Swap, 30Y Swap, 5x10Y forward Swap and 5x10Y Swaption), respectively, each considering three collateralization schemes (no collateral, with VM, with VM and IM) and three moneyness (OTM/ATM/ITM).

Looking at uncollateralized exposures (top panels in figs. \ref{fig:irs_15Y_results}-\ref{fig:swaption_results}), the jagged shape observed for spot-starting Swaps (figs.~\ref{fig:irs_15Y_results} and \ref{fig:irs_30Y_results}) is due to the different coupons frequency (receive semi-annual floating coupons, pay annual fixed coupons) which determines semi-annual jumps in the future simulated mark-to-market values at cash flow dates. 
The EPE is larger than the ENE, in absolute terms, except for the out-of-the-money (OTM) 15Y Swap, due to the forward rates structure which causes expected floating leg values greater than those of the fixed leg. This is evident for in-the-money (ITM) Swaps for which the ENE is also almost flat, given the low probability to observe negative future simulated mark-to-market values. Longer maturities are clearly riskier, due to the greater number of coupons to be exchanged. 
Similar exposure's shapes are observed for the 5x10Y forward Swaps (fig. \ref{fig:irs_fwd_results}), starting from $t = 5$ years. Here we observe the asymmetric effect of forward rates on opposite transactions: the OTM payer forward Swap displays larger EPE and smaller ENE compared to the OTM receiver one. 
The 5x10Y physically settled European Swaptios (fig.~\ref{fig:swaption_results}) are written on the same forward Swaps, before the expiry the exposure is always positive and greater than the one of the corresponding underlying forward Swap as the price is always positive. After the expiry, OTM paths are excluded as the exercise do not take place, determining smaller EPE and ENE (in absolute value) with respect to the corresponding underlying forward Swap.         
 
Looking at collateralized exposures with VM (middle panels in figs. \ref{fig:irs_15Y_results}-\ref{fig:swaption_results}), we observe a reduction of one order of magnitude, since the VM tracks the instrument's future simulated mark-to-market values with a delay equal to the MPoR's length (2 days). This collateral friction causes an imperfect collateralzation, leading to a material residual exposure characterized by spikes at cash flow dates, when the mark to market suddely changes but the VM is adjusted two days later. In general, when only floating coupons are received, one can expect to observe downward spikes due to the fact that the counterparty makes a payment for which the bank still has not returned VM. Conversely, upward spikes arise when fixed cash flows occur and are counterbalanced by the downward ones stemming from floating coupons. The magnitude of these spikes is determined by the simulated forward rates structure, e.g.~the large upward spikes displayed at the early stage of payer Swaps' life are due to negative rates (see sec.~\ref{sec:Spikes analysis} for details). We point out that the negative exposure arising before the expiry of the Swaption is due to the fact that for some paths the MPoR produces an overcollateralization.  

Collateralized exposures with both VM and IM (bottom panels in figs. \ref{fig:irs_15Y_results}-\ref{fig:swaption_results}), have been discussed in sec. \ref{sec:exposure results}.

\begin{figure}[H]
	\begin{subfigure}[t]{0.3\textwidth}
		\caption{15Y Swap OTM.}
		\includegraphics[width=\linewidth]{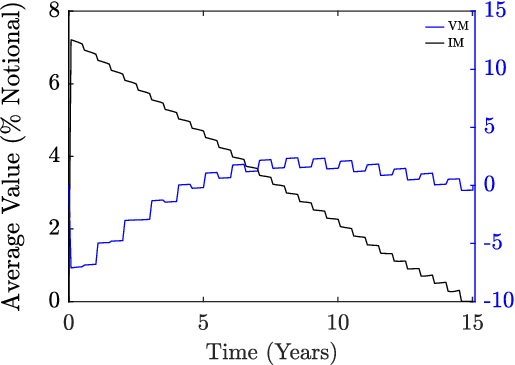}
		\label{fig:irs_15Y_OTM_meanVMandIM}
	\end{subfigure}
	\hfill
	\begin{subfigure}[t]{0.3\textwidth}
		\caption{15Y Swap ATM.}
		\includegraphics[width=\linewidth]{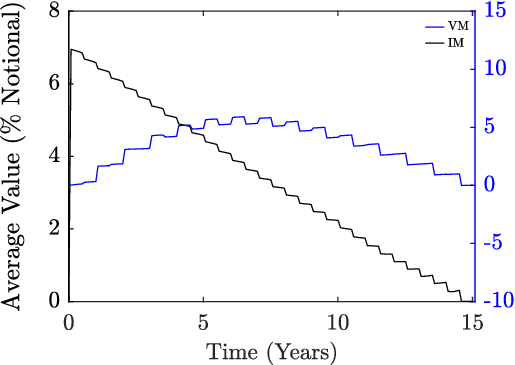}
		\label{fig:irs_15Y_ATM_meanVMandIM}
	\end{subfigure}
	\hfill
	\begin{subfigure}[t]{0.3\textwidth}
		\caption{15Y Swap ITM.}
		\includegraphics[width=\linewidth]{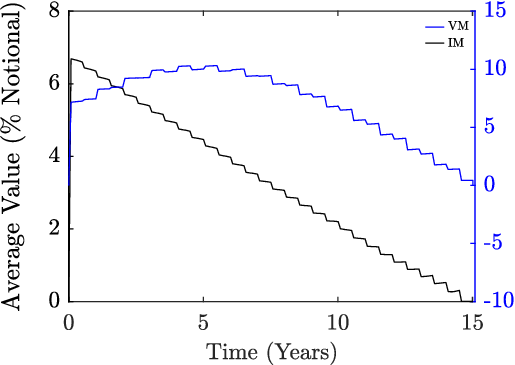}
		\label{fig:irs_15Y_ITM_meanVMandIM}
	\end{subfigure}\\
	
	\begin{subfigure}[t]{0.3\textwidth}
		\caption{30Y Swap OTM.}
		\includegraphics[width=\linewidth]{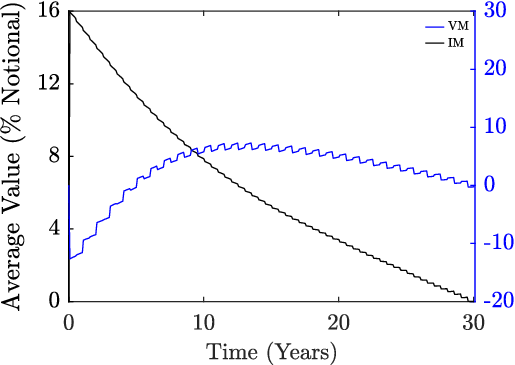}
		\label{fig:irs_30Y_OTM_meanVMandIM}
	\end{subfigure}
	\hfill
	\begin{subfigure}[t]{0.3\textwidth}
		\caption{30Y Swap ATM.}
		\includegraphics[width=\linewidth]{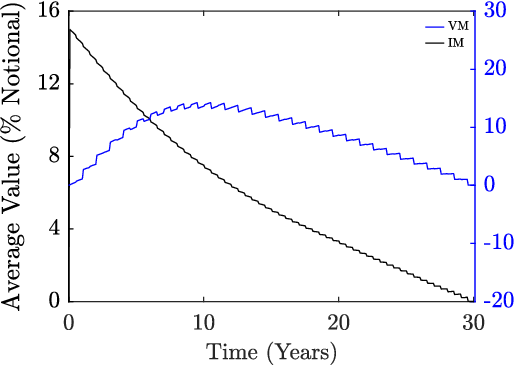}
		\label{fig:irs_30Y_ATM_meanVMandIM}
	\end{subfigure}
	\hfill
	\begin{subfigure}[t]{0.3\textwidth}
		\caption{30Y Swap ITM}
		\includegraphics[width=\linewidth]{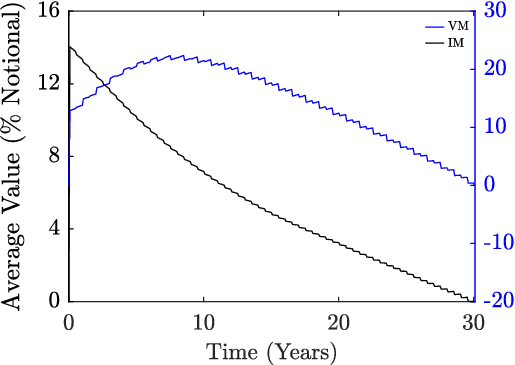}
		\label{fig:irs_30Y_ITM_meanVMandIM}
	\end{subfigure}\\
	
	\begin{subfigure}[t]{0.3\textwidth}
		\caption{5x10Y R Fwd Swap OTM.}
		\includegraphics[width=\linewidth]{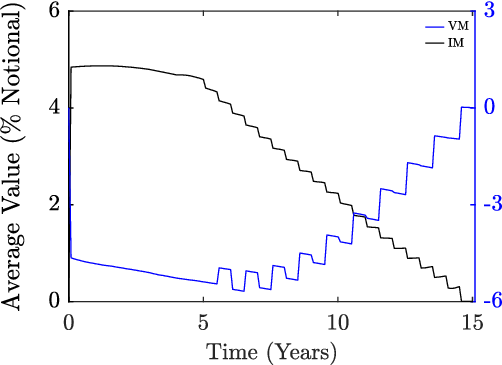}
		\label{fig:irs_fwd_receiver_OTM_meanVMandIM}
	\end{subfigure}
	\hfill
	\begin{subfigure}[t]{0.3\textwidth}
		\caption{5x10Y P Fwd Swap ATM.}
		\includegraphics[width=\linewidth]{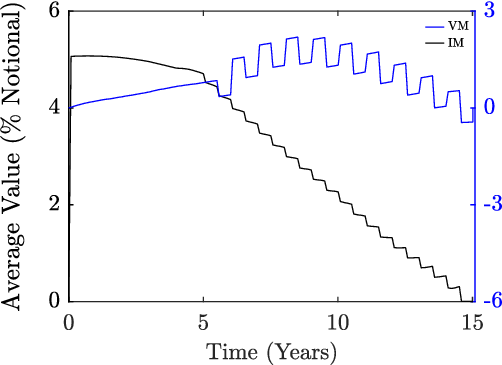}
		\label{fig:irs_fwd_ATM_meanVMandIM}
	\end{subfigure}
	\hfill
	\begin{subfigure}[t]{0.3\textwidth}
		\caption{5x10Y P Fwd Swap OTM.}
		\includegraphics[width=\linewidth]{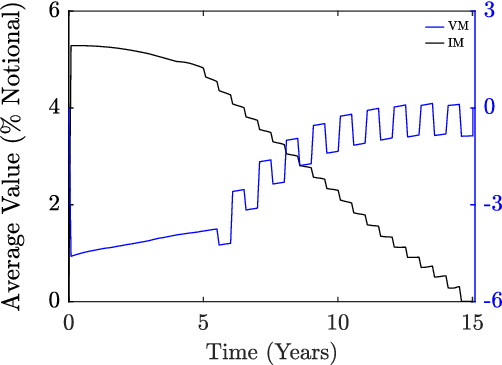}
		\label{fig:irs_fwd_payer_OTM_meanVMandIM}
	\end{subfigure}\\
	
	\begin{subfigure}[t]{0.3\textwidth}
		\caption{5x10Y R Swaption OTM.}
		\includegraphics[width=\linewidth]{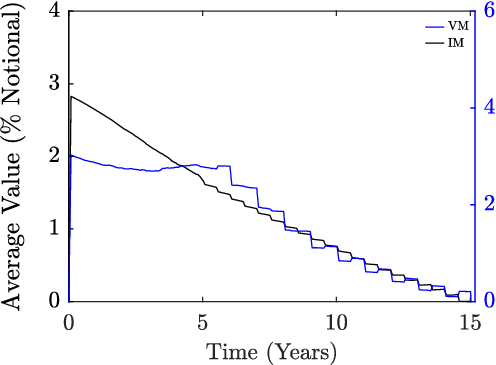}
		\label{fig:swpt_receiver_OTM_meanVMandIM}
	\end{subfigure}
	\hfill
	\begin{subfigure}[t]{0.3\textwidth}
		\caption{5x10Y P Swaption ATM.}
		\includegraphics[width=\linewidth]{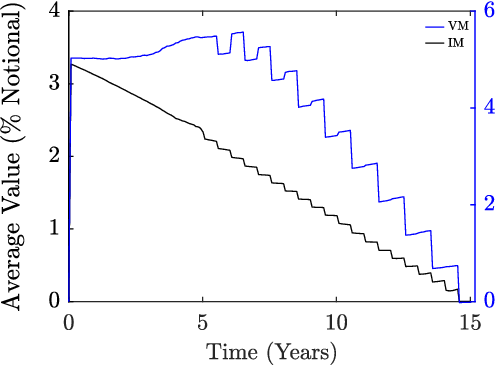}
		\label{fig:swpt_ATM_meanVMandIM}
	\end{subfigure}
	\hfill
	\begin{subfigure}[t]{0.3\textwidth}
		\caption{5x10Y P Swaption OTM.}
		\includegraphics[width=\linewidth]{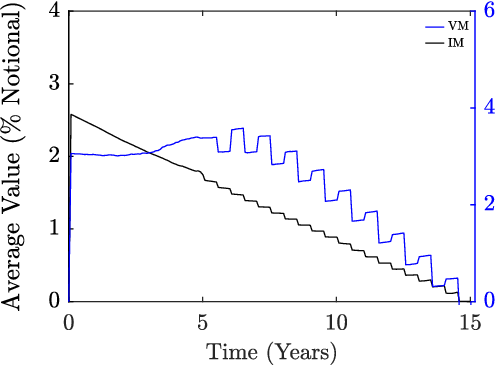}
		\label{fig:swpt_payer_OTM_meanVMandIM}
	\end{subfigure}
	\caption{Expected VM profiles (blue lines, right-hand scale) and IM profiles (black lines, left-hand scale) for instruments in tab.~\ref{tab:instruments}. Other settings as in fig. \ref{fig:irs_15Y_results}.}
	\label{fig:mean_EPE_VM_IM_results}
\end{figure}

In order to appreciate the distinct effects of VM and IM on the total EPE/ENE discussed before, we show in fig.~\ref{fig:mean_EPE_VM_IM_results} the expected VM and IM profiles. 
We observe that the decreasing profile with downward steps at cash flow dates of the IM, implies an incomplete suppression of the exposure close to maturity. Furthermore, it is possible to notice how, in absolute terms, payer (receiver) instruments show larger spikes in ENE (EPE) as, on average, fixed rates are greater than floating rates.     


\subsection{G2++ Model Calibration}
\label{app:G2++ model calibration}

In this section we report additional details regarding the 7 different calibrations used in sec.~\ref{sec:G2++ Model Calibration} to assess the G2++ model calibration risk.

The 7 alternative calibrations are defined as follows.
\begin{enumerate}
\item \emph{Baseline calibration}: calibration of the full ATM price matrix (parameters denoted with $p$). This calibration comprises 182 market quotes.
\item Calibration of the full ATM price matrix imposing flat volatility G2++ parameters (parameters denoted with $p_1$). This calibration comprises 182 market quotes (same as baseline calibration).
\item Calibration of the ATM price matrix excluding the expiries greater than 15Y since we tested instruments with 15Y maturities (parameters denoted with $p_2$). This calibration comprises 143 market quotes.
\item Calibration of the full price cube (parameters denoted with $p_3$). This calibration comprises 30 ATM and 300 smile market quotes.
\item Calibration of the price cube excluding strikes outside the interval ATM $\pm$ 0.005 (parameters denoted with $p_4$). This calibration comprises 30 ATM and 120 smile market quotes.
\item Calibration of the price cube excluding the expiries greater than 15Y, similarly to $p_2$ (parameters denoted with $p_5$). This calibration comprises 20 ATM and 200 smile market quotes.
\item Calibration of the price cube excluding both the expiries greater than 15Y and strikes outside the interval ATM $\pm 0.005$, similarly to $p_2$ plus $p_4$ (parameters denoted with $p_6$). This calibration comprises 20 ATM and 80 smile market quotes.
\end{enumerate}

We summarize the above calibrations in tab.~\ref{tab:G2++ model calibrations}.
\begin{table}[!htbp]
\centering 
\begin{tabular}{c l l l l l l}
	\toprule
	
	\multicolumn{1}{c}{G2++} & \multicolumn{1}{c}{ATM/} & \multicolumn{1}{c}{$\xi$} & \multicolumn{1}{c}{$\mathfrak{T}$} & \multicolumn{1}{c}{Delta} & \multicolumn{1}{c}{\multirow{2}*{Volatility}} & \multicolumn{1}{c}{Total}\\ 
	\multicolumn{1}{c}{parameters} & \multicolumn{1}{c}{Cube} & \multicolumn{1}{c}{(years)} & \multicolumn{1}{c}{(years)} & \multicolumn{1}{c}{$K^{ATM} \pm$} & & \multicolumn{1}{c}{market quotes} \\
	\midrule
\multirow{1}*{$p$} & \multirow{1}*{ATM} & Full & Full &  \multirow{1}*{- -}  & 		    \multirow{1}*{Time dep.} & 182 ATM\\

\multirow{1}*{$p_1$} & \multirow{1}*{ATM} & Full & Full &  \multirow{1}*{- -}  & 		    \multirow{1}*{Flat} & 182 ATM\\

\multirow{1}*{$p_2$} & \multirow{1}*{ATM} & Partial & Full &  \multirow{1}*{- -}  & 		    \multirow{1}*{Time dep.} & 143 ATM\\

\multirow{1}*{$p_3$} & \multirow{1}*{Cube} & \multirow{1}*{Full} & Full &  Full   &   \multirow{1}*{Time dep.} & 30 ATM, 300 smile\\

\multirow{1}*{$p_4$} & \multirow{1}*{Cube} & \multirow{1}*{Full}  & Full   &  \multirow{1}*{Partial}   &   \multirow{1}*{Time dep.} & 30 ATM, 120 smile\\

\multirow{1}*{$p_5$} & \multirow{1}*{Cube} & \multirow{1}*{Partial}  & Full &  Full   &   \multirow{1}*{Time dep.} & 20 ATM, 200 smile\\

\multirow{1}*{$p_6$} & \multirow{1}*{Cube} & \multirow{1}*{Partial} & Full & \multirow{1}*{Partial}   &   \multirow{1}*{Time dep.} & 20 ATM, 80 smile \\
 	
	\bottomrule
\end{tabular}
\caption{Details of the different calibrations considered, with baseline calibration denoted with $p$. With respect to ATM calibrations, full refers to the set of expiries $\xi$ = \{2, 3, 4, 5, 6, 7, 8, 9, 10, 12, 15, 20, 25, 30\} and tenors $\mathfrak{T}$ = \{2, 3, 4, 5, 6, 7, 8, 9, 10, 15, 20, 25, 30\}, while partial refers to the set of expiries $\xi$ = \{2, 3, 4, 5, 6, 7, 8, 9, 10, 12, 15\}. With respect to cube calibration, full refers to the set of expiries $\xi$ = \{2, 5, 10, 15, 20, 30\}, tenors $\mathfrak{T}$ = \{2, 5, 10, 20, 30\} and shifts to ATM strike rates Delta $K^{ATM}$ = \{$0$, $\pm0.0025$, $\pm0.005$, $\pm0.01$, $\pm0.015$, $\pm0.02$\}, while partial refers to the set of expiries $\xi$ = \{2, 5, 10, 15\} and shifts to ATM strike rates Delta $K^{ATM}$ = \{$0$, $\pm0.0025$, $\pm0.005$\}.}
\label{tab:G2++ model calibrations}
\end{table}
The resulting calibrated parameters, computed according to eqs.~\ref{A1} and \ref{A2}, are reported in tab.~\ref{tab:calibrations_parameters}.
\begin{table}[!htbp]
\centering
\begin{tabular}{c r r r r r r r}
	\toprule
	\multicolumn{1}{c}{G2++} & \multicolumn{1}{c}{\multirow{2}*{$p$}} & \multicolumn{1}{c}{\multirow{2}*{$p_1$}} & \multicolumn{1}{c}{\multirow{2}*{$p_2$}} & \multicolumn{1}{c}{\multirow{2}*{$p_3$}} & \multicolumn{1}{c}{\multirow{2}*{$p_4$}} & \multicolumn{1}{c}{\multirow{2}*{$p_5$}} & \multicolumn{1}{c}{\multirow{2}*{$p_6$}}  \\
	parameter & & & & & & & \\
	\midrule
		  $a$          & 1.1664 & 1.1664 & 1.1799 & 1.0785 & 1.0679 & 1.1221 & 1.0626\\
          $\sigma$ & 0.0501 & 0.0501 & 0.0511 & 0.0443 & 0.0428 & 0.0457 & 0.0429\\
          $b$          & 0.0304 & 0.0304 & 0.0286 & 0.0300 & 0.0289 & 0.0271 & 0.0284\\
          $\eta$     & 0.0084 & 0.0084 & 0.0083 & 0.0082 & 0.0082 & 0.0080 & 0.0082\\
          $\rho$    & -1.0000 & -1.0000 & -1.0000 & -1.0000 & -1.0000 & -1.0000 & -1.0000\\
          \midrule
          $\Gamma_1$ & 0.9530 & 1 & 0.9509 & 0.9712 & 0.9548 & 0.9664 & 0.9527\\
          $\Gamma_2$ & 0.9781 & 1 & 0.9700 & 0.9712 & 0.9548 & 0.9664 & 0.9527\\
          $\Gamma_3$ & 1.0895 & 1 & 1.0851 & 0.9712 & 0.9548 & 0.9664 & 0.9527\\
          $\Gamma_4$ & 1.0709 & 1 & 1.0646 & 1.0323 & 1.0821 & 1.0299 & 1.0780\\
          $\Gamma_5$ & 1.0032 & 1 & 0.9934 & 1.0323 & 1.0821 & 1.0299 & 1.0780\\
          $\Gamma_6$ & 1.0776 & 1 & 1.0672 & 1.0323 & 1.0821 & 1.0299 & 1.0780\\
          $\Gamma_7$ & 1.0488 & 1 & 1.0377 & 1.0323 & 1.0821 & 1.0299 & 1.0780\\
          $\Gamma_8$ & 1.0186 & 1 & 1.0056 & 1.0323 & 1.0821 & 1.0299 & 1.0780\\
          $\Gamma_9$ & 1.1000 & 1 & 1.0864 & 1.0543 & 1.0613 & 1.0435 & 1.0552\\
          $\Gamma_{10}$ & 0.9608 & 1 & 0.9523 & 1.0543 & 1.0613 & 1.0435 & 1.0552\\
          $\Gamma_{11}$ & 1.0114 & 1 & 0.9954 & 0.9896 & 1.0064 & 0.9692 & 0.9984\\
          $\Gamma_{12}$ & 0.9553 & 1 & & 0.9554 & 0.9571 &  & \\
          $\Gamma_{13}$ & 0.9629 & 1 & & 0.9554 & 0.9571 &  & \\
          $\Gamma_{14}$ & 0.9340 & 1 & & 0.9521 & 0.9445 &  & \\
	\bottomrule
\end{tabular}
\caption{G2++ model parameters for the different calibrations considered (see tab.~\ref{tab:G2++ model calibrations}).}
\label{tab:calibrations_parameters}
\end{table}

In the following tab.~\ref{tab:mid_calib_errors} we show the accuracies of the 7 different calibrations, both with respect to the calibration set and to the complete market data set. We observe that, in general, the calibration accuracy worsens when less liquid OTM market quotes (i.e.~calibrations $p_3$-$p_6$) are included in the calibration set. As expected, the pricing error is much larger than the calibration error, since much more market points are included and the pricing error is higher on the extreme strikes far from the ATM.
\begin{table}[!htbp]
	\centering
	\begin{tabular}{l l c c c c c c c}
		\toprule
		&   & \multicolumn{1}{c}{$p$} & \multicolumn{1}{c}{$p_1$} & \multicolumn{1}{c}{$p_2$} & \multicolumn{1}{c}{$p_3$} & \multicolumn{1}{c}{$p_4$} & \multicolumn{1}{c}{$p_5$} & \multicolumn{1}{c}{$p_6$} \\
		\midrule
		Calibration & RMSRE 	& 2.91 & 3.62 & 3.22 & 16.92 & 6.10 & 18.17 & 7.17 \\
		errors (\%) & Imp. Vol. & 0.49 & 0.66 & 0.55 & 1.89  & 0.81 & 1.90  & 0.93 \\  
		\midrule
		Pricing 	& RMSRE 	& 20.32 & 20.50 & 19.09 & 21.51 & 20.20 & 20.73 & 19.24 \\
		errors (\%) & Imp. Vol. & 1.49  & 1.63  & 1.53  & 1.55  & 1.59  & 1.67  & 1.63 \\  
		\bottomrule
	\end{tabular}
	\caption{Accuracies of the different calibrations tested. Calibration errors refer to the subset of market quotes used for calibration (see tab.~\ref{tab:G2++ model calibrations}). Pricing errors refer to the full market data set including both the Swaption ATM matrix (tab.~\ref{tab:mkt_swpt_prices}) and the Swaption cube (tab.~\ref{tab:mkt_swpt_prices_cube}). The RMSRE is the Root Mean Square Relative Error, i.e. the square root of the objective function used for  G2++ calibration (see app.~\ref{app:G2++ calibration procedure}). Imp.~Vol.~ refers to  the average absolute difference between the shifted-Black volatilities implied in G2++ prices and those implied in market prices.}
	\label{tab:mid_calib_errors}
\end{table}

\begin{figure}[!htbp]			
	\begin{subfigure}{0.49\linewidth}
		\centering
		\caption{$p$ ATM.}
		\includegraphics[width=0.9\linewidth]{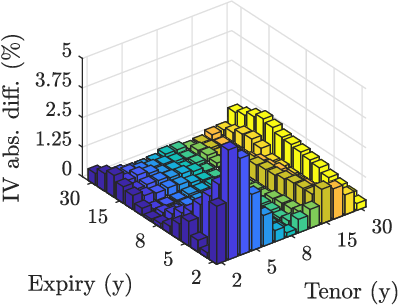}	
		\label{fig:p_ATM_calibration_errors}
	\end{subfigure}
	\hfill
	\begin{subfigure}{0.49\linewidth}
		\centering
		\caption{$p$ cube.}
		\includegraphics[width=0.9\linewidth]{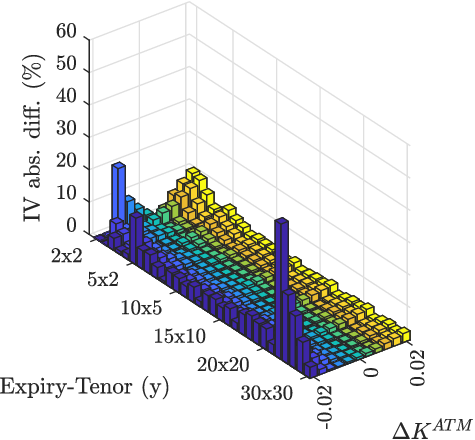}	
		\label{fig:p_cube_calibration_errors}
	\end{subfigure} \\
	\begin{subfigure}{0.49\linewidth}
		\centering
		\caption{$p$ and $p_3$ 5x10Y smile.}
		\includegraphics[width=0.9\linewidth]{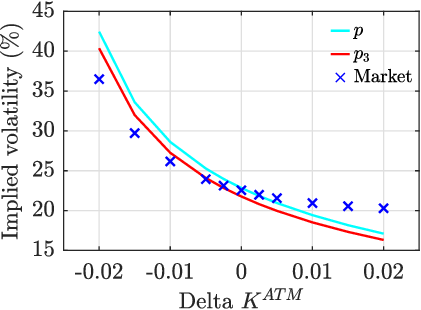}	
		\label{fig:p_5x10Y_calibration_errors}
	\end{subfigure}
	\hfill
	\begin{subfigure}{0.49\linewidth}
		\centering
		\caption{$p$ and $p_3$ 30x10Y smile.}
		\includegraphics[width=0.9\linewidth]{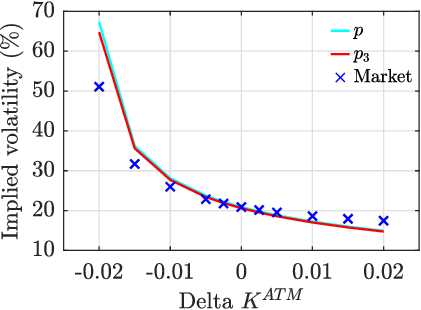}	
		\label{fig:p_30x10Y_calibration_errors}
	\end{subfigure} \\
	\caption{Baseline calibration ($p$) accuracy. Top panels: absolute differences between shifted-Black volatilities implied in G2++ prices and those implied in the market Swaption ATM matrix (panel a, see tab. \ref{tab:mkt_swpt_prices}) and in the market Swaption cube (panel b, see tab. \ref{tab:mkt_swpt_prices_cube}). Scales on the axis are set consistently with figs. \ref{fig:ATM_calibration_errors} and \ref{fig:cube_calibration_errors}. Bottom panels: details for the 5x10Y (panel c) and 30x10Y (panel d).}
	\label{fig:p_ATM_cube_calibration_errors}
\end{figure}

In fig.~\ref{fig:p_ATM_cube_calibration_errors} we focus on the baseline calibration ($p$) accuracy showing the  differences between the shifted-Black volatilities implied in G2++ prices and those implied in market prices. 
Regarding the ATM accuracy (fig. \ref{fig:p_ATM_calibration_errors}), the larger errors (5\%) can be observed for short expiry-tenor combinations, where it is more difficult for the G2++ model to calibrate smaller prices (see tab.~\ref{tab:mkt_swpt_prices}) with small Vega sensitivities. Excluding this section of the volatility surface, the remaining points show calibration errors below 1.5\%, which can be considered a good accuracy, achieved thanks to the time-dependent volatility in our G2++ model. 
Regarding the OTM accuracy (fig. \ref{fig:p_cube_calibration_errors}), the larger errors are observed for extreme strikes, as expected since these quotes are illiquid.
The calibration accuracy on the smile can be appreciated in figs. \ref{fig:p_5x10Y_calibration_errors} and \ref{fig:p_30x10Y_calibration_errors}, where we also compare calibrations $p$ and $p_3$. Similar results are obtained for the other calibrations. As expected, the G2++ model is able to capture the market skew in the ATM region, while the precision worsens for OTM quotes. The two calibrations $p$ and $p_3$ yield similar results since their average absolute volatility difference is equal to respectively $2.05\%$ and $1.89\%$ for the 5x10Y smile section, and to respectively $2.83\%$ and $2.57\%$ for the 30x10Y smile section.

\begin{figure}[!htbp]			
	\begin{subfigure}{0.49\linewidth}
		\centering
		\caption{$p_1$.}
		\includegraphics[width=0.9\linewidth]{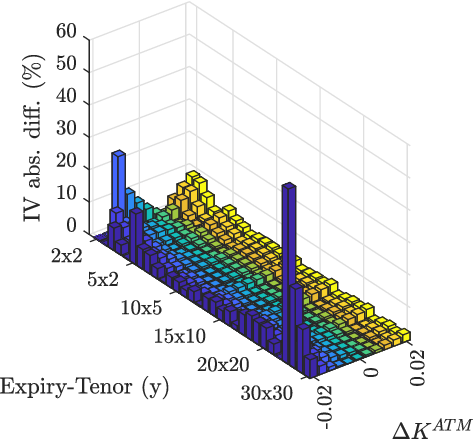}	
		\label{fig:p1_cube_calibration_errors}
	\end{subfigure}
	\hfill
	\begin{subfigure}{0.49\linewidth}
		\centering
		\caption{$p_2$.}
		\includegraphics[width=0.9\linewidth]{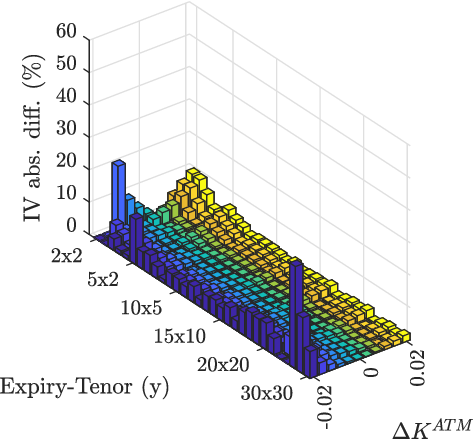}	
		\label{fig:p2_cube_calibration_errors}
	\end{subfigure} \\
	\begin{subfigure}{0.49\linewidth}
		\centering
		\caption{$p_3$.}
		\includegraphics[width=0.9\linewidth]{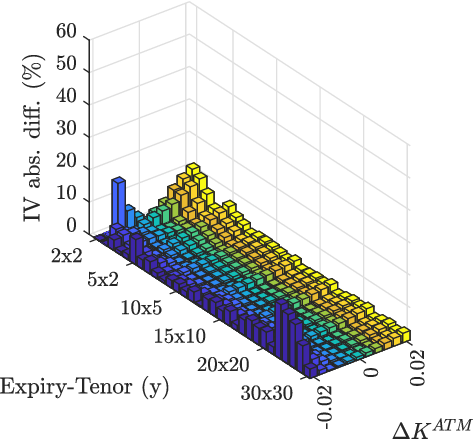}	
		\label{fig:p3_cube_calibration_errors}
	\end{subfigure}
	\hfill
	\begin{subfigure}{0.49\linewidth}
		\centering
		\caption{$p_4$.}
		\includegraphics[width=0.9\linewidth]{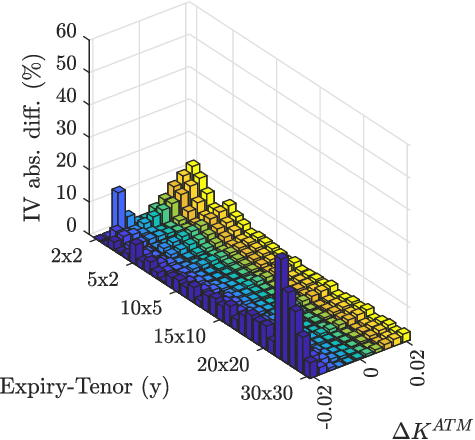}	
		\label{fig:p4_cube_calibration_errors}
	\end{subfigure} \\
	\begin{subfigure}{0.49\linewidth}
		\centering
		\caption{$p_5$.}
		\includegraphics[width=0.9\linewidth]{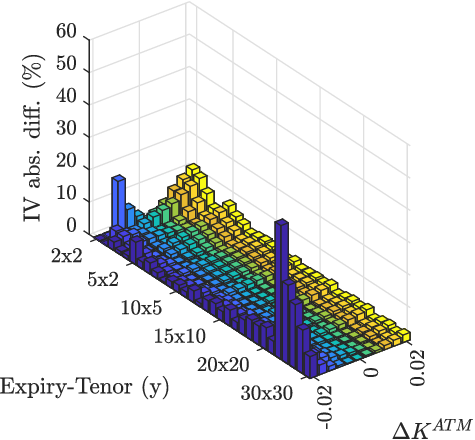}	
		\label{fig:p5_cube_calibration_errors}
	\end{subfigure}
	\hfill
	\begin{subfigure}{0.49\linewidth}
		\centering
		\caption{$p_6$.}
		\includegraphics[width=0.9\linewidth]{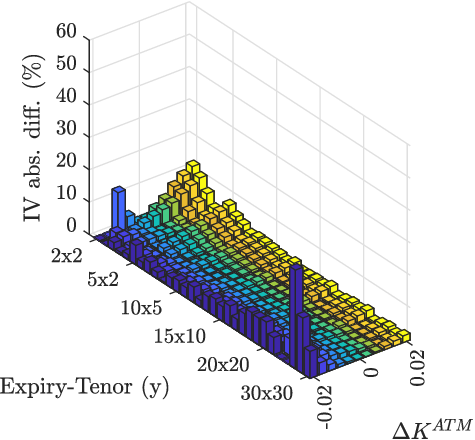}	
		\label{fig:p6_cube_calibration_errors}
	\end{subfigure} \\
	\caption{Alternative calibrations ($p_1,\cdots,p_6$) accuracy for the Swaption cube.}
	\label{fig:cube_calibration_errors}
\end{figure}

Finally, we report the accuracies for all the alternative calibrations $p_1,\cdots,p_6$ in the following fig. 	\ref{fig:ATM_calibration_errors} for the full Swaption ATM matrix and fig. \ref{fig:cube_calibration_errors} for the full Swaption cube.

\begin{figure}[!htbp]			
	\begin{subfigure}{0.49\linewidth}
		\centering
		\caption{$p_1$.}
		\includegraphics[width=0.9\linewidth]{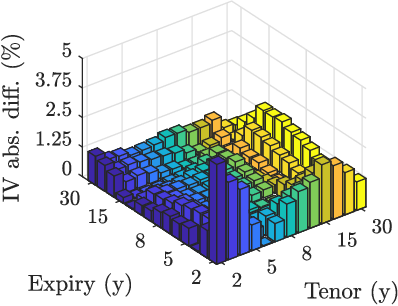}	
		\label{fig:p1_ATM_calibration_errors}
	\end{subfigure}
	\hfill
	\begin{subfigure}{0.49\linewidth}
		\centering
		\caption{$p_2$.}
		\includegraphics[width=0.9\linewidth]{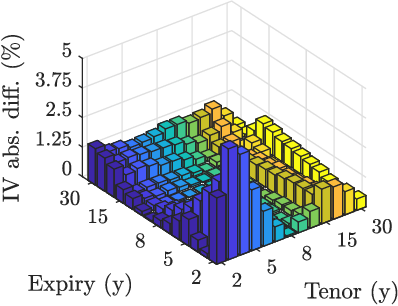}	
		\label{fig:p2_ATM_calibration_errors}
	\end{subfigure} \\
	\begin{subfigure}{0.49\linewidth}
		\centering
		\caption{$p_3$.}
		\includegraphics[width=0.9\linewidth]{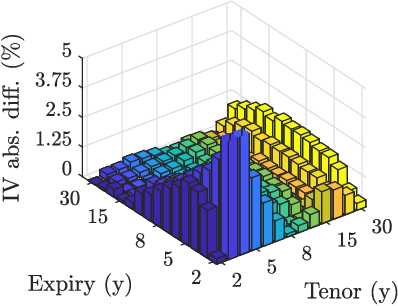}	
		\label{fig:p3_ATM_calibration_errors}
	\end{subfigure}
	\hfill
	\begin{subfigure}{0.49\linewidth}
		\centering
		\caption{$p_4$.}
		\includegraphics[width=0.9\linewidth]{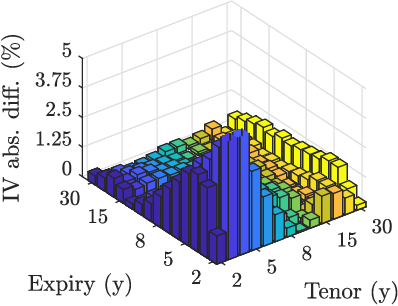}	
		\label{fig:p4_ATM_calibration_errors}
	\end{subfigure} \\
	\begin{subfigure}{0.49\linewidth}
		\centering
		\caption{$p_5$.}
		\includegraphics[width=0.9\linewidth]{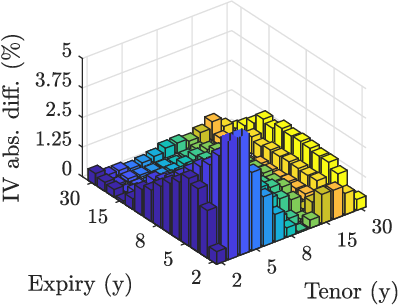}	
		\label{fig:p5_ATM_calibration_errors}
	\end{subfigure}
	\hfill
	\begin{subfigure}{0.49\linewidth}
		\centering
		\caption{$p_6$.}
		\includegraphics[width=0.9\linewidth]{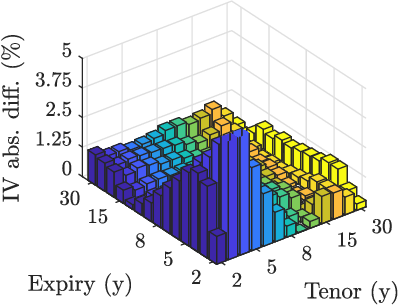}	
		\label{fig:p6_ATM_calibration_errors}
	\end{subfigure} \\
	\caption{Alternative calibrations ($p_1,\cdots,p_6$) accuracy for the Swaption ATM matrix.}
	\label{fig:ATM_calibration_errors}
\end{figure}

\newpage
\subsection{Time Simulation Grid}
\label{app:Time Simulation Grid}

In this section we report additional details regarding the construction of the time simulation grid discussed in sec. \ref{sec:Time Simulation Grid}.

\begin{figure}[!htbp]
	\begin{subfigure}{0.49\linewidth}
		\centering
		\caption{EE no collateral and Swap legs.}			
		\includegraphics[width=0.9\linewidth]{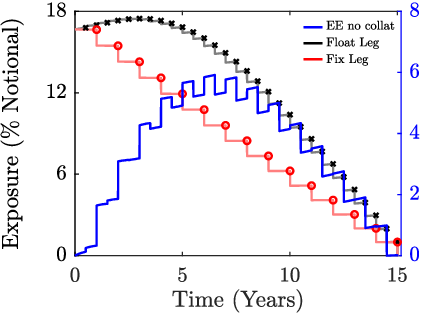}
		\label{fig:irs_15Y_legs_EE}		
	\end{subfigure}
	\hfill
	\begin{subfigure}{0.49\linewidth}
		\centering
		\caption{Jump in EE no collateral.}	
		\includegraphics[width=0.9\linewidth]{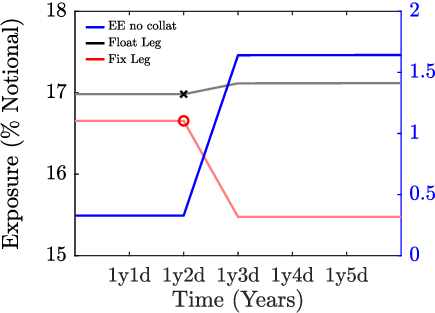}
		\label{fig:irs_15Y_spikes_noMargins}
	\end{subfigure}	\\
	\begin{subfigure}{0.49\linewidth}
		\centering	
		\caption{EE with VM and Swap legs.}		
		\includegraphics[width=0.9\linewidth]{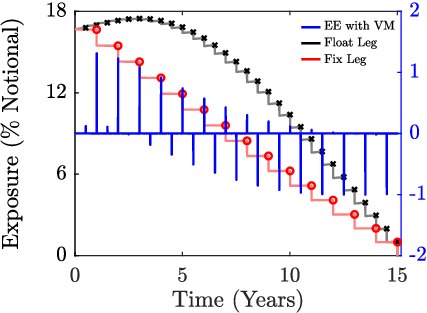}
		\label{fig:irs_15Y_spikes_EEwVM}				
	\end{subfigure}
	\hfill
	\begin{subfigure}{0.49\linewidth}
		\centering
		\caption{Spike in EE with VM.}	
		\includegraphics[width=0.9\linewidth]{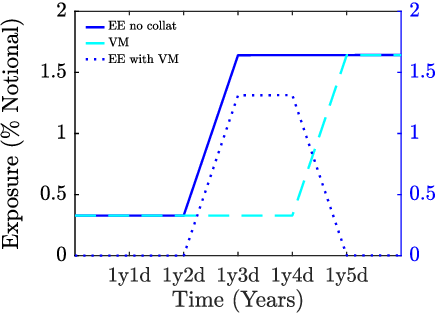}	
		\label{fig:irs_15Y_spikes_wVM}
	\end{subfigure}\\	
	\caption{Expected exposure (EE) profiles for 15Y ATM payer Swap EUR 100 Mio nominal amount, on a daily grid for two  collateralization schemes (top: no collateral, bottom: VM). 
		L.h.s. panels displays the total EE (blue line, r.h.s.) and also the separate EEs (l.h.s) for floating leg (black line) and fixed legs (red line). Black crosses: floating cash flow dates (semi-annual frequency); red circles: fixed cash flow dates (annual frequency). 
		R.h.s. panels focus on the jump in EE just after fixed and floating coupons payment taking place at $t=$ 1y2d. 
		In particular, the bottom r.h.s. panel displays the spike in EE with VM (blue dotted line) occurring just after fixed and floating coupons payment at $t=$ 1y2d captured by the VM (cyan dashed line) MPoR days later. Other model parameters as in tab.~\ref{tab:model_setup}. Quantities expressed as a percentage of the nominal amount.}
	\label{fig:irs_15Y_spikes}
\end{figure}

We show in fig.~\ref{fig:irs_15Y_spikes} the results of a further investigation on the nature of the exposure's spikes by focusing on the 15Y ATM payer Swap's Expected Exposure (EE)\footnote{The Expected Exposure (EE) is defined as: $\mathcal{H}(t,t_i) = P(t;t_i) \frac{1}{N_{MC}} \sum_{m=1}^{N_{MC}} H_{m}(t_i)$.} without collateral and with VM.
Left-hand side panels show that both the jagged shape of uncollalteralized EE (fig.~\ref{fig:irs_15Y_legs_EE}) and the spikes in EE with VM (fig.~\ref{fig:irs_15Y_spikes_noMargins}) are determined by sudden changes in the average values of Swap legs right after coupon payments. 
Right-hand side panels show a focus around $t=1$y, when both fixed and floating coupons take place: after these cash flow, the average value of floating leg increases and fixed leg one decreases, resulting in a positive jump in uncollateralized EE at $t =$ 1y3d (fig.~\ref{fig:irs_15Y_spikes_EEwVM}). This jump is captured by VM at $t =$ 1y5d, this delay of 2 days due to MPoR causes an upward spike in EE with VM (fig.~\ref{fig:irs_15Y_spikes_wVM}). The direction and magnitude depend on the simulated forward rates structure and on whether fixed or floating coupon payments take place. 

Focusing on semi-annual floating coupons paid by the counterparty two different behaviours can be observed during the lifetime of the Swap. The first three floating coupons determine positive jumps in uncollateralized EE and upward spikes in EE with VM. This is due to the fact that the simulated forward rates are on average negative until 2.5 years implying that, until that time, these coupons are actually paid by B. Once forward rates revert to positive values, negative jumps in uncollateralized EE and downward spikes in EE with VM arise in correspondence of the remaining floating coupons (figs.~\ref{fig:irs_15Y_legs_EE} and \ref{fig:irs_15Y_spikes_noMargins}). 

When also annual fixed coupons paid by the bank take place, positive jumps in uncollateralized EE and upward spikes with decreasing magnitude in EE with VM can be observed. Negative rates cause wide spikes in correspondence of the first two fixed and floating coupons due to the simultaneous increase in the average value of floating leg and decrease in the fixed leg one, in bank's perspective, right after coupons payment (figs.~\ref{fig:irs_15Y_spikes_EEwVM} and \ref{fig:irs_15Y_spikes_wVM}).

\par 
Below we report the process followed to build a \emph{parsimonious time simulation grid} capable of capturing the spikes in collateralized exposure in a reasonable computational time.
\begin{figure}[!htbp]			
	\centering
	\includegraphics[width=1\linewidth]{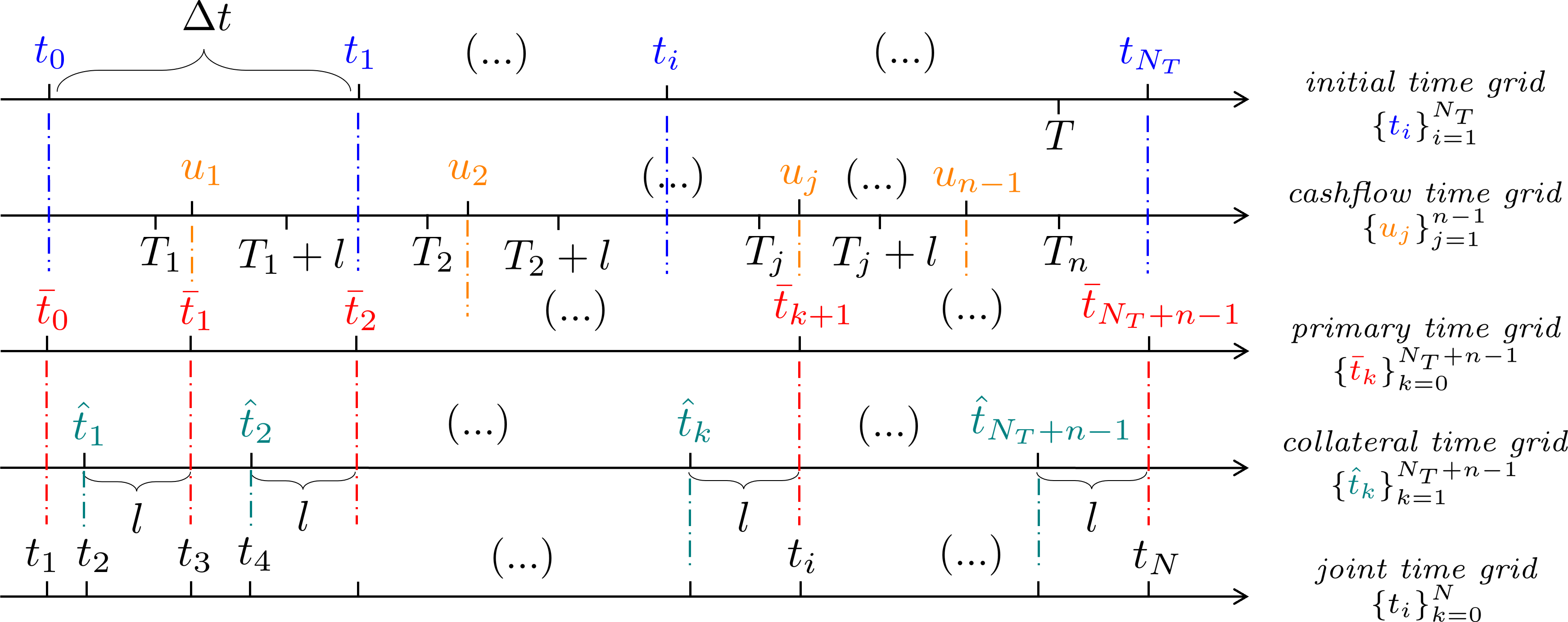}	
	\caption{Picture of the time grid construction described in app.~\ref{app:Time Simulation Grid}.}
	\label{fig:grids}
\end{figure}
\begin{enumerate}
	
	\item[a.] We select an \emph{initial time grid} $\left\{t_i\right\}_{i=0}^{N_T}$ evenly spaced with constant $\Delta t = t_i - t_{i-1}$, where $t_0$ is the valuation date, consistent with the market data set used, and $t_{N_T}>T$, where $T$ is the last cash flow date of the instrument (or portfolio) considered.
	
	\item[b.] As discussed in sec. \ref{sec:Spikes analysis}, in order to capture the spikes arising in collateralized exposure, we enrich the time grid above adding a second set of points $\left\{u_j\right\}_{j=1}^{n-1}$, called \emph{cash flow time grid}, such that $u_j\in (T_j,T_j + l]$, where $\left\{T_j\right\}_{j=1}^n$ are the instruments' (or portfolio's) cash flow dates\footnote{For the instruments considered in this paper, the fixed cash flow dates are a subset of the floating ones, otherwise both fixed and floating cash flow dates should be added to the time grid.}, $T_n = T$ and $l$ is the length of the MPoR (in days). We do not include the last cash flow date $u_n = T$ since we impose null exposure for $t\geq T$.
	
	\item[c.] Joining the previous time grids we obtain the \emph{primary time grid} $\left\{\bar{t}_k\right\}_{k=0}^{N_T+n-1} = \left\{t_i\right\}_{i=0}^{N_T} \cup \left\{u_j\right\}_{j=1}^{n-1}$ contains $N_T + n$ points.   
	
	\item[d.] Then, in order to compute VM and IM for collateralized exposures, we add to the previous grid a third set of points $\left\{\hat{t}_k\right\}_{k=1}^{N_T+n-1}$, called \emph{collateral time grid}, where $\hat{t}_k = \bar{t}_k - l$ is a look-back point at which we compute the collateral available at $\bar{t}_k$ taking into account the MPoR (see app.~\ref{app:collateral management} for details).
	
	\item[e.] The final \emph{joint time grid}\footnote{We abuse the notation naming as $t_i$ the points of both the initial and the joint grids, since the initial grid is only the starting point of our construction and is never used.} $\left\{t_i\right\}_{i=0}^{N_S} = \left\{\bar{t}_k\right\}_{k=0}^{N_T+n-1} \cup \left\{\hat{t}_k\right\}_{k=1}^{N_T+n-1}$ is our choice for the Monte Carlo time simulation. In general, this time grid includes $N_S=2(N_T+n-1)+1$ points. Clearly the number of points may be lower when one or more points from the different grids coincide. 
\end{enumerate} 
\par
\begin{figure}[!htbp]			
	\begin{subfigure}{0.49\linewidth}
		\centering
		\caption{Std grid, EPE/ENE no collat.}
		\includegraphics[width=0.9\linewidth]{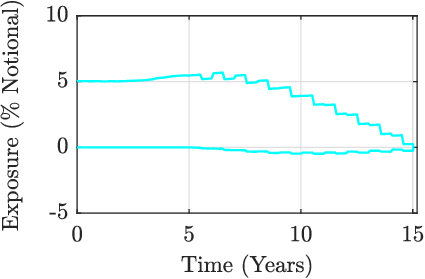}	
		\label{fig:swpt5x10_ATM_standard_grid_noColl}
	\end{subfigure}
\hfill
	\begin{subfigure}{0.49\linewidth}
		\centering
		\caption{Joint grid, EPE/ENE no collat.}
		\includegraphics[width=0.9\linewidth]{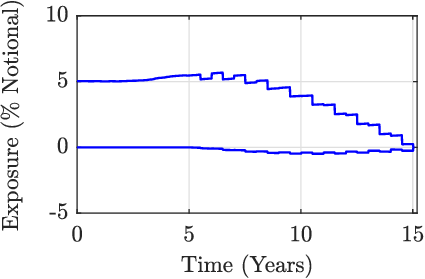}	
		\label{fig:swpt5x10_ATM_joint_grid_noColl}
	\end{subfigure} \\
	\begin{subfigure}{0.49\linewidth}
		\centering
		\caption{Std grid, EPE/ENE VM.}
		\includegraphics[width=0.9\linewidth]{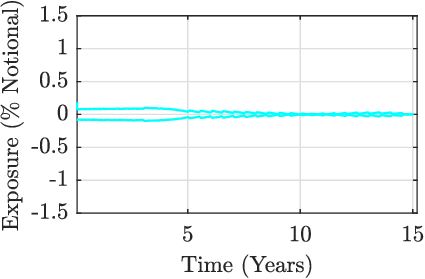}	
		\label{fig:swpt5x10_ATM_standard_grid_VM}
	\end{subfigure}
\hfill
	\begin{subfigure}{0.49\linewidth}
		\centering
		\caption{Joint grid, EPE/ENE VM.}
		\includegraphics[width=0.9\linewidth]{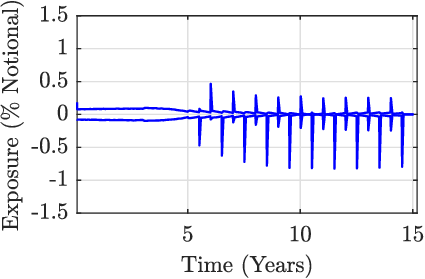}	
		\label{fig:swpt5x10_ATM_joint_grid_VM}
	\end{subfigure} \\
	\begin{subfigure}{0.49\linewidth}
		\centering
		\caption{Std grid, EPE/ENE VM and IM.}
		\includegraphics[width=0.9\linewidth]{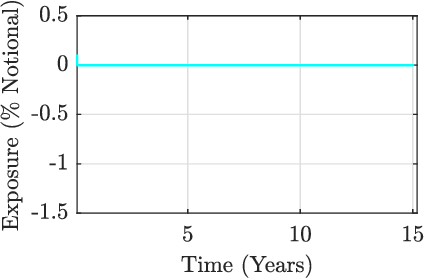}	
		\label{fig:swpt5x10_ATM_standard_grid_IM}
	\end{subfigure}
\hfill
	\begin{subfigure}{0.49\linewidth}
		\centering
		\caption{Joint grid, EPE/ENE VM and IM.}
		\includegraphics[width=0.9\linewidth]{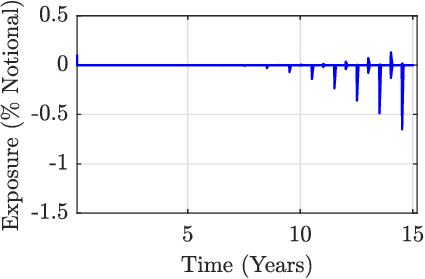}	
		\label{fig:swpt5x10_ATM_joint_grid_IM}
	\end{subfigure} \\
\caption{EPE/ENE profiles for 5x10Y ATM physically settled European payer Swaption, EUR 100 Mio nominal amount, obtained with standard grid (left-hand side) and joint grid (right-hand side) with monthly granularity. The \emph{standard time grid} is built by primary + collateral time grids (does not include the cash flow time grid). Other parameters as in fig. \ref{fig:irs_swpt_EPE_ENE_DailyGrid}. The joint grid $\Delta t =1M$ exposures are very similar to $\Delta t =1D$ exposures in fig. \ref{fig:irs_swpt_EPE_ENE_DailyGrid} (right-hand side).}
\label{fig:swpt5x10_ATM_joint_vs_standard_grid}
\end{figure}

We show in fig.~\ref{fig:swpt5x10_ATM_joint_vs_standard_grid} the exposure profiles for the 5x10Y Swaption obtained with different grids for the three collateralization schemes considered. For testing purposes we compare the joint time grid with a \emph{standard time grid}, obtained adding the primary time grid and its corresponding collateral time grid, which does not include the cash flow time grid. The results are similar to those obtained for the 15Y Swap reported in sec. \ref{sec:Time Simulation Grid}.

\newpage
\subsection{Monte Carlo Convergence}
\label{app:XVAs MC convergence}

In this section we report additional details regarding the XVA convergence with respect to the number of Monte Carlo scenarios discussed in sec. \ref{sec:XVAs MC convergence}.

We show in fig.~\ref{fig:swpt_5x10Y_ATM_convergenceXVA} the XVA convergence diagrams for the 15Y Swap, for the three collateralization schemes considered. The results are similar to those obtained for the 5x10Y Swaption reported in sec.\ref{sec:XVAs MC convergence}.
\begin{figure}[!htbp]
	\begin{subfigure}{0.49\linewidth}
		\centering
		\caption{CVA no collateral.}
		\includegraphics[width=0.9\linewidth]{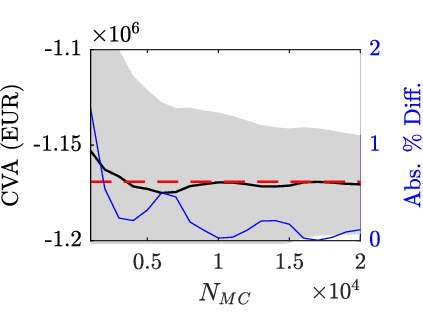}
		\label{fig:irs_15Y_ATM_CVAconvergence_noMargins}
	\end{subfigure}
\hfill
	\begin{subfigure}{0.49\linewidth}
		\centering
		\caption{DVA no collateral.}
		\includegraphics[width=0.9\linewidth]{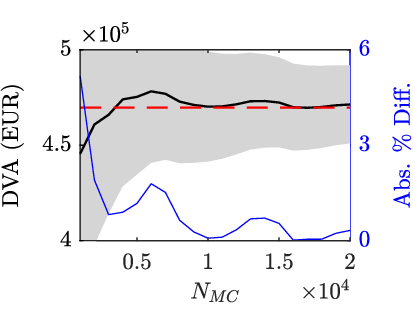}
		\label{fig:irs_15Y_ATM_DVAconvergence_noMargins}
	\end{subfigure}\\
	\begin{subfigure}{0.49\linewidth}
		\centering
		\caption{CVA with VM.}
		\includegraphics[width=0.9\linewidth]{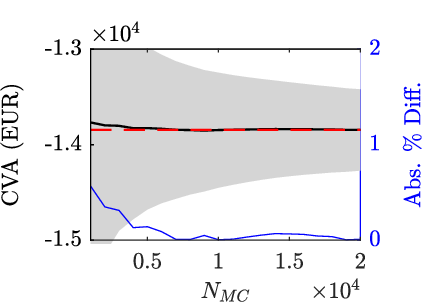}
		\label{fig:irs_15Y_ATM_CVAconvergence_VM}
	\end{subfigure}
\hfill
	\begin{subfigure}{0.49\linewidth}
		\centering	
		\caption{DVA with VM.}
		\includegraphics[width=0.9\linewidth]{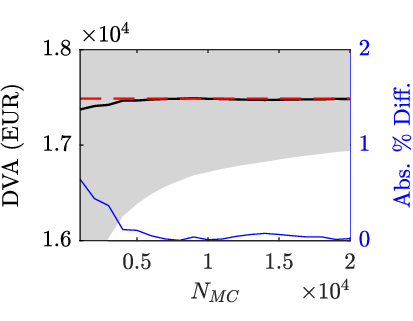}
		\label{fig:irs_15Y_ATM_DVAconvergence_VM}
	\end{subfigure}\\
	\begin{subfigure}{0.49\linewidth}
		\centering
		\caption{CVA with VM and IM.}
		\includegraphics[width=0.9\linewidth]{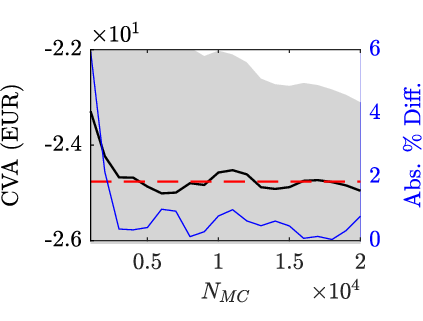}
		\label{fig:irs_15Y_ATM_CVAconvergence_VMandIM}
	\end{subfigure}
\hfill
	\begin{subfigure}{0.49\linewidth}
		\centering
		\caption{DVA with VM and IM.}
		\includegraphics[width=0.9\linewidth]{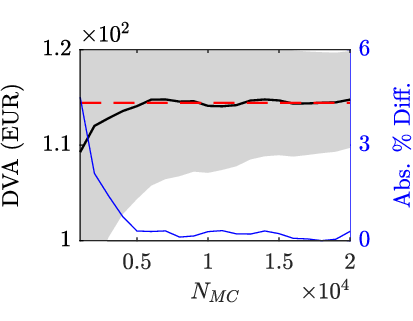}
		\label{fig:irs_15Y_ATM_DVAconvergence_VMandIM}
	\end{subfigure}\\	
\caption{CVA (l.h.s) and DVA (r.h.s) convergence diagrams versus number of MC scenarios for the 15Y ATM payer Swap, EUR 100 Mio nominal amount, and the three collateralization schemes considered (top: no collateral, mid: VM, bottom: VM and IM). Left-hand scale, black line: simulated XVA; grey area: $3\sigma$ confidence interval; dashed red line: ``exact'' value proxies (we omit their small confidence interval). Right-hand scale, blue line: convergence rate in terms of absolute percentage difference w.r.t. ``exact'' values. Model parameters other than $N_{MC}$ as in tab.~\ref{tab:model_setup}.}
\label{fig:irs_15Y_ATM_convergenceXVA}
\end{figure}

\newpage
\subsection{XVA Sensitivities to CSA Parameters}
\label{app:XVA Sensitivities to CSA Parameters}
In this section we report additional details regarding the XVA sensitivities with respect to CSA parameters discussed in sec. \ref{sec:XVA Sensitivities to CSA Parameters}.

\begin{figure}[!htbp]
	\begin{subfigure}{0.49\linewidth}
		\centering
		\caption{CVA with VM for different K, no MTA.}	
		\includegraphics[width=0.9\linewidth]{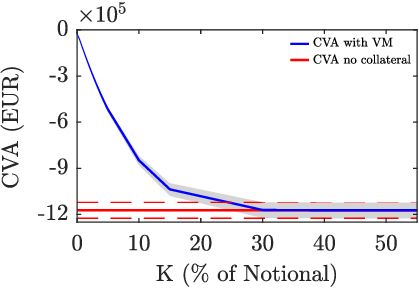}
		\label{fig:irs_15Y_CVA_mta0}
	\end{subfigure}
	\hfill
	\begin{subfigure}{0.49\linewidth}
		\centering
		\caption{CVA with VM for different MTA, no K.}	
		\includegraphics[width=0.9\linewidth]{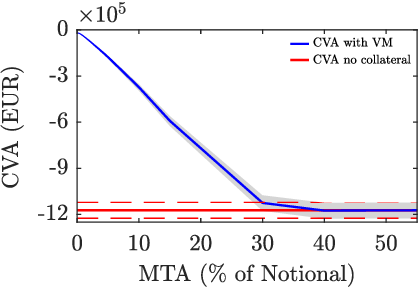}
		\label{fig:irs_15Y_CVA_k0}
	\end{subfigure} \\	
	\begin{subfigure}{0.49\linewidth}
		\centering
		\caption{CVA with VM and IM for different K on VM only, no MTA.}	
		\includegraphics[width=0.9\linewidth]{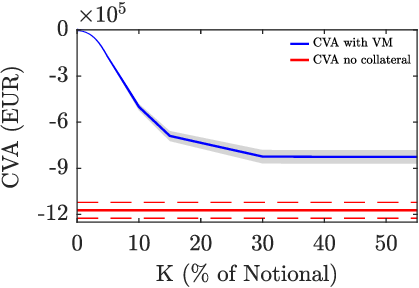}
		\label{fig:irs_15Y_CVA_mta0_kVM_IM}
	\end{subfigure}
	\hfill
	\begin{subfigure}{0.49\linewidth}
		\centering
		\caption{CVA with VM and IM for different MTA on VM only, no K.}	
		\includegraphics[width=0.9\linewidth]{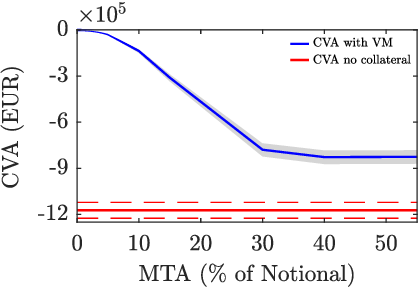}
		\label{fig:irs_15Y_CVA_k0_mtaVM_IM}
	\end{subfigure} \\	
	\begin{subfigure}{0.49\linewidth}
		\centering
		\caption{CVA with VM and IM for different K on VM and IM, no MTA.}	
		\includegraphics[width=0.9\linewidth]{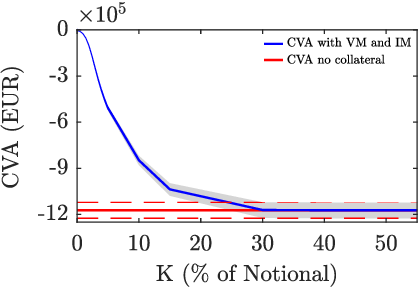}
		\label{fig:irs_15Y_CVA_mta0_kVM_kIM}
	\end{subfigure}
	\hfill
	\begin{subfigure}{0.49\linewidth}
		\centering
		\caption{CVA with VM and IM for different MTA on VM and IM, no K.}	
		\includegraphics[width=0.9\linewidth]{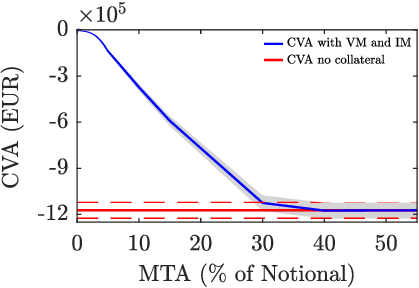}
		\label{fig:irs_15Y_CVA_k0_mtaVM_mtaIM}
	\end{subfigure} 
	\caption{15Y ATM payer Swap EUR 100 Mio nominal amount, collateralized CVA convergence to uncollateralized case with respect to threshold (K, left-hand side) and minimum transfer amount (MTA, right-hand side), keeping other model parameters as in tab.~\ref{tab:model_setup}. Top panels: CVA with VM, mid panels: CVA with VM and IM with K/MTA only on VM, bottom panels: CVA and with VM and IM with K/MTA on both margins. Grey areas: collateralized CVA $3\sigma$ confidence intervals, dashed lines: $3\sigma$ uncollateralized CVA confidence intervals. Convergence to uncollateralized CVA is reached for top and bottom panels for smaller values of K compared to MTA, to confirm that K introduces higher friction.} 
	\label{fig:irs_15Y_k_mta_CVA_convergence}
\end{figure}

\begin{figure}[!htbp]
	\begin{subfigure}{0.49\linewidth}
		\centering
		\caption{DVA with VM for different K, no MTA.}	
		\includegraphics[width=0.9\linewidth]{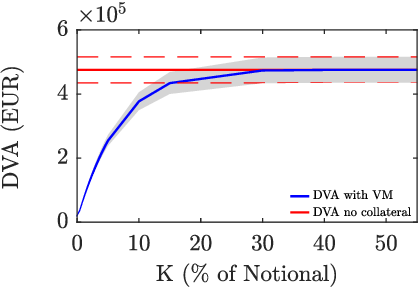}
		\label{fig:irs_15Y_DVA_mta0}
	\end{subfigure}
	\hfill
	\begin{subfigure}{0.49\linewidth}
		\centering
		\caption{DVA with VM for different MTA, no K.}	
		\includegraphics[width=0.9\linewidth]{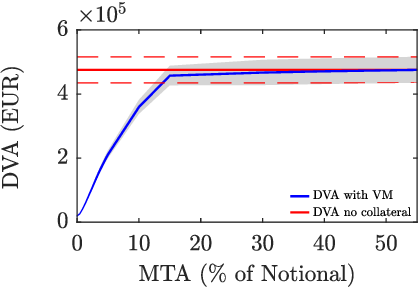}
		\label{fig:irs_15Y_DVA_k0}
	\end{subfigure} \\	
	\begin{subfigure}{0.49\linewidth}
		\centering
		\caption{DVA with VM and IM for different K on VM only, no MTA.}	
		\includegraphics[width=0.9\linewidth]{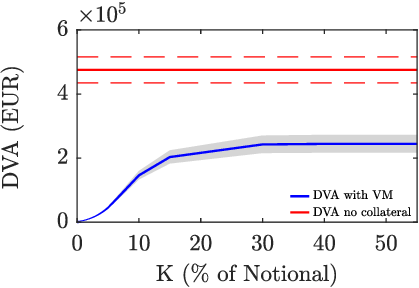}
		\label{fig:irs_15Y_DVA_mta0_kVM_IM}
	\end{subfigure}
	\hfill
	\begin{subfigure}{0.49\linewidth}
		\centering
		\caption{DVA with VM and IM for different MTA on VM only, no K.}	
		\includegraphics[width=0.9\linewidth]{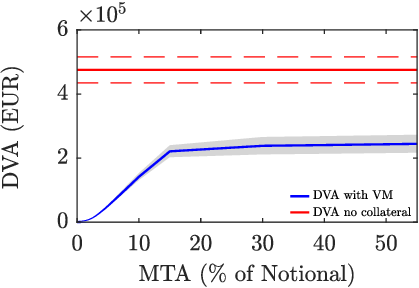}
		\label{fig:irs_15Y_DVA_k0_mtaVM_IM}
	\end{subfigure} \\	
	\begin{subfigure}{0.49\linewidth}
		\centering
		\caption{DVA with VM and IM for different K on VM and IM, no MTA.}	
		\includegraphics[width=0.9\linewidth]{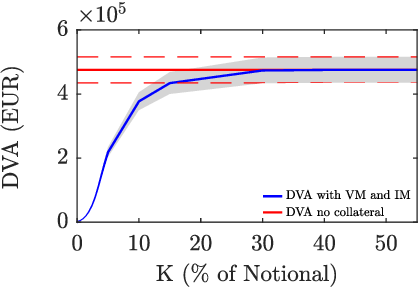}
		\label{fig:irs_15Y_DVA_mta0_kVM_kIM}
	\end{subfigure}
	\hfill
	\begin{subfigure}{0.49\linewidth}
		\centering
		\caption{DVA with VM and IM for different MTA on VM and IM, no K.}	
		\includegraphics[width=0.9\linewidth]{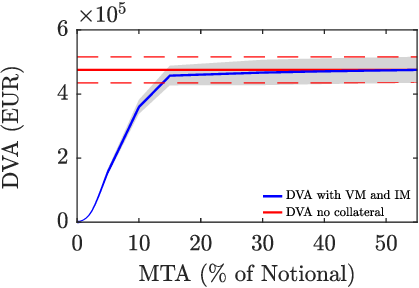}
		\label{fig:irs_15Y_DVA_k0_mtaVM_mtaIM}
	\end{subfigure} 
	\caption{15Y ATM payer Swap EUR 100 Mio nominal amount, collateralized DVA convergence to uncollateralized case with respect to threshold (K, left-hand side) and minimum transfer amount (MTA, right-hand side), keeping other model parameters as in tab.~\ref{tab:model_setup}. Top panels: DVA with VM, mid panels: DVA with VM and IM with K/MTA only on VM, bottom panels: DVA and with VM and IM with K/MTA on both margins. Grey areas: collateralized DVA $3\sigma$ confidence intervals, dashed lines: $3\sigma$ uncollateralized DVA confidence intervals. Analogous considerations of fig.\ref{fig:irs_15Y_k_mta_CVA_convergence}.} 
	\label{fig:irs_15Y_k_mta_DVA_convergence}
\end{figure}

Convergence diagrams for the 15Y ATM payer Swap are shown in figs.~\ref{fig:irs_15Y_k_mta_CVA_convergence}
and \ref{fig:irs_15Y_k_mta_DVA_convergence}.
In particular, left-hand side panels show XVA convergence with respect to K, with $\text{MTA}= 0$ EUR and $l=2$ days; conversely, right-hand side panels show the convergence with respect to MTA, with $\text{K}= 0$ EUR and $l=2$ days.
As can be seen, case 1.~(top panels) and case 3.~(bottom panels) display similar results except for small values of K and MTA, thus IM is ineffective when significant frictions are considered. Instead, case 2.~(middle panels) displays collateralized XVA not converging to uncollateralized ones, this means that without frictions IM is effective in reducing residual credit exposure.  
Furthermore, the results suggest that K leads to a faster convergence to uncollateralized figures compared to MTA; e.g.~in case 1.~$\text{K}= 5$ Mio EUR leads to an increase in absolute terms of approx.~770\% in CVA and 280\% in DVA with respect to the reference case (see tab.~\ref{tab:model_setup}), while $\text{MTA}= 5$ Mio EUR leads to an increase of approx.~3600\% in CVA and 280\% in DVA. As expected, K introduces an higher degree of friction since determining the maximum amount of allowed unsecured exposure; on the other hand, MTA governs only the minimum amount for each margin call, therefore significant impacts can be observed only for large values (see eqs.~\ref{eq:vm} and \ref{eq:im}). The 5x10Y ATM physically settled European payer Swaption displays same results and is not reported.

\begin{figure}[!htbp]
	\begin{subfigure}{0.49\linewidth}
		\centering
		\caption{EPE and ENE Swap with VM, no MTA.}	
		\includegraphics[width=0.9\linewidth]{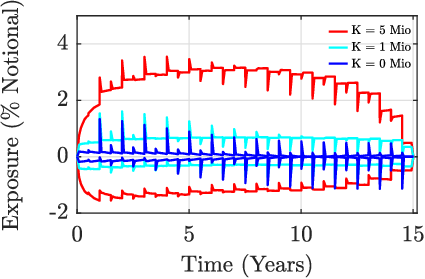}
		\label{fig:irs_15Y_sensitivity_K_VM}
	\end{subfigure}
	\hfill
	\begin{subfigure}{0.49\linewidth}
		\centering
		\caption{EPE and ENE Swaption with VM, no MTA.}
		\includegraphics[width=0.9\linewidth]{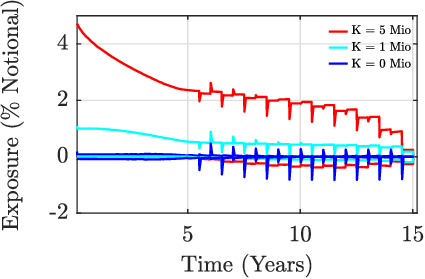}
		\label{fig:swpt_5x10_sensitivity_K_VM}
	\end{subfigure} \\
	\begin{subfigure}{0.49\linewidth}
		\centering
		\caption{EPE and ENE Swap with VM and IM, K on VM only, no MTA.}
		\includegraphics[width=0.9\linewidth]{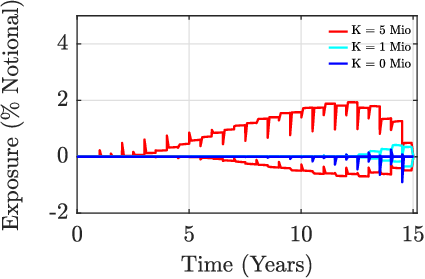}	
		\label{fig:irs_15Y_sensitivity_K_VM_IM}
	\end{subfigure}
	\hfill
	\begin{subfigure}{0.49\linewidth}
		\centering
		\caption{EPE and ENE Swaption with VM and IM, K on VM only, no MTA.}
		\includegraphics[width=0.9\linewidth]{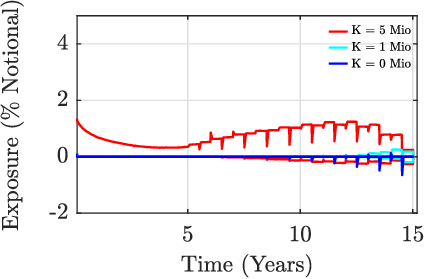}	
		\label{fig:swpt_5x10_sensitivity_K_VM_IM}
	\end{subfigure} \\
	\begin{subfigure}{0.49\linewidth}
		\centering
		\caption{EPE and ENE Swap with VM and IM, K on VM and IM, no MTA.}
		\includegraphics[width=0.9\linewidth]{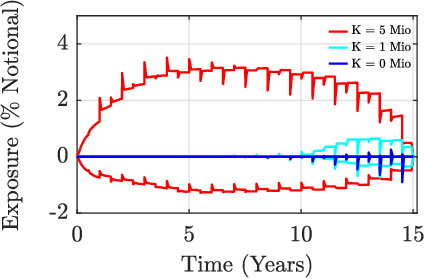}	
		\label{fig:irs_15Y_sensitivity_K_VM_K_IM}
	\end{subfigure}
	\hfill
	\begin{subfigure}{0.49\linewidth}
		\centering
		\caption{EPE and ENE Swaption with VM and IM, K on VM and IM, no MTA.}
		\includegraphics[width=0.9\linewidth]{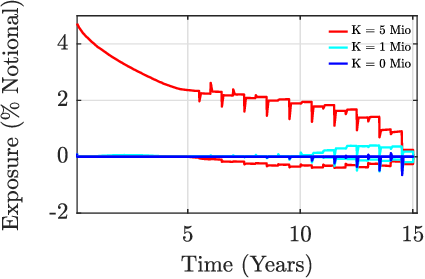}	
		\label{fig:swpt_5x10_sensitivity_K_VM_K_IM}
	\end{subfigure}
	
	\caption{15Y ATM payer Swap (left-hand side) and 5x10Y ATM physically settled European payer Swaption (right-hand side) EUR 100 Mio nominal amount, collateralized EPE/ENE profiles for different values of threshold: K=0 Mio (blue lines), K = 1 Mio (cyan lines) and K = 5 Mio (red lines). Top panels: exposure with VM, mid panels: exposure with VM and IM with K only on VM, bottom panels: exposure with VM and IM with K on both margins. The residual exposure increases with K and IM becomes ineffective for large K. To enhance Swaption's plots readability, we excluded collateralized EPE for time step $t_0$ as not mitigated by collateral, in particular EPE($t_0$) corresponds to Swaption's price: 5030423 EUR (approx. $5\%$ of the nominal amount). Other model parameters as in tab.~\ref{tab:model_setup}. Quantities expressed as a percentage of the nominal amount.} 
	\label{fig:exposure_sensitivity_K}
\end{figure}

A focus on the effects of K on collateralized exposure is shown in fig.~\ref{fig:exposure_sensitivity_K}, which displays EPE/ENE for the 15Y Swap (left-hand side panels) and the 5x10Y ATM physically settled European payer Swaption (right-hand side panels) for the three collateralization schemes above, for different values of K (with $\text{MTA}= 0$ EUR and $l=2$ days).
Focusing on the Swap with VM (top panel), the average EPE (ENE) over the time steps, expressed as a percentage of the nominal amount, is equal to $0.13\%$, $0.63\%$ and $2.56\%$ ($-0.13\%$, $-0.33\%$ and $-1.12\%$), respectively for threshold values of 0, 1 and 5 EUR Mio. By adding IM with no threshold (middle panel), large part of residual exposure is suppressed: average EPE (ENE) is equal to $0.00\%$, $0.04\%$ and $0.91\%$ ($-0.02\%$, $-0.03\%$ and $-0.25\%$) of the nominal amount. By considering threshold also on IM (bottom panel), average EPE (ENE) is equal to $0.00\%$, $0.14\%$ and $2.46\%$ ($-0.02\%$, $-0.07\%$ and $-0.97\%$) of the nominal amount meaning that, as K increases, IM looses its effectiveness and EPE/ENE approach to those with only VM.    

\clearpage
\subsection{XVA Model Risk}
\label{app:Model Risk}

In this section we report additional details regarding the model risk analyses discussed in sec. \ref{sec:Model Risk}. In particular, we show in the following tables \ref{tab:swap_AVA_MoRi_noCollateral_app} and \ref{tab:swpt_AVA_MoRi_noCollateral_app} the full XVA distributions obtained from the alternative combinations considered of model and parametrizations for the 15Y Swap and the 5x10Y Swaption, respectively.
\begin{center}
	\begin{small}
		\begin{longtable}{c c c c c r r r r}
			
			\toprule
			\multirow{3}*{$M_j$} & \multirow{3}*{Model} & \multicolumn{4}{c}{Parameters} & \multicolumn{1}{c}{\multirow{3}*{CVA (\euro)}}   & \multicolumn{1}{c}{\multirow{3}*{DVA (\euro)}}   & \multicolumn{1}{c}{\multirow{3}*{XVA (\euro)}}  \\
			\cline{3-6} 
			&       & G2++ & Time & $\Delta t$ & $N_{MC}$  &       &       &  \\
			& & calib. & grid & & & & & \\
			\midrule 
			\endfirsthead
			
			\multicolumn{9}{c}%
			{{\bfseries \tablename\ \thetable{} -- continued from previous page}} \\
			\toprule
			\multicolumn{6}{c}{XVA framework} &       &       &  \\
			\cline{1-6}
			\multirow{3}*{$M_j$} & \multirow{3}*{Model} & \multicolumn{4}{c}{Parameters} & \multicolumn{1}{c}{\multirow{3}*{CVA}}   & \multicolumn{1}{c}{\multirow{3}*{DVA}}   & \multicolumn{1}{c}{\multirow{3}*{XVA}}  \\
			\cline{3-6} 
			&       & G2++ & Time & $\Delta t$ & $N_{MC}$  &       &       &  \\
			& & pars. & grid & & & & & \\
			\midrule 
			\endhead
			
			\midrule \multicolumn{9}{r}{{Continued on next page}} \\
			\endfoot
			
			\midrule
			\multicolumn{6}{l}{XVA$(t_0;M)$} & -1172938 & 475324 & -697614 \\
			\multicolumn{6}{l}{XVA$(t_0;M_{24})$} & -1175336 & 486628 & -688709 \\
			\multicolumn{6}{l}{AVA$^{\text{MoRi}}(t_0)$} & \multicolumn{1}{c}{2398} & & \\
			\bottomrule
			\caption{XVA distribution for the 15Y ATM payer Swap, EUR 100 Mio nominal amount, no collateral. Each row represents a different combination of model and parameters, sorted by CVA values and labelled $M_j$, with $j = 1,...,235$. Label $M$ denotes the CVA* obtained using the XVA framework set according to our acceptable compromise between accuracy and performance. Label $M_{24}$ denotes the prudent CVA** at $90\%$ confidence level.} 
			\label{tab:swap_AVA_MoRi_noCollateral_app} \\
			\endlastfoot
			
			$M_{1}$ & MC    & $p_{2}$ & Joint & 1M    &  6000 &  -1180939 & 495052 & -685887 \\
			$M_{2}$ & MC    & $p_{2}$ & Joint & 1M    &  7000 &  -1180459 & 493762 & -686697 \\
			$M_{3}$ & MC    & $p_{2}$ & Joint & 1D    &  5000 &  -1179138 & 491171 & -687967 \\
			$M_{4}$ & MC    & $p_{2}$ & Std.  & 1M    &  5000 &  -1178802 & 492103 & -686700 \\
			$M_{5}$ & MC    & $p_{2}$ & Joint & 1M    &  5000 &  -1178799 & 492086 & -686713 \\
			$M_{6}$ & MC    & $p_{2}$ & Std.  & 3M    &  5000 &  -1178040 & 493718 & -684323 \\
			$M_{7}$ & MC    & $p_{2}$ & Joint & 3M    &  5000 &  -1178034 & 493664 & -684370 \\
			$M_{8}$ & MC    & $p_{2}$ & Joint & 1M    &  4000 &  -1177455 & 490733 & -686721 \\
			$M_{9}$ & MC    & $p_{2}$ & Joint & 1M    &  14000 &  -1177447 & 489830 & -687617 \\
			$M_{10}$ & MC    & $p_{2}$ & Joint & 1M    &  13000 &  -1177380 & 489714 & -687667 \\
			$M_{11}$ & MC    & $p_{2}$ & Joint & 1M    &  8000 &  -1177282 & 489528 & -687754 \\
			$M_{12}$ & MC    & $p_{2}$ & Joint & 1M    &  15000 &  -1176990 & 489027 & -687963 \\
			$M_{13}$ & MC    & $p_{1}$ & Joint & 1M    &  6000 &  -1176833 & 476424 & -700408 \\
			$M_{14}$ & MC    & $p_{5}$ & Joint & 1M    &  6000 &  -1176578 & 496362 & -680216 \\
			$M_{15}$ & MC    & $p_{1}$ & Joint & 1M    &  7000 &  -1176339 & 475149 & -701191 \\
			$M_{16}$ & MC    & $p_{2}$ & Joint & 1M    &  20000 &  -1176284 & 487958 & -688326 \\
			$M_{17}$ & MC    & $p_{2}$ & Joint & 1M    &  9000 &  -1176227 & 487748 & -688480 \\
			$M_{18}$ & MC    & $p_{2}$ & Joint & 1M    &  12000 &  -1176226 & 488065 & -688161 \\
			$M_{19}$ & MC    & $p_{2}$ & Joint & 6M    &  5000 &  -1176210 & 494975 & -681235 \\
			$M_{20}$ & MC    & $p_{2}$ & Std.  & 6M    &  5000 &  -1176190 & 495046 & -681144 \\
			$M_{21}$ & MC    & $p_{5}$ & Joint & 1M    &  7000 &  -1176075 & 495036 & -681038 \\
			$M_{22}$ & MC    & $p_{2}$ & Joint & 1M    &  19000 &  -1176016 & 487524 & -688493 \\
			$M_{23}$ & MC    & $p_{2}$ & Joint & 1M    &  11000 &  -1175363 & 486943 & -688421 \\
			$M_{24}$ & MC    & $p_{2}$ & Joint & 1M    &  18000 &  -1175336** & 486628 & -688709 \\
			$M_{25}$ & MC    & $p_{2}$ & Joint & 1M    &  10000 &  -1175271 & 486806 & -688464 \\
			$M_{26}$ & MC    & $p_{2}$ & Joint & 1M    &  16000 &  -1175235 & 486505 & -688730 \\
			$M_{27}$ & MC    & $p_{1}$ & Joint & 1D    &  5000 &  -1175116 & 472673 & -702444 \\
			$M_{28}$ & MC    & $p$ & Joint & 1M    &  6000 &  -1175033 & 478215 & -696818 \\
			$M_{29}$ & MC    & $p_{2}$ & Joint & 1M    &  17000 &  -1174977 & 486176 & -688801 \\
			$M_{30}$ & MC    & $p_{5}$ & Joint & 1D    &  5000 &  -1174892 & 492603 & -682289 \\
			$M_{31}$ & MC    & $p_{1}$ & Std.  & 1M    &  5000 &  -1174735 & 473544 & -701191 \\
			$M_{32}$ & MC    & $p_{1}$ & Joint & 1M    &  5000 &  -1174733 & 473529 & -701205 \\
			$M_{33}$ & MC    & $p$ & Joint & 1M    &  7000 &  -1174536 & 476932 & -697604 \\
			$M_{34}$ & MC    & $p_{5}$ & Std.  & 1M    &  5000 &  -1174462 & 493436 & -681025 \\
			$M_{35}$ & MC    & $p_{5}$ & Joint & 1M    &  5000 &  -1174460 & 493422 & -681039 \\
			$M_{36}$ & MC    & $p_{1}$ & Std.  & 3M    &  5000 &  -1173889 & 475045 & -698845 \\
			$M_{37}$ & MC    & $p_{1}$ & Joint & 3M    &  5000 &  -1173886 & 474994 & -698892 \\
			$M_{38}$ & MC    & $p_{5}$ & Std.  & 3M    &  5000 &  -1173522 & 494865 & -678657 \\
			$M_{39}$ & MC    & $p_{5}$ & Joint & 3M    &  5000 &  -1173521 & 494816 & -678705 \\
			$M_{40}$ & MC    & $p_{1}$ & Joint & 1M    &  14000 &  -1173445 & 471363 & -702082 \\
			$M_{41}$ & MC    & $p_{1}$ & Joint & 1M    &  4000 &  -1173421 & 472209 & -701213 \\
			$M_{42}$ & MC    & $p_{1}$ & Joint & 1M    &  13000 &  -1173384 & 471254 & -702130 \\
			$M_{43}$ & MC    & $p$ & Joint & 1D    &  5000 &  -1173296 & 474469 & -698826 \\
			$M_{44}$ & MC    & $p_{1}$ & Joint & 1M    &  8000 &  -1173265 & 471050 & -702216 \\
			$M_{45}$ & MC    & $p_{5}$ & Joint & 1M    &  4000 &  -1173125 & 492094 & -681032 \\
			$M_{46}$ & MC    & $p_{5}$ & Joint & 1M    &  14000 &  -1173032 & 491076 & -681956 \\
			$M_{47}$ & MC    & $p_{1}$ & Joint & 1M    &  15000 &  -1173001 & 470583 & -702418 \\
			$M_{48}$ & MC    & $p_{5}$ & Joint & 1M    &  13000 &  -1172958 & 490951 & -682007 \\
			$M_{49}$ & MC    & $p$ & Std.  & 1M    &  5000 &  -1172940 & 475340 & -697601 \\
			$M$ & MC    & $p$ & Joint & 1M    &  5000 &  -1172938* & 475324 & -697614 \\
			$M_{50}$ & MC    & $p_{5}$ & Joint & 1M    &  8000 &  -1172874 & 490777 & -682096 \\
			$M_{51}$ & MC    & $p_{5}$ & Joint & 1M    &  15000 &  -1172577 & 490274 & -682303 \\
			$M_{52}$ & MC    & $p_{1}$ & Joint & 1M    &  20000 &  -1172320 & 469551 & -702769 \\
			$M_{53}$ & MC    & $p_{1}$ & Joint & 1M    &  12000 &  -1172248 & 469638 & -702610 \\
			$M_{54}$ & MC    & $p_{1}$ & Joint & 1M    &  9000 &  -1172241 & 469324 & -702917 \\
			$M_{55}$ & MC    & $p$ & Std.  & 3M    &  5000 &  -1172144 & 476839 & -695304 \\
			$M_{56}$ & MC    & $p$ & Joint & 3M    &  5000 &  -1172139 & 476789 & -695350 \\
			$M_{57}$ & MC    & $p_{6}$ & Joint & 1M    &  6000 &  -1172136 & 467902 & -704234 \\
			$M_{58}$ & MC    & $p_{2}$ & Joint & 1M    &  3000 &  -1172097 & 482443 & -689654 \\
			$M_{59}$ & MC    & $p_{1}$ & Joint & 1M    &  19000 &  -1172051 & 469120 & -702931 \\
			$M_{60}$ & MC    & $p_{1}$ & Joint & 6M    &  5000 &  -1171950 & 476154 & -695796 \\
			$M_{61}$ & MC    & $p_{1}$ & Std.  & 6M    &  5000 &  -1171926 & 476220 & -695706 \\
			$M_{62}$ & MC    & $p_{5}$ & Joint & 1M    &  20000 &  -1171866 & 489199 & -682667 \\
			$M_{63}$ & MC    & $p_{5}$ & Joint & 1M    &  12000 &  -1171791 & 489286 & -682505 \\
			$M_{64}$ & MC    & $p_{5}$ & Joint & 1M    &  9000 &  -1171783 & 488952 & -682830 \\
			$M_{65}$ & MC    & $p$ & Joint & 1M    &  14000 &  -1171638 & 473143 & -698495 \\
			$M_{66}$ & MC    & $p$ & Joint & 1M    &  4000 &  -1171633 & 474013 & -697620 \\
			$M_{67}$ & MC    & $p_{5}$ & Joint & 1M    &  19000 &  -1171603 & 488773 & -682830 \\
			$M_{68}$ & MC    & $p$ & Joint & 1M    &  13000 &  -1171579 & 473037 & -698542 \\
			$M_{69}$ & MC    & $p_{6}$ & Joint & 1M    &  7000 &  -1171562 & 466531 & -705031 \\
			$M_{70}$ & MC    & $p_{5}$ & Joint & 6M    &  5000 &  -1171464 & 495878 & -675586 \\
			$M_{71}$ & MC    & $p$ & Joint & 1M    &  8000 &  -1171461 & 472833 & -698628 \\
			$M_{72}$ & MC    & $p_{4}$ & Joint & 1M    &  6000 &  -1171458 & 464810 & -706648 \\
			$M_{73}$ & MC    & $p_{5}$ & Std.  & 6M    &  5000 &  -1171435 & 495940 & -675495 \\
			$M_{74}$ & MC    & $p_{1}$ & Joint & 1M    &  11000 &  -1171404 & 468543 & -702861 \\
			$M_{75}$ & MC    & $p_{1}$ & Joint & 1M    &  18000 &  -1171388 & 468249 & -703139 \\
			$M_{76}$ & MC    & $p_{1}$ & Joint & 1M    &  10000 &  -1171303 & 468399 & -702903 \\
			$M_{77}$ & MC    & $p_{1}$ & Joint & 1M    &  16000 &  -1171285 & 468126 & -703160 \\
			$M_{78}$ & MC    & $p$ & Joint & 1M    &  15000 &  -1171196 & 472365 & -698831 \\
			$M_{79}$ & MC    & $p_{1}$ & Joint & 1M    &  17000 &  -1171039 & 467810 & -703229 \\
			$M_{80}$ & MC    & $p_{5}$ & Joint & 1M    &  11000 &  -1170953 & 488196 & -682758 \\
			$M_{81}$ & MC    & $p_{5}$ & Joint & 1M    &  18000 &  -1170921 & 487874 & -683047 \\
			$M_{82}$ & MC    & $p_{4}$ & Joint & 1M    &  7000 &  -1170886 & 463447 & -707439 \\
			$M_{83}$ & MC    & $p_{5}$ & Joint & 1M    &  10000 &  -1170848 & 488041 & -682808 \\
			$M_{84}$ & MC    & $p_{5}$ & Joint & 1M    &  16000 &  -1170818 & 487748 & -683070 \\
			$M_{85}$ & MC    & $p_{6}$ & Joint & 1D    &  5000 &  -1170601 & 464453 & -706149 \\
			$M_{86}$ & MC    & $p_{5}$ & Joint & 1M    &  17000 &  -1170575 & 487437 & -683138 \\
			$M_{87}$ & MC    & $p$ & Joint & 1M    &  20000 &  -1170511 & 471327 & -699184 \\
			$M_{88}$ & MC    & $p$ & Joint & 1M    &  12000 &  -1170445 & 471424 & -699021 \\
			$M_{89}$ & MC    & $p$ & Joint & 1M    &  9000 &  -1170432 & 471101 & -699331 \\
			$M_{90}$ & MC    & $p$ & Joint & 6M    &  5000 &  -1170271 & 478016 & -692255 \\
			$M_{91}$ & MC    & $p$ & Std.  & 6M    &  5000 &  -1170249 & 478016 & -692233 \\
			$M_{92}$ & MC    & $p$ & Joint & 1M    &  19000 &  -1170243 & 470898 & -699345 \\
			$M_{93}$ & MC    & $p_{6}$ & Joint & 1M    &  5000 &  -1170109 & 465102 & -705007 \\
			$M_{94}$ & MC    & $p_{6}$ & Std.  & 1M    &  5000 &  -1170109 & 465114 & -704995 \\
			$M_{95}$ & MC    & $p_{4}$ & Joint & 1D    &  5000 &  -1169919 & 461368 & -708551 \\
			$M_{96}$ & MC    & $p$ & Joint & 1M    &  11000 &  -1169599 & 470326 & -699273 \\
			$M_{97}$ & MC    & $p$ & Joint & 1M    &  18000 &  -1169580 & 470026 & -699554 \\
			$M_{98}$ & MC    & $p$ & Joint & 1M    &  10000 &  -1169500 & 470184 & -699316 \\
			$M_{99}$ & MC    & $p$ & Joint & 1M    &  16000 &  -1169479 & 469905 & -699574 \\
			$M_{100}$ & MC    & $p_{4}$ & Std.  & 1M    &  5000 &  -1169437 & 462032 & -707405 \\
			$M_{101}$ & MC    & $p_{4}$ & Joint & 1M    &  5000 &  -1169436 & 462020 & -707417 \\
			$M_{102}$ & MC    & $p$ & Joint & 1M    &  17000 &  -1169232 & 469590 & -699642 \\
			$M_{103}$ & MC    & $p_{6}$ & Joint & 3M    &  5000 &  -1169055 & 466167 & -702888 \\
			$M_{104}$ & MC    & $p_{6}$ & Std.  & 3M    &  5000 &  -1169053 & 466208 & -702845 \\
			$M_{105}$ & MC    & $p_{6}$ & Joint & 1M    &  4000 &  -1168865 & 463880 & -704985 \\
			$M_{106}$ & MC    & $p_{6}$ & Joint & 1M    &  14000 &  -1168705 & 462802 & -705903 \\
			$M_{107}$ & MC    & $p_{6}$ & Joint & 1M    &  13000 &  -1168642 & 462690 & -705952 \\
			$M_{108}$ & MC    & $p_{6}$ & Joint & 1M    &  8000 &  -1168534 & 462496 & -706038 \\
			$M_{109}$ & MC    & $p_{4}$ & Joint & 3M    &  5000 &  -1168401 & 463090 & -705312 \\
			$M_{110}$ & MC    & $p_{4}$ & Std.  & 3M    &  5000 &  -1168400 & 463131 & -705269 \\
			$M_{111}$ & MC    & $p_{2}$ & Joint & 1M    &  2000 &  -1168398 & 477193 & -691205 \\
			$M_{112}$ & MC    & $p_{6}$ & Joint & 1M    &  15000 &  -1168280 & 462045 & -706234 \\
			$M_{113}$ & MC    & $p_{1}$ & Joint & 1M    &  3000 &  -1168215 & 464164 & -704051 \\
			$M_{114}$ & MC    & $p_{4}$ & Joint & 1M    &  4000 &  -1168197 & 460800 & -707396 \\
			$M_{115}$ & MC    & $p_{4}$ & Joint & 1M    &  14000 &  -1168053 & 459747 & -708306 \\
			$M_{116}$ & MC    & $p_{4}$ & Joint & 1M    &  13000 &  -1167992 & 459638 & -708355 \\
			$M_{117}$ & MC    & $p_{4}$ & Joint & 1M    &  8000 &  -1167879 & 459438 & -708441 \\
			$M_{118}$ & MC    & $p_{5}$ & Joint & 1M    &  3000 &  -1167735 & 483754 & -683981 \\
			$M_{119}$ & MC    & $p_{4}$ & Joint & 1M    &  15000 &  -1167630 & 458995 & -708636 \\
			$M_{120}$ & MC    & $p_{6}$ & Joint & 1M    &  20000 &  -1167600 & 461016 & -706584 \\
			$M_{121}$ & MC    & $p_{6}$ & Joint & 1M    &  12000 &  -1167500 & 461072 & -706428 \\
			$M_{122}$ & MC    & $p_{6}$ & Joint & 1M    &  9000 &  -1167459 & 460714 & -706745 \\
			$M_{123}$ & MC    & $p_{6}$ & Joint & 1M    &  19000 &  -1167340 & 460603 & -706737 \\
			$M_{124}$ & MC    & $p_{4}$ & Joint & 1M    &  20000 &  -1166954 & 457971 & -708983 \\
			$M_{125}$ & MC    & $p_{4}$ & Joint & 1M    &  12000 &  -1166856 & 458028 & -708828 \\
			$M_{126}$ & MC    & $p_{6}$ & Joint & 6M    &  5000 &  -1166852 & 466787 & -700065 \\
			$M_{127}$ & MC    & $p_{6}$ & Std.  & 6M    &  5000 &  -1166818 & 466835 & -699984 \\
			$M_{128}$ & MC    & $p_{4}$ & Joint & 1M    &  9000 &  -1166814 & 457672 & -709142 \\
			$M_{129}$ & MC    & $p_{6}$ & Joint & 1M    &  11000 &  -1166714 & 460051 & -706663 \\
			$M_{130}$ & MC    & $p_{4}$ & Joint & 1M    &  19000 &  -1166694 & 457558 & -709136 \\
			$M_{131}$ & MC    & $p_{6}$ & Joint & 1M    &  18000 &  -1166685 & 459743 & -706942 \\
			$M_{132}$ & MC    & $p_{6}$ & Joint & 1M    &  10000 &  -1166585 & 459869 & -706716 \\
			$M_{133}$ & MC    & $p_{6}$ & Joint & 1M    &  16000 &  -1166584 & 459619 & -706965 \\
			$M_{134}$ & MC    & $p$ & Joint & 1M    &  3000 &  -1166416 & 465953 & -700463 \\
			$M_{135}$ & MC    & $p_{6}$ & Joint & 1M    &  17000 &  -1166374 & 459346 & -707027 \\
			$M_{136}$ & MC    & $p_{4}$ & Joint & 6M    &  5000 &  -1166222 & 463715 & -702508 \\
			$M_{137}$ & MC    & $p_{4}$ & Std.  & 6M    &  5000 &  -1166190 & 463763 & -702426 \\
			$M_{138}$ & MC    & $p_{4}$ & Joint & 1M    &  11000 &  -1166069 & 457007 & -709063 \\
			$M_{139}$ & MC    & $p_{4}$ & Joint & 1M    &  18000 &  -1166042 & 456703 & -709339 \\
			$M_{140}$ & MC    & $p_{4}$ & Joint & 1M    &  16000 &  -1165941 & 456579 & -709362 \\
			$M_{141}$ & MC    & $p_{4}$ & Joint & 1M    &  10000 &  -1165941 & 456827 & -709114 \\
			$M_{142}$ & MC    & $p_{4}$ & Joint & 1M    &  17000 &  -1165731 & 456306 & -709425 \\
			$M_{143}$ & MC    & $p_{1}$ & Joint & 1M    &  2000 &  -1164611 & 459057 & -705554 \\
			$M_{144}$ & MC    & $p_{5}$ & Joint & 1M    &  2000 &  -1164068 & 478544 & -685524 \\
			$M_{145}$ & MC    & $p_{6}$ & Joint & 1M    &  3000 &  -1163698 & 455890 & -707808 \\
			$M_{146}$ & MC    & $p_{4}$ & Joint & 1M    &  3000 &  -1163058 & 452857 & -710201 \\
			$M_{147}$ & MC    & $p_{3}$ & Joint & 1M    &  6000 &  -1162901 & 463378 & -699523 \\
			$M_{148}$ & MC    & $p$ & Joint & 1M    &  2000 &  -1162820 & 460853 & -701967 \\
			$M_{149}$ & MC    & $p_{3}$ & Joint & 1M    &  7000 &  -1162353 & 462051 & -700302 \\
			$M_{150}$ & MC    & $p_{3}$ & Joint & 1D    &  5000 &  -1161376 & 459916 & -701459 \\
			$M_{151}$ & MC    & $p_{3}$ & Joint & 1M    &  5000 &  -1160878 & 460594 & -700284 \\
			$M_{152}$ & MC    & $p_{3}$ & Std.  & 1M    &  5000 &  -1160878 & 460607 & -700271 \\
			$M_{153}$ & MC    & $p_{6}$ & Joint & 1M    &  2000 &  -1160247 & 450976 & -709271 \\
			$M_{154}$ & MC    & $p_{3}$ & Joint & 3M    &  5000 &  -1159813 & 461715 & -698098 \\
			$M_{155}$ & MC    & $p_{3}$ & Std.  & 3M    &  5000 &  -1159810 & 461757 & -698053 \\
			$M_{156}$ & MC    & $p_{3}$ & Joint & 1M    &  4000 &  -1159624 & 459356 & -700269 \\
			$M_{157}$ & MC    & $p_{4}$ & Joint & 1M    &  2000 &  -1159622 & 447965 & -711657 \\
			$M_{158}$ & MC    & $p_{3}$ & Joint & 1M    &  14000 &  -1159525 & 458362 & -701162 \\
			$M_{159}$ & MC    & $p_{3}$ & Joint & 1M    &  13000 &  -1159464 & 458254 & -701209 \\
			$M_{160}$ & MC    & $p_{3}$ & Joint & 1M    &  8000 &  -1159347 & 458052 & -701294 \\
			$M_{161}$ & MC    & $p_{3}$ & Joint & 1M    &  15000 &  -1159099 & 457610 & -701488 \\
			$M_{162}$ & MC    & $p_{2}$ & Joint & 1M    &  1000 &  -1158473 & 461577 & -696896 \\
			$M_{163}$ & MC    & $p_{3}$ & Joint & 1M    &  20000 &  -1158427 & 456595 & -701832 \\
			$M_{164}$ & MC    & $p_{3}$ & Joint & 1M    &  12000 &  -1158330 & 456654 & -701677 \\
			$M_{165}$ & MC    & $p_{3}$ & Joint & 1M    &  9000 &  -1158296 & 456310 & -701986 \\
			$M_{166}$ & MC    & $p_{3}$ & Joint & 1M    &  19000 &  -1158166 & 456181 & -701985 \\
			$M_{167}$ & MC    & $p_{3}$ & Joint & 6M    &  5000 &  -1157590 & 462411 & -695179 \\
			$M_{168}$ & MC    & $p_{3}$ & Std.  & 6M    &  5000 &  -1157555 & 462462 & -695093 \\
			$M_{169}$ & MC    & $p_{3}$ & Joint & 1M    &  11000 &  -1157538 & 455626 & -701912 \\
			$M_{170}$ & MC    & $p_{3}$ & Joint & 1M    &  18000 &  -1157516 & 455331 & -702185 \\
			$M_{171}$ & MC    & $p_{3}$ & Joint & 1M    &  10000 &  -1157416 & 455455 & -701961 \\
			$M_{172}$ & MC    & $p_{3}$ & Joint & 1M    &  16000 &  -1157415 & 455208 & -702207 \\
			$M_{173}$ & MC    & $p_{3}$ & Joint & 1M    &  17000 &  -1157201 & 454931 & -702270 \\
			$M_{174}$ & An G2 & $p$ & Std.  & 6M    & NA    & -1157012 & 471340 & -685672 \\
			$M_{175}$ & An G2 & $p$ & Std.  & 1D    & NA    & -1156083 & 469562 & -686520 \\
			$M_{176}$ & MC    & $p_{1}$ & Joint & 1M    &  1000 &  -1154822 & 443763 & -711059 \\
			$M_{177}$ & An G2 & $p_{2}$ & Std.  & 6M    & NA    & -1154548 & 468701 & -685847 \\
			$M_{178}$ & MC    & $p_{3}$ & Joint & 1M    &  3000 &  -1154503 & 451461 & -703042 \\
			$M_{179}$ & An G2 & $p_{1}$ & Std.  & 1D    & NA    & -1154118 & 470605 & -683513 \\
			$M_{180}$ & An G2 & $p_{1}$ & Std.  & 6M    & NA    & -1154058 & 471569 & -682489 \\
			$M_{181}$ & MC    & $p_{5}$ & Joint & 1M    &  1000 &  -1153882 & 462597 & -691285 \\
			$M_{182}$ & An G2 & $p_{2}$ & Std.  & 1D    & NA    & -1153616 & 466927 & -686688 \\
			$M_{183}$ & MC    & $p$ & Joint & 1M    &  1000 &  -1153010 & 445537 & -707473 \\
			$M_{184}$ & An G2 & $p$ & Std.  & 1M    & NA    & -1152567 & 468721 & -683846 \\
			$M_{185}$ & MC    & $p_{3}$ & Joint & 1M    &  2000 &  -1151043 & 446555 & -704488 \\
			$M_{186}$ & An G2 & $p_{1}$ & Std.  & 1M    & NA    & -1150498 & 469695 & -680804 \\
			$M_{187}$ & An G2 & $p_{2}$ & Std.  & 1M    & NA    & -1150104 & 466088 & -684016 \\
			$M_{188}$ & MC    & $p_{6}$ & Joint & 1M    &  1000 &  -1150030 & 435226 & -714804 \\
			$M_{189}$ & MC    & $p_{4}$ & Joint & 1M    &  1000 &  -1149463 & 432315 & -717147 \\
			$M_{190}$ & An G2 & $p$ & Std.  & 3M    & NA    & -1147871 & 468884 & -678987 \\
			$M_{191}$ & An G2 & $p_{1}$ & Std.  & 3M    & NA    & -1145463 & 469566 & -675897 \\
			$M_{192}$ & An G2 & $p_{2}$ & Std.  & 3M    & NA    & -1145393 & 466229 & -679164 \\
			$M_{193}$ & MC    & $p_{3}$ & Joint & 1M    &  1000 &  -1141007 & 431102 & -709905 \\
			$M_{194}$ & An G2 & $p_{4}$ & Std.  & 6M    & NA    & -1137687 & 450245 & -687443 \\
			$M_{195}$ & An G2 & $p_{6}$ & Std.  & 6M    & NA    & -1137168 & 449645 & -687523 \\
			$M_{196}$ & An G2 & $p_{4}$ & Std.  & 1D    & NA    & -1136472 & 448453 & -688019 \\
			$M_{197}$ & An G2 & $p_{6}$ & Std.  & 1D    & NA    & -1135933 & 447834 & -688099 \\
			$M_{198}$ & An G2 & $p_{4}$ & Std.  & 1M    & NA    & -1133010 & 447623 & -685387 \\
			$M_{199}$ & An G2 & $p_{6}$ & Std.  & 1M    & NA    & -1132472 & 447006 & -685467 \\
			$M_{200}$ & An G2 & $p_{3}$ & Std.  & 6M    & NA    & -1132469 & 445589 & -686880 \\
			$M_{201}$ & An G2 & $p_{3}$ & Std.  & 1D    & NA    & -1131772 & 444247 & -687526 \\
			$M_{202}$ & An G2 & $p_{4}$ & Std.  & 3M    & NA    & -1128443 & 447824 & -680619 \\
			$M_{203}$ & An G2 & $p_{3}$ & Std.  & 1M    & NA    & -1128282 & 443400 & -684881 \\
			$M_{204}$ & An G2 & $p_{6}$ & Std.  & 3M    & NA    & -1127904 & 447202 & -680702 \\
			$M_{205}$ & An G2 & $p_{5}$ & Std.  & 6M    & NA    & -1127358 & 439799 & -687559 \\
			$M_{206}$ & An G2 & $p_{5}$ & Std.  & 1D    & NA    & -1126588 & 438425 & -688162 \\
			$M_{207}$ & MC    & $p_{2}$ & Joint & 12M   &  5000 &  -1125419 & 502368 & -623051 \\
			$M_{208}$ & An G2 & $p_{3}$ & Std.  & 3M    & NA    & -1123553 & 443463 & -680091 \\
			$M_{209}$ & An G2 & $p_{5}$ & Std.  & 1M    & NA    & -1123116 & 437590 & -685526 \\
			$M_{210}$ & MC    & $p_{2}$ & Std.  & 12M   &  5000 &  -1122131 & 505070 & -617062 \\
			$M_{211}$ & MC    & $p_{5}$ & Joint & 12M   &  5000 &  -1121395 & 504027 & -617368 \\
			$M_{212}$ & MC    & $p_{1}$ & Joint & 12M   &  5000 &  -1121235 & 483889 & -637346 \\
			$M_{213}$ & MC    & $p$ & Joint & 12M   &  5000 &  -1119629 & 485660 & -633968 \\
			$M_{214}$ & An G2 & $p_{5}$ & Std.  & 3M    & NA    & -1118401 & 437648 & -680753 \\
			$M_{215}$ & MC    & $p_{1}$ & Std.  & 12M   &  5000 &  -1117732 & 486341 & -631392 \\
			$M_{216}$ & MC    & $p_{5}$ & Std.  & 12M   &  5000 &  -1117587 & 506193 & -611394 \\
			$M_{217}$ & MC    & $p_{6}$ & Joint & 12M   &  5000 &  -1117259 & 475920 & -641339 \\
			$M_{218}$ & MC    & $p_{4}$ & Joint & 12M   &  5000 &  -1116552 & 472804 & -643748 \\
			$M_{219}$ & MC    & $p$ & Std.  & 12M   &  5000 &  -1116254 & 488095 & -628159 \\
			$M_{220}$ & MC    & $p_{6}$ & Std.  & 12M   &  5000 &  -1113132 & 477249 & -635882 \\
			$M_{221}$ & MC    & $p_{4}$ & Std.  & 12M   &  5000 &  -1112484 & 474157 & -638327 \\
			$M_{222}$ & MC    & $p_{3}$ & Joint & 12M   &  5000 &  -1108077 & 471424 & -636653 \\
			$M_{223}$ & MC    & $p_{3}$ & Std.  & 12M   &  5000 &  -1103933 & 472938 & -630995 \\
			$M_{224}$ & An G2 & $p$ & Std.  & 12M   & NA    & -1091604 & 454023 & -637581 \\
			$M_{225}$ & An G2 & $p_{2}$ & Std.  & 12M   & NA    & -1089152 & 451377 & -637775 \\
			$M_{226}$ & An G2 & $p_{1}$ & Std.  & 12M   & NA    & -1088311 & 453997 & -634314 \\
			$M_{227}$ & An Blk & NA    & Std.  & 6M    & NA    & -1077346 & 449890 & -627456 \\
			$M_{228}$ & An Blk & NA    & Std.  & 1D    & NA    & -1075398 & 448170 & -627229 \\
			$M_{229}$ & An G2 & $p_{4}$ & Std.  & 12M   & NA    & -1072815 & 433189 & -639626 \\
			$M_{230}$ & An G2 & $p_{6}$ & Std.  & 12M   & NA    & -1072281 & 432567 & -639714 \\
			$M_{231}$ & An Blk & NA    & Std.  & 1M    & NA    & -1072231 & 447369 & -624862 \\
			$M_{232}$ & An G2 & $p_{3}$ & Std.  & 12M   & NA    & -1067816 & 428825 & -638990 \\
			$M_{233}$ & An Blk & NA    & Std.  & 3M    & NA    & -1067726 & 447358 & -620369 \\
			$M_{234}$ & An G2 & $p_{5}$ & Std.  & 12M   & NA    & -1062852 & 423141 & -639711 \\
			$M_{235}$ & An Blk & NA    & Std.  & 12M   & NA    & -1016082 & 432373 & -583709 \\
			
		\end{longtable}
	\end{small}
\end{center}

\begin{center}
	\begin{small}
		\begin{longtable}{c c c c c r r r r}
			
			\toprule
			\multirow{3}*{$M_j$} & \multirow{3}*{Model} & \multicolumn{4}{c}{Parameters} & \multicolumn{1}{c}{\multirow{3}*{CVA (\euro)}}   & \multicolumn{1}{c}{\multirow{3}*{DVA (\euro)}}   & \multicolumn{1}{c}{\multirow{3}*{XVA (\euro)}}  \\
			\cline{3-6} 
			&       & G2++ & Time & $\Delta t$ & $N_{MC}$  &       &       &  \\
			& & calib. & grid & & & & & \\
			\midrule 
			\endfirsthead
			
			\multicolumn{9}{c}%
			{{\bfseries \tablename\ \thetable{} -- continued from previous page}} \\
			\toprule
			\multicolumn{6}{c}{XVA framework} &       &       &  \\
			\cline{1-6}
			\multirow{3}*{$M_j$} & \multirow{3}*{Model} & \multicolumn{4}{c}{Parameters} & \multicolumn{1}{c}{\multirow{3}*{CVA}}   & \multicolumn{1}{c}{\multirow{3}*{DVA}}   & \multicolumn{1}{c}{\multirow{3}*{XVA}}  \\
			\cline{3-6} 
			&       & G2++ & Time & $\Delta t$ & $N_{MC}$  &       &       &  \\
			& & pars. & grid & & & & & \\
			\midrule 
			\endhead
			
			\midrule \multicolumn{9}{r}{{Continued on next page}} \\
			\endfoot
			
			\midrule
			\multicolumn{6}{l}{XVA$(t_0;M)$} & -818308 & 36418 & -781890 \\
			\multicolumn{6}{l}{XVA$(t_0;M_{20})$} & -823724 & 37859 & -785865 \\
			\multicolumn{6}{l}{AVA$^{\text{MoRi}}(t_0)$} & \multicolumn{1}{c}{5416} & & \\
			\bottomrule
			\caption{XVA distribution for the 5x10Y ATM physically settled European payer Swaption, EUR 100 Mio nominal amount, no collateral. Table structure as of tab.~\ref{tab:swap_AVA_MoRi_noCollateral_app}.} 
			\label{tab:swpt_AVA_MoRi_noCollateral_app} \\
			\endlastfoot
			
			$M_{1}$ & MC    & $p_{2}$ & Joint & 1M    & 7000  & -829878 & 37102 & -792776 \\
			$M_{2}$ & MC    & $p_{1}$ & Joint & 1M    & 7000  & -828711 & 36314 & -792398 \\
			$M_{3}$ & MC    & $p_{2}$ & Joint & 1M    & 6000  & -827745 & 37057 & -790688 \\
			$M_{4}$ & MC    & $p_{5}$ & Joint & 1M    & 7000  & -827025 & 35741 & -791284 \\
			$M_{5}$ & MC    & $p_{1}$ & Joint & 1M    & 6000  & -826577 & 36157 & -790420 \\
			$M_{6}$ & MC    & $p_{2}$ & Joint & 1M    & 8000  & -825596 & 37692 & -787903 \\
			$M_{7}$ & MC    & $p_{2}$ & Joint & 1D    & 5000  & -825122 & 37418 & -787704 \\
			$M_{8}$ & MC    & $p_{2}$ & Joint & 1M    & 14000 & -825026 & 38559 & -786468 \\
			$M_{9}$ & MC    & $p_{5}$ & Joint & 1M    & 6000  & -824992 & 35582 & -789409 \\
			$M_{10}$ & MC    & $p_{1}$ & Joint & 1M    & 8000  & -824473 & 36855 & -787617 \\
			$M_{11}$ & MC    & $p_{2}$ & Joint & 1M    & 13000 & -824300 & 38920 & -785379 \\
			$M_{12}$ & MC    & $p_{2}$ & Joint & 1M    & 15000 & -824267 & 38472 & -785795 \\
			$M_{13}$ & MC    & $p$ & Joint & 1M    & 7000  & -824079 & 36192 & -787887 \\
			$M_{14}$ & MC    & $p_{1}$ & Joint & 1D    & 5000  & -824044 & 36480 & -787564 \\
			$M_{15}$ & MC    & $p_{2}$ & Joint & 1M    & 20000 & -824031 & 38280 & -785752 \\
			$M_{16}$ & MC    & $p_{6}$ & Joint & 1M    & 7000  & -823887 & 34017 & -789870 \\
			$M_{17}$ & MC    & $p_{2}$ & Joint & 1M    & 5000  & -823830 & 37473 & -786357 \\
			$M_{18}$ & MC    & $p_{2}$ & Std.  & 1M    & 5000  & -823816 & 37482 & -786334 \\
			$M_{19}$ & MC    & $p_{2}$ & Joint & 1M    & 19000 & -823745 & 38066 & -785678 \\
			$M_{20}$ & MC    & $p_{1}$ & Joint & 1M    & 14000 & -823724** & 37859 & -785865 \\
			$M_{21}$ & MC    & $p_{2}$ & Joint & 1M    & 9000  & -823583 & 38807 & -784776 \\
			$M_{22}$ & MC    & $p_{2}$ & Joint & 1M    & 18000 & -823104 & 37864 & -785240 \\
			$M_{23}$ & MC    & $p_{1}$ & Joint & 1M    & 15000 & -823100 & 37803 & -785297 \\
			$M_{24}$ & MC    & $p_{1}$ & Joint & 1M    & 13000 & -823012 & 38058 & -784955 \\
			$M_{25}$ & MC    & $p_{2}$ & Joint & 3M    & 5000  & -822973 & 37608 & -785366 \\
			$M_{26}$ & MC    & $p_{1}$ & Joint & 1M    & 20000 & -822950 & 37715 & -785235 \\
			$M_{27}$ & MC    & $p_{2}$ & Joint & 1M    & 12000 & -822932 & 38890 & -784042 \\
			$M_{28}$ & MC    & $p_{2}$ & Std.  & 3M    & 5000  & -822913 & 37631 & -785283 \\
			$M_{29}$ & MC    & $p_{4}$ & Joint & 1M    & 7000  & -822802 & 33959 & -788843 \\
			$M_{30}$ & MC    & $p_{5}$ & Joint & 1D    & 5000  & -822742 & 35780 & -786962 \\
			$M_{31}$ & MC    & $p_{1}$ & Joint & 1M    & 5000  & -822656 & 36530 & -786126 \\
			$M_{32}$ & MC    & $p_{1}$ & Std.  & 1M    & 5000  & -822641 & 36538 & -786103 \\
			$M_{33}$ & MC    & $p_{1}$ & Joint & 1M    & 19000 & -822612 & 37530 & -785082 \\
			$M_{34}$ & MC    & $p_{2}$ & Joint & 1M    & 16000 & -822600 & 38104 & -784496 \\
			$M_{35}$ & MC    & $p_{5}$ & Joint & 1M    & 8000  & -822576 & 36309 & -786267 \\
			$M_{36}$ & MC    & $p_{2}$ & Joint & 1M    & 11000 & -822421 & 38332 & -784089 \\
			$M_{37}$ & MC    & $p_{1}$ & Joint & 1M    & 9000  & -822417 & 37912 & -784505 \\
			$M_{38}$ & MC    & $p_{5}$ & Joint & 1M    & 14000 & -822374 & 37377 & -784996 \\
			$M_{39}$ & MC    & $p_{2}$ & Joint & 1M    & 17000 & -822280 & 38015 & -784264 \\
			$M_{40}$ & MC    & $p_{2}$ & Joint & 1M    & 10000 & -822075 & 38564 & -783510 \\
			$M_{41}$ & MC    & $p$ & Joint & 1M    & 6000  & -822059 & 36034 & -786025 \\
			$M_{42}$ & MC    & $p_{6}$ & Joint & 1M    & 6000  & -821993 & 33614 & -788380 \\
			$M_{43}$ & MC    & $p_{1}$ & Joint & 1M    & 18000 & -821898 & 37356 & -784542 \\
			$M_{44}$ & MC    & $p_{2}$ & Joint & 1M    & 4000  & -821800 & 38013 & -783787 \\
			$M_{45}$ & MC    & $p_{1}$ & Joint & 1M    & 12000 & -821699 & 37990 & -783708 \\
			$M_{46}$ & MC    & $p_{5}$ & Joint & 1M    & 13000 & -821603 & 37729 & -783874 \\
			$M_{47}$ & MC    & $p_{1}$ & Joint & 3M    & 5000  & -821585 & 36657 & -784928 \\
			$M_{48}$ & MC    & $p_{5}$ & Joint & 1M    & 15000 & -821540 & 37328 & -784213 \\
			$M_{49}$ & MC    & $p_{1}$ & Std.  & 3M    & 5000  & -821524 & 36678 & -784845 \\
			$M_{50}$ & MC    & $p_{5}$ & Joint & 1M    & 5000  & -821480 & 35823 & -785657 \\
			$M_{51}$ & MC    & $p_{5}$ & Std.  & 1M    & 5000  & -821464 & 35831 & -785633 \\
			$M_{52}$ & MC    & $p_{1}$ & Joint & 1M    & 16000 & -821427 & 37419 & -784008 \\
			$M_{53}$ & MC    & $p_{2}$ & Joint & 6M    & 5000  & -821393 & 37782 & -783612 \\
			$M_{54}$ & MC    & $p_{5}$ & Joint & 1M    & 20000 & -821341 & 37298 & -784044 \\
			$M_{55}$ & MC    & $p_{2}$ & Std.  & 6M    & 5000  & -821257 & 37816 & -783442 \\
			$M_{56}$ & MC    & $p_{1}$ & Joint & 1M    & 11000 & -821251 & 37459 & -783791 \\
			$M_{57}$ & MC    & $p_{1}$ & Joint & 1M    & 17000 & -821084 & 37366 & -783718 \\
			$M_{58}$ & MC    & $p_{5}$ & Joint & 1M    & 19000 & -821076 & 37102 & -783973 \\
			$M_{59}$ & MC    & $p_{4}$ & Joint & 1M    & 6000  & -820946 & 33560 & -787386 \\
			$M_{60}$ & MC    & $p_{1}$ & Joint & 1M    & 10000 & -820925 & 37706 & -783219 \\
			$M_{61}$ & MC    & $p_{5}$ & Joint & 3M    & 5000  & -820658 & 35933 & -784725 \\
			$M_{62}$ & MC    & $p_{5}$ & Joint & 1M    & 9000  & -820615 & 37509 & -783107 \\
			$M_{63}$ & MC    & $p_{5}$ & Std.  & 3M    & 5000  & -820594 & 35952 & -784642 \\
			$M_{64}$ & MC    & $p_{1}$ & Joint & 1M    & 4000  & -820549 & 36990 & -783559 \\
			$M_{65}$ & MC    & $p_{5}$ & Joint & 1M    & 18000 & -820345 & 36915 & -783430 \\
			$M_{66}$ & MC    & $p_{5}$ & Joint & 1M    & 12000 & -820138 & 37762 & -782376 \\
			$M_{67}$ & MC    & $p$ & Joint & 1M    & 8000  & -819809 & 36733 & -783075 \\
			$M_{68}$ & MC    & $p_{5}$ & Joint & 1M    & 16000 & -819793 & 36986 & -782807 \\
			$M_{69}$ & MC    & $p_{6}$ & Joint & 1D    & 5000  & -819754 & 33865 & -785889 \\
			$M_{70}$ & MC    & $p_{1}$ & Joint & 6M    & 5000  & -819705 & 36818 & -782887 \\
			$M_{71}$ & MC    & $p$ & Joint & 1D    & 5000  & -819616 & 36367 & -783249 \\
			$M_{72}$ & MC    & $p_{6}$ & Joint & 1M    & 8000  & -819600 & 34404 & -785197 \\
			$M_{73}$ & MC    & $p_{5}$ & Joint & 1M    & 11000 & -819593 & 37114 & -782478 \\
			$M_{74}$ & MC    & $p_{5}$ & Joint & 1M    & 4000  & -819572 & 36293 & -783280 \\
			$M_{75}$ & MC    & $p_{1}$ & Std.  & 6M    & 5000  & -819568 & 36850 & -782718 \\
			$M_{76}$ & MC    & $p_{5}$ & Joint & 1M    & 17000 & -819508 & 36946 & -782563 \\
			$M_{77}$ & MC    & $p_{6}$ & Joint & 1M    & 14000 & -819309 & 35411 & -783897 \\
			$M_{78}$ & MC    & $p$ & Joint & 1M    & 14000 & -819182 & 37741 & -781441 \\
			$M_{79}$ & MC    & $p_{5}$ & Joint & 1M    & 10000 & -819155 & 37303 & -781852 \\
			$M_{80}$ & MC    & $p_{5}$ & Joint & 6M    & 5000  & -819145 & 36069 & -783076 \\
			$M_{81}$ & MC    & $p_{5}$ & Std.  & 6M    & 5000  & -819004 & 36097 & -782907 \\
			$M_{82}$ & MC    & $p_{6}$ & Joint & 1M    & 15000 & -818616 & 35308 & -783307 \\
			$M_{83}$ & MC    & $p_{4}$ & Joint & 1D    & 5000  & -818613 & 33815 & -784797 \\
			$M_{84}$ & MC    & $p_{6}$ & Joint & 1M    & 13000 & -818575 & 35642 & -782933 \\
			$M_{85}$ & MC    & $p$ & Joint & 1M    & 15000 & -818526 & 37677 & -780849 \\
			$M_{86}$ & MC    & $p_{4}$ & Joint & 1M    & 8000  & -818497 & 34343 & -784154 \\
			$M_{87}$ & MC    & $p_{6}$ & Joint & 1M    & 5000  & -818475 & 33896 & -784578 \\
			$M_{88}$ & MC    & $p$ & Joint & 1M    & 13000 & -818459 & 37938 & -780521 \\
			$M_{89}$ & MC    & $p_{6}$ & Std.  & 1M    & 5000  & -818457 & 33901 & -784556 \\
			$M_{90}$ & MC    & $p$ & Joint & 1M    & 20000 & -818405 & 37577 & -780828 \\
			$M$ & MC    & $p$ & Joint & 1M    & 5000  & -818308* & 36418 & -781890 \\
			$M_{91}$ & MC    & $p$ & Std.  & 1M    & 5000  & -818293 & 36426 & -781867 \\
			$M_{92}$ & MC    & $p_{6}$ & Joint & 1M    & 20000 & -818260 & 35355 & -782905 \\
			$M_{93}$ & MC    & $p_{4}$ & Joint & 1M    & 14000 & -818123 & 35416 & -782707 \\
			$M_{94}$ & MC    & $p$ & Joint & 1M    & 19000 & -818061 & 37390 & -780671 \\
			$M_{95}$ & MC    & $p_{6}$ & Joint & 1M    & 19000 & -818030 & 35229 & -782801 \\
			$M_{96}$ & MC    & $p$ & Joint & 1M    & 9000  & -817778 & 37790 & -779988 \\
			$M_{97}$ & MC    & $p_{6}$ & Joint & 3M    & 5000  & -817586 & 33997 & -783589 \\
			$M_{98}$ & MC    & $p_{6}$ & Std.  & 3M    & 5000  & -817519 & 34012 & -783508 \\
			$M_{99}$ & MC    & $p_{6}$ & Joint & 1M    & 9000  & -817423 & 35444 & -781979 \\
			$M_{100}$ & MC    & $p_{4}$ & Joint & 1M    & 15000 & -817408 & 35304 & -782103 \\
			$M_{101}$ & MC    & $p_{4}$ & Joint & 1M    & 13000 & -817403 & 35653 & -781750 \\
			$M_{102}$ & MC    & $p$ & Joint & 3M    & 5000  & -817396 & 36544 & -780851 \\
			$M_{103}$ & MC    & $p$ & Joint & 1M    & 18000 & -817336 & 37215 & -780121 \\
			$M_{104}$ & MC    & $p$ & Std.  & 3M    & 5000  & -817335 & 36565 & -780770 \\
			$M_{105}$ & MC    & $p_{4}$ & Joint & 1M    & 5000  & -817334 & 33847 & -783487 \\
			$M_{106}$ & MC    & $p_{6}$ & Joint & 1M    & 18000 & -817333 & 34928 & -782406 \\
			$M_{107}$ & MC    & $p_{4}$ & Std.  & 1M    & 5000  & -817317 & 33852 & -783465 \\
			$M_{108}$ & MC    & $p$ & Joint & 1M    & 12000 & -817132 & 37870 & -779261 \\
			$M_{109}$ & MC    & $p_{4}$ & Joint & 1M    & 20000 & -817049 & 35360 & -781688 \\
			$M_{110}$ & MC    & $p_{6}$ & Joint & 1M    & 12000 & -817033 & 35624 & -781409 \\
			$M_{111}$ & MC    & $p_{6}$ & Joint & 1M    & 4000  & -816868 & 34090 & -782778 \\
			$M_{112}$ & MC    & $p$ & Joint & 1M    & 16000 & -816860 & 37294 & -779566 \\
			$M_{113}$ & MC    & $p_{6}$ & Joint & 1M    & 16000 & -816840 & 34927 & -781913 \\
			$M_{114}$ & MC    & $p_{4}$ & Joint & 1M    & 19000 & -816824 & 35240 & -781584 \\
			$M_{115}$ & MC    & $p$ & Joint & 1M    & 11000 & -816663 & 37339 & -779324 \\
			$M_{116}$ & MC    & $p_{6}$ & Joint & 1M    & 11000 & -816585 & 35087 & -781498 \\
			$M_{117}$ & MC    & $p_{6}$ & Joint & 1M    & 17000 & -816522 & 34935 & -781587 \\
			$M_{118}$ & MC    & $p$ & Joint & 1M    & 17000 & -816521 & 37229 & -779292 \\
			$M_{119}$ & MC    & $p_{4}$ & Joint & 3M    & 5000  & -816449 & 33949 & -782500 \\
			$M_{120}$ & MC    & $p_{4}$ & Std.  & 3M    & 5000  & -816383 & 33964 & -782419 \\
			$M_{121}$ & MC    & $p$ & Joint & 1M    & 10000 & -816317 & 37584 & -778733 \\
			$M_{122}$ & MC    & $p_{4}$ & Joint & 1M    & 9000  & -816299 & 35379 & -780921 \\
			$M_{123}$ & MC    & $p$ & Joint & 1M    & 4000  & -816207 & 36877 & -779330 \\
			$M_{124}$ & MC    & $p_{4}$ & Joint & 1M    & 18000 & -816137 & 34918 & -781218 \\
			$M_{125}$ & MC    & $p_{6}$ & Joint & 1M    & 10000 & -816030 & 35187 & -780844 \\
			$M_{126}$ & MC    & $p_{6}$ & Joint & 6M    & 5000  & -815994 & 34122 & -781872 \\
			$M_{127}$ & MC    & $p_{4}$ & Joint & 1M    & 12000 & -815880 & 35643 & -780237 \\
			$M_{128}$ & MC    & $p_{6}$ & Std.  & 6M    & 5000  & -815853 & 34145 & -781707 \\
			$M_{129}$ & MC    & $p_{4}$ & Joint & 1M    & 4000  & -815783 & 34045 & -781738 \\
			$M_{130}$ & MC    & $p$ & Joint & 6M    & 5000  & -815745 & 36705 & -779040 \\
			$M_{131}$ & MC    & $p_{4}$ & Joint & 1M    & 16000 & -815630 & 34919 & -780711 \\
			$M_{132}$ & MC    & $p$ & Std.  & 6M    & 5000  & -815610 & 36737 & -778873 \\
			$M_{133}$ & MC    & $p_{4}$ & Joint & 1M    & 11000 & -815447 & 35022 & -780425 \\
			$M_{134}$ & MC    & $p_{4}$ & Joint & 1M    & 17000 & -815323 & 34923 & -780401 \\
			$M_{135}$ & MC    & $p_{3}$ & Joint & 1M    & 7000  & -815304 & 34288 & -781015 \\
			$M_{136}$ & MC    & $p_{4}$ & Joint & 1M    & 10000 & -814914 & 35123 & -779791 \\
			$M_{137}$ & MC    & $p_{4}$ & Joint & 6M    & 5000  & -814855 & 34075 & -780780 \\
			$M_{138}$ & MC    & $p_{4}$ & Std.  & 6M    & 5000  & -814715 & 34098 & -780617 \\
			$M_{139}$ & MC    & $p_{2}$ & Joint & 1M    & 3000  & -814475 & 38609 & -775866 \\
			$M_{140}$ & MC    & $p_{3}$ & Joint & 1M    & 6000  & -813366 & 34098 & -779268 \\
			$M_{141}$ & MC    & $p_{1}$ & Joint & 1M    & 3000  & -813215 & 37596 & -775619 \\
			$M_{142}$ & MC    & $p_{5}$ & Joint & 1M    & 3000  & -812118 & 36649 & -775470 \\
			$M_{143}$ & MC    & $p_{3}$ & Joint & 1D    & 5000  & -811111 & 34306 & -776804 \\
			$M_{144}$ & MC    & $p_{3}$ & Joint & 1M    & 8000  & -811082 & 34735 & -776347 \\
			$M_{145}$ & MC    & $p_{3}$ & Joint & 1M    & 14000 & -810799 & 35936 & -774863 \\
			$M_{146}$ & MC    & $p_{3}$ & Joint & 1M    & 15000 & -810107 & 35850 & -774257 \\
			$M_{147}$ & MC    & $p_{3}$ & Joint & 1M    & 13000 & -810069 & 36170 & -773900 \\
			$M_{148}$ & MC    & $p_{3}$ & Joint & 1M    & 20000 & -809851 & 35866 & -773985 \\
			$M_{149}$ & MC    & $p_{3}$ & Joint & 1M    & 5000  & -809796 & 34339 & -775457 \\
			$M_{150}$ & MC    & $p_{3}$ & Std.  & 1M    & 5000  & -809779 & 34345 & -775434 \\
			$M_{151}$ & MC    & $p_{6}$ & Joint & 1M    & 3000  & -809690 & 34801 & -774889 \\
			$M_{152}$ & MC    & $p_{3}$ & Joint & 1M    & 19000 & -809608 & 35731 & -773877 \\
			$M_{153}$ & MC    & $p_{3}$ & Joint & 1M    & 9000  & -808960 & 35830 & -773129 \\
			$M_{154}$ & MC    & $p_{3}$ & Joint & 1M    & 18000 & -808883 & 35433 & -773451 \\
			$M_{155}$ & MC    & $p$ & Joint & 1M    & 3000  & -808845 & 37481 & -771364 \\
			$M_{156}$ & MC    & $p_{3}$ & Joint & 3M    & 5000  & -808817 & 34443 & -774373 \\
			$M_{157}$ & MC    & $p_{3}$ & Std.  & 3M    & 5000  & -808750 & 34459 & -774291 \\
			$M_{158}$ & MC    & $p_{4}$ & Joint & 1M    & 3000  & -808727 & 34760 & -773967 \\
			$M_{159}$ & MC    & $p_{3}$ & Joint & 1M    & 12000 & -808519 & 36137 & -772382 \\
			$M_{160}$ & MC    & $p_{3}$ & Joint & 1M    & 16000 & -808350 & 35475 & -772875 \\
			$M_{161}$ & MC    & $p_{3}$ & Joint & 1M    & 4000  & -808084 & 34658 & -773426 \\
			$M_{162}$ & MC    & $p_{3}$ & Joint & 1M    & 17000 & -808062 & 35490 & -772572 \\
			$M_{163}$ & MC    & $p_{3}$ & Joint & 1M    & 11000 & -808044 & 35484 & -772560 \\
			$M_{164}$ & MC    & $p_{2}$ & Joint & 1M    & 2000  & -807665 & 39059 & -768606 \\
			$M_{165}$ & MC    & $p_{3}$ & Joint & 1M    & 10000 & -807608 & 35556 & -772052 \\
			$M_{166}$ & MC    & $p_{3}$ & Joint & 6M    & 5000  & -807093 & 34568 & -772526 \\
			$M_{167}$ & MC    & $p_{3}$ & Std.  & 6M    & 5000  & -806950 & 34592 & -772359 \\
			$M_{168}$ & MC    & $p_{1}$ & Joint & 1M    & 2000  & -806693 & 38005 & -768688 \\
			$M_{169}$ & MC    & $p_{5}$ & Joint & 1M    & 2000  & -805282 & 37647 & -767635 \\
			$M_{170}$ & MC    & $p_{6}$ & Joint & 1M    & 2000  & -803347 & 35619 & -767728 \\
			$M_{171}$ & MC    & $p_{4}$ & Joint & 1M    & 2000  & -802571 & 35597 & -766974 \\
			$M_{172}$ & MC    & $p$ & Joint & 1M    & 2000  & -802335 & 37890 & -764445 \\
			$M_{173}$ & MC    & $p_{3}$ & Joint & 1M    & 3000  & -800851 & 35316 & -765534 \\
			$M_{174}$ & MC    & $p_{2}$ & Joint & 1M    & 1000  & -796135 & 43262 & -752873 \\
			$M_{175}$ & MC    & $p_{1}$ & Joint & 1M    & 1000  & -795153 & 42120 & -753033 \\
			$M_{176}$ & MC    & $p_{3}$ & Joint & 1M    & 2000  & -794334 & 35981 & -758353 \\
			$M_{177}$ & MC    & $p_{5}$ & Joint & 1M    & 1000  & -793878 & 41184 & -752694 \\
			$M_{178}$ & MC    & $p_{6}$ & Joint & 1M    & 1000  & -791351 & 39532 & -751819 \\
			$M_{179}$ & MC    & $p_{4}$ & Joint & 1M    & 1000  & -791203 & 39558 & -751645 \\
			$M_{180}$ & MC    & $p$ & Joint & 1M    & 1000  & -790761 & 41999 & -748761 \\
			$M_{181}$ & MC    & $p_{3}$ & Joint & 1M    & 1000  & -783083 & 39570 & -743513 \\
			$M_{182}$ & MC    & $p_{2}$ & Joint & 12M   & 5000  & -781668 & 42427 & -739241 \\
			$M_{183}$ & MC    & $p_{5}$ & Joint & 12M   & 5000  & -779825 & 40692 & -739134 \\
			$M_{184}$ & MC    & $p_{1}$ & Joint & 12M   & 5000  & -779456 & 41485 & -737971 \\
			$M_{185}$ & MC    & $p_{2}$ & Std.  & 12M   & 5000  & -778402 & 43162 & -735240 \\
			$M_{186}$ & MC    & $p_{6}$ & Joint & 12M   & 5000  & -776537 & 38731 & -737806 \\
			$M_{187}$ & MC    & $p_{5}$ & Std.  & 12M   & 5000  & -776467 & 41329 & -735138 \\
			$M_{188}$ & MC    & $p_{1}$ & Std.  & 12M   & 5000  & -776173 & 42178 & -733995 \\
			$M_{189}$ & MC    & $p$ & Joint & 12M   & 5000  & -775964 & 41357 & -734606 \\
			$M_{190}$ & MC    & $p_{4}$ & Joint & 12M   & 5000  & -775356 & 38685 & -736671 \\
			$M_{191}$ & MC    & $p_{6}$ & Std.  & 12M   & 5000  & -773149 & 39248 & -733901 \\
			$M_{192}$ & MC    & $p$ & Std.  & 12M   & 5000  & -772699 & 42047 & -730651 \\
			$M_{193}$ & MC    & $p_{4}$ & Std.  & 12M   & 5000  & -771981 & 39202 & -732778 \\
			$M_{194}$ & MC    & $p_{3}$ & Joint & 12M   & 5000  & -767683 & 39235 & -728448 \\
			$M_{195}$ & MC    & $p_{3}$ & Std.  & 12M   & 5000  & -764292 & 39784 & -724508 \\
			
		\end{longtable}
	\end{small}
\end{center}

\section{Market Data}
\label{app:Market Data}

In this appendix we report the full market data set used for all the numerical calculations reported in this paper. All the data refer to end of day 28 December 2018, which is also the valuation date.

\subsection{Interest Rate and Credit Curves}

\begin{table}[h]
  \centering
    \begin{tabular}{r c  r c  r c  r c  r c}
    \toprule
    {$t$(d)}     & {$P(0;t)$}   & {$t$(d)}     & {$P(0;t)$}   & {$t$(d)}     & {$P(0;t)$}   &{$t$(d)}     & {$P(0;t)$}   & {$t$(d)}     & {$P(0;t)$} \\
    \midrule
    0     & 1.0000 & 311   & 1.0020 & 738   & 1.0036 & 5484  & 0.8355 & 10597 & 0.6644 \\
    6     & 1.0000 & 339   & 1.0022 & 917   & 1.0033 & 5849  & 0.8195 & 10965 & 0.6551 \\
    12    & 1.0001 & 370   & 1.0024 & 1102  & 1.0022 & 6214  & 0.8039 & 11329 & 0.6461 \\
    19    & 1.0001 & 402   & 1.0025 & 1466  & 0.9977 & 6580  & 0.7890 & 11693 & 0.6373 \\
    26    & 1.0002 & 430   & 1.0027 & 1831  & 0.9902 & 6947  & 0.7747 & 12058 & 0.6287 \\
    38    & 1.0003 & 461   & 1.0028 & 2197  & 0.9800 & 7311  & 0.7612 & 12424 & 0.6201 \\
    66    & 1.0005 & 493   & 1.0030 & 2562  & 0.9675 & 7675  & 0.7484 & 12789 & 0.6117 \\
    95    & 1.0006 & 522   & 1.0031 & 2929  & 0.9531 & 8041  & 0.7361 & 13156 & 0.6034 \\
    125   & 1.0008 & 552   & 1.0032 & 3293  & 0.9374 & 8406  & 0.7246 & 13520 & 0.5953 \\
    157   & 1.0010 & 584   & 1.0033 & 3658  & 0.9208 & 8771  & 0.7136 & 13885 & 0.5873 \\
    186   & 1.0012 & 614   & 1.0034 & 4023  & 0.9037 & 9138  & 0.7031 & 14250 & 0.5796 \\
    217   & 1.0014 & 644   & 1.0035 & 4388  & 0.8863 & 9502  & 0.6932 & 14615 & 0.5721 \\
    248   & 1.0016 & 675   & 1.0035 & 4753  & 0.8690 & 9867  & 0.6834 & 18268 & 0.5046 \\
    278   & 1.0018 & 705   & 1.0036 & 5120  & 0.8520 & 10232 & 0.6738 & 21920 & 0.4461 \\
    \bottomrule
    \end{tabular}
    \caption{EURIBOR 6M Zero Coupon Bond curve $\mathcal{C}_{6M}(t)$ built using IR instruments on EURIBOR 6M quoted on the market at valuation date $t=$ 28 December 2018. Times are measured in days since the valuation date. This curve is used to compute EURIBOR 6M forward rates.}
  \label{tab:EURIBOR6M_ZCB}
\end{table}
\begin{table}[H]
  \centering
    \begin{tabular}{r c  r c  r c  r c  r c}
    \toprule
    {$t$(d)}     & {$P(0;t)$}   & {$t$(d)}     & {$P(0;t)$}   & {$t$(d)}     & {$P(0;t)$}   &{$t$(d)}     & {$P(0;t)$}   & {$t$(d)}     & {$P(0;t)$} \\
    \midrule
    0     & 1.0000 & 248   & 1.0025 & 675   & 1.0059 & 5120  & 0.8689 & 10232 & 0.6942 \\
    3     & 1.0000 & 278   & 1.0028 & 705   & 1.0061 & 5484  & 0.8530 & 10597 & 0.6850 \\
    5     & 1.0001 & 311   & 1.0031 & 738   & 1.0062 & 5849  & 0.8375 & 10965 & 0.6758 \\
    6     & 1.0001 & 339   & 1.0033 & 1102  & 1.0063 & 6214  & 0.8224 & 12789 & 0.6325 \\
    12    & 1.0001 & 370   & 1.0036 & 1466  & 1.0033 & 6580  & 0.8079 & 14615 & 0.5930 \\
    19    & 1.0002 & 402   & 1.0039 & 1831  & 0.9975 & 6947  & 0.7939 & 18268 & 0.5250 \\
    26    & 1.0003 & 430   & 1.0042 & 2197  & 0.9888 & 7311  & 0.7807 & 21920 & 0.4655 \\
    38    & 1.0004 & 461   & 1.0044 & 2562  & 0.9778 & 7675  & 0.7682 &       &  \\
    66    & 1.0007 & 493   & 1.0047 & 2929  & 0.9648 & 8041  & 0.7562 &       &  \\
    95    & 1.0009 & 522   & 1.0049 & 3293  & 0.9501 & 8406  & 0.7448 &       &  \\
    125   & 1.0012 & 552   & 1.0052 & 3658  & 0.9346 & 8771  & 0.7340 &       &  \\
    157   & 1.0016 & 584   & 1.0054 & 4023  & 0.9183 & 9138  & 0.7235 &       &  \\
    186   & 1.0019 & 614   & 1.0056 & 4388  & 0.9018 & 9502  & 0.7135 &       &  \\
    217   & 1.0022 & 644   & 1.0058 & 4753  & 0.8853 & 9867  & 0.7037 &       &  \\
    \bottomrule
    \end{tabular}
    \caption{EONIA Zero Coupon Bond term structure $\mathcal{C}(t)$ built using EONIA Overnight Index Swaps (OIS) quoted on the market at valuation date $t=$ 28 December 2018. Times are measured in days since the valuation date. This curve is used to compute discount factors for future cash flows.}
  \label{tab:OIS_ZCB}
\end{table}

\begin{table}[H]
	\centering
	\begin{tabular}{r c c}
		\toprule
		{$t$ (d)} & bank & counterparty \\
		\midrule
		180   & 91    & 24 \\
		360   & 105   & 29 \\
		720   & 125   & 48 \\
		1080  & 146   & 72 \\
		1440  & 163   & 99 \\
		1800  & 181   & 126 \\
		2520  & 200   & 159 \\
		3600  & 216   & 183 \\
		5400  & 219   & 195 \\
		7200  & 222   & 202 \\
		10800 & 227   & 213 \\
		\bottomrule
	\end{tabular}%
	\caption{CDS spread term structure for the bank and the counterparty (in bps). This curve is used to compute default and survival probabilities for XVA calculations.}
	\label{tab:CDS spread term structure}%
\end{table}
 
\subsection{Swaption Cube}

\begin{table}[h]
  \centering
    \begin{tabular}{c|ccccccccccccc}
    \toprule
$\xi$/$\mathfrak{T}$ & 2     & 3     & 4     & 5     & 6     & 7     & 8     & 9     & 10    & 15    & 20    & 25    & 30 \\  
    \midrule
    2     & 95    & 155   & 216   & 273   & 331   & 386   & 437   & 488   & 537   & 746   & 933   & 1108  & 1270 \\
    3     & 143   & 222   & 297   & 373   & 447   & 518   & 585   & 651   & 713   & 975   & 1221  & 1445  & 1653 \\
    4     & 185   & 279   & 371   & 460   & 548   & 632   & 715   & 793   & 865   & 1170  & 1459  & 1723  & 1976 \\
    5     & 221   & 328   & 432   & 534   & 634   & 731   & 824   & 913   & 996   & 1338  & 1666  & 1964  & 2250 \\
    6     & 248   & 367   & 481   & 594   & 705   & 810   & 912   & 1008  & 1101  & 1472  & 1832  & 2156  & 2471 \\
    7     & 271   & 399   & 523   & 644   & 763   & 876   & 986   & 1095  & 1200  & 1598  & 1980  & 2331  & 2674 \\
    8     & 290   & 427   & 560   & 690   & 818   & 940   & 1058  & 1171  & 1281  & 1705  & 2106  & 2487  & 2853 \\
    9     & 306   & 451   & 591   & 727   & 861   & 992   & 1116  & 1236  & 1355  & 1801  & 2227  & 2623  & 3007 \\
    10    & 320   & 473   & 620   & 765   & 904   & 1041  & 1170  & 1297  & 1427  & 1902  & 2344  & 2755  & 3164 \\
    12    & 342   & 503   & 663   & 816   & 967   & 1116  & 1257  & 1397  & 1528  & 2042  & 2516  & 2973  & 3390 \\
    15    & 368   & 544   & 713   & 879   & 1042  & 1198  & 1351  & 1505  & 1658  & 2225  & 2748  & 3230  & 3678 \\
    20    & 395   & 587   & 768   & 950   & 1131  & 1299  & 1469  & 1635  & 1794  & 2434  & 2961  & 3467  & 3927 \\
    25    & 417   & 620   & 812   & 1003  & 1188  & 1365  & 1537  & 1711  & 1878  & 2556  & 3086  & 3606  & 4082 \\
    30    & 432   & 640   & 842   & 1042  & 1229  & 1407  & 1571  & 1737  & 1906  & 2601  & 3136  & 3667  & 4164 \\
	\bottomrule
    \end{tabular}
\caption{EUR ATM Swaption Straddle prices with expiries ($\xi$) and tenors ($\mathfrak{T}$) on rows and columns respectively (in years), quoted on the market at the valuation date 28 December 2018. The nominal is 10000 EUR. Swaption Straddles consist of a long payer Swaption plus a long receiver Swaption with the same expiry, tenor and ATM strike. The payoff of these instruments is ``physically settled LCH", meaning that, upon exercise, the option's holder receives the underlying Swap which is, in turn, cleared at the London Clearing House. Since for these instruments the call-put parity relationship holds, i.e. the price of the ATM Swaption is equal to half the price of the corresponding Straddle.}
  \label{tab:mkt_swpt_prices}%
\end{table}%
\begin{table}[h]
  \centering
  \resizebox{\linewidth}{!}{
    \begin{tabular}{cc|ccccccccccc}
    \toprule
    $\xi$ & $\mathfrak{T}$ & $-0.02$ & $-0.015$ & $-0.01$ & $-0.005$ & $-0.0025$ & $0$ & $+0.0025$ & $+0.005$ & $+0.01$ & $+0.015$ & $+0.02$ \\
    \midrule
    2     & 2     & 0     & 0     & 0     & 9     & 23   & 47 & 30    & 18    & 7     & 3     & 1 \\
    2     & 5     & 0     & 0     & 6     & 39    & 78   & 137 & 89    & 57    & 22    & 9     & 3 \\
    2     & 10    & 0     & 2     & 17    & 84    & 158 & 269 & 175   & 111   & 44    & 18    & 8 \\
    2     & 20    & 1     & 5     & 32    & 147   & 274  & 467 & 286   & 167   & 52    & 16    & 5 \\
    2     & 30    & 1     & 6     & 42    & 197   & 370  & 635 & 381   & 215   & 59    & 15    & 4 \\
    5     & 2     & 4     & 12    & 30    & 62    & 84   & 111 & 90    & 74    & 49    & 33    & 22 \\
    5     & 5     & 12    & 32    & 77    & 153   & 205 &  267 & 216   & 174   & 112   & 71    & 45 \\
    5     & 10    & 20    & 58    & 139   & 282   & 381 &  498 & 401   & 321   & 203   & 127   & 79 \\
    5     & 20    & 30    & 89    & 219   & 457   & 627  & 833 & 649   & 500   & 288   & 162   & 91 \\
    5     & 30    & 33    & 105   & 278   & 603   & 838  & 1125 & 869   & 662   & 372   & 204   & 111 \\
    10    & 2     & 26    & 45    & 73    & 111   & 135  & 160 & 140   & 122   & 92    & 69    & 52 \\
    10    & 5     & 57    & 104   & 172   & 265   & 320 &  383& 332   & 288   & 215   & 159   & 118 \\
    10    & 10    & 106   & 193   & 320   & 492   & 597 &  714 & 615   & 529   & 387   & 280   & 202 \\
    10    & 20    & 120   & 253   & 466   & 770   & 959  & 1172 & 991   & 834   & 583   & 404   & 279 \\
    10    & 30    & 138   & 305   & 593   & 1015  & 1280 & 1582 & 1333  & 1117  & 777   & 536   & 370 \\
    15    & 2     & 40    & 63    & 94    & 135   & 158  & 184 & 163   & 144   & 112   & 87    & 68 \\
    15    & 5     & 87    & 144   & 221   & 319   & 377  & 440 & 389   & 344   & 267   & 206   & 159 \\
    15    & 10    & 163   & 270   & 415   & 601   & 710  & 829 & 731   & 643   & 495   & 379   & 289 \\
    15    & 20    & 191   & 358   & 613   & 950   & 1151 & 1374 & 1194  & 1034  & 771   & 573   & 426 \\
    15    & 30    & 209   & 417   & 763   & 1235  & 1520 & 1839 & 1592  & 1375  & 1020  & 756   & 561 \\
    20    & 2     & 50    & 75    & 108   & 148   & 172  & 198 & 176   & 157   & 125   & 98    & 78 \\
    20    & 5     & 109   & 169   & 251   & 353   & 411  & 475 & 424   & 379   & 300   & 238   & 189 \\
    20    & 10    & 194   & 305   & 463   & 660   & 773  & 897 & 800   & 712   & 563   & 444   & 351 \\
    20    & 20    & 235   & 410   & 687   & 1041  & 1250 & 1481 & 1300  & 1139  & 872   & 668   & 514 \\
    20    & 30    & 250   & 467   & 838   & 1335  & 1633 & 1964 & 1716  & 1498  & 1141  & 871   & 670 \\
    30    & 2     & 65    & 92    & 126   & 167   & 191  & 216 & 194   & 175   & 141   & 113   & 91 \\
    30    & 5     & 139   & 201   & 289   & 396   & 456  & 521 & 471   & 425   & 345   & 281   & 229 \\
    30    & 10    & 219   & 332   & 498   & 705   & 824  & 953 & 859   & 774   & 630   & 515   & 424 \\
    30    & 20    & 290   & 470   & 755   & 1120  & 1333 & 1568 & 1386  & 1225  & 958   & 754   & 601 \\
    30    & 30    & 310   & 537   & 920   & 1434  & 1741 & 2082 & 1834  & 1617  & 1262  & 994   & 792 \\
	\midrule
    \multicolumn{2}{c|}{$\omega$}  & $-1$ & $-1$ & $-1$ &$-1$ &$-1$ &$\pm1$ & $+1$ & $+1$ & $+1$ & $+1$ & $+1$\\
	\bottomrule   
    \end{tabular}}
    \caption{EUR ATM Swaption price cube (physically settled LCH, EUR 10000 nominal amount) quoted on the market at the valuation date 28 December 2018. The first column shows expiries ($\xi$) and tenors ($\mathfrak{T}$) in years. The central column shows ATM Swaption prices obtained by dividing by two the corresponding Straddle prices shown in tab.~\ref{tab:mkt_swpt_prices}. The first row shows strikes as shifts w.r.t. ATM strike rates. The following rows show prices of receiver (left-hand side of the ATM) and payer (right-hand side of the ATM) Swaptions. The last row shows $\omega = -1$ for receiver and $\omega = +1$ for payer Swaptions, respectively.}
  \label{tab:mkt_swpt_prices_cube}%
\end{table}

\end{appendices}

\end{document}